\documentclass[prd,aps,
longbibliography,
preprintnumbers,
nofootinbib,
floatfix,
superscriptaddress]{revtex4}
\usepackage{graphicx}
\usepackage{epsfig}
\usepackage{rotating}
\usepackage{amssymb}
\usepackage{dsfont}
\usepackage{psfrag}
\usepackage{amsmath,euscript,array,mathrsfs}
\usepackage{bbold}
\usepackage{bm}
\usepackage{multirow}
\usepackage{dcolumn}
\RequirePackage{doi}
\usepackage[skip=10pt plus1pt, indent=20pt]{parskip}
\usepackage{float}
\floatstyle{plaintop}
\usepackage{bbding}
\usepackage{xcolor}
\usepackage{orcidlink}
\usepackage{hyperref}
\usepackage{tikz}
\usepackage{longtable}
\usepackage{supertabular}
\usepackage[normalem]{ulem}

\newcommand{\FveLargestMass}{9}
\newcommand{\FveSmallestMass}{5}

\newcommand{\ChiSmallestBeta}{1.6}

\newcommand{\NuSmallestBeta}{1.3}

\newcommand{\beq}{\begin{equation}}
\newcommand{\eeq}{\end{equation}}
\newcommand{\beqs}{\begin{eqnarray}}
\newcommand{\eeqs}{\end{eqnarray}}
\newcommand{\Tr}{{\rm Tr}}

\newcommand{\gsim}{\mathrel{\raisebox{-
.6ex}{$\stackrel{\textstyle>}{\sim}$}}}

\newcommand{\be}{\begin{equation}}
\newcommand{\ee}{\end{equation}}
\newcommand{\ba}{\begin{array}}
\newcommand{\ea}{\end{array}}

\newcommand{\orcidauthorBENNETT}{0000-0002-1678-6701}
\newcommand{\orcidauthorBOYLE}{0000-0002-8960-1587}
\newcommand{\orcidauthorLUCINI}{0000-0001-8974-8266}
\newcommand{\orcidauthorPIAI}{0000-0002-2251-0111} 
\newcommand{\orcidauthorFORZANO}{0000-0003-0985-8858}
\newcommand{\orcidauthorVADACCHINO}{0000-0002-5783-5602}
\newcommand{\orcidauthorHILL}{0000-0003-2383-940X}
\newcommand{\orcidauthorHONG}{0000-0002-3923-4184}
\newcommand{\orcidauthorDELDEBBIO}{0000-0003-4246-3305}
\newcommand{\orcidauthorLIN}{0000-0003-3743-0840}
\newcommand{\orcidauthorLEE}{0000-0002-4616-2422}
\newcommand{\orcidauthorPROVATAS}{0000-0002-3132-9621}
\newcommand{\orcidauthorSIMONETTI}{0009-0002-3921-2687}

\makeatletter
\def\@collaboration@present#1#2#3{%
 \par
 \begingroup
  \frontmatter@collaboration@above
  \@author@present{\ignorespaces#2\unskip}{#3}%
  \par
 \endgroup
 \set@listcomma@list#1%
}%
\makeatother

\begin{document}

\author{Ed Bennett\,\orcidlink{\orcidauthorBENNETT}}
\email{E.J.Bennett@swansea.ac.uk}
\affiliation{Swansea Academy of Advanced Computing, Swansea University (Bay Campus), Fabian Way, Swansea SA1 8EN, United Kingdom}
\affiliation{Centre for Quantum Fields and Gravity, Faculty  of Science and Engineering, Swansea University, Singleton Park, SA2 8PP, Swansea, United Kingdom}

\author{Peter A. Boyle\,\orcidlink{\orcidauthorBOYLE}}
\email{pboyle@bnl.gov}
\affiliation{School of Physics and Astronomy, The University of Edinburgh, Peter Guthrie Tait Road, Edinburgh EH9 3FD, United Kingdom}
\affiliation{Physics Department, Brookhaven National Laboratory, Upton, NY 11973, USA}

\author{Luigi Del Debbio\,\orcidlink{\orcidauthorDELDEBBIO}}
\email{luigi.del.debbio@ed.ac.uk}
\affiliation{School of Physics and Astronomy, 
The University of Edinburgh, Peter Guthrie Tait Road, Edinburgh EH9 3FD, United Kingdom}

\author{Niccolò Forzano\,\orcidlink{\orcidauthorFORZANO}}
\email{ niccolo.forzano@swansea.ac.uk}
\affiliation{Department of Physics, Faculty of Science and Engineering, Swansea University, Singleton Park, SA2 8PP, Swansea, United Kingdom}

\author{Ryan~C. Hill\,\orcidlink{\orcidauthorHILL}}
\email{ryan.hill@ed.ac.uk}
\affiliation{School of Physics and Astronomy, The University of Edinburgh, Peter Guthrie Tait Road, Edinburgh EH9 3FD, United Kingdom}

\author{Deog~Ki Hong\,\orcidlink{\orcidauthorHONG}}
\email{dkhong@pusan.ac.kr}
\affiliation{Department of Physics, Pusan National University, Busan 46241, Korea}
\affiliation{Extreme Physics Institute, Pusan National University, Busan 46241, Korea}

\author{Jong-Wan Lee\,\orcidlink{\orcidauthorLEE}}
\email{j.w.lee@ibs.re.kr}
\affiliation{Particle Theory  and Cosmology Group, Center for Theoretical Physics of the Universe, Institute for Basic Science (IBS), Daejeon, 34126, Korea }

\author{C.-J. David Lin\,\orcidlink{\orcidauthorLIN}}
\email{dlin@nycu.edu.tw}
\affiliation{Institute of Physics, National Yang Ming Chiao Tung University, 1001 Ta-Hsueh Road, Hsinchu 30010, Taiwan}
\affiliation{Centre for High Energy Physics, Chung-Yuan Christian University, Chung-Li 32023, Taiwan}
\affiliation{Physics Division, National Centre for Theoretical Sciences, Taipei 106319, Taiwan}

\author{Biagio Lucini\,\orcidlink{\orcidauthorLUCINI}}
\email{b.lucini@qmul.ac.uk}
\affiliation{Swansea Academy of Advanced Computing, Swansea University (Bay Campus), Fabian Way, Swansea SA1 8EN, United Kingdom}
\affiliation{Department of Mathematics, Faculty of Science and Engineering, Swansea University (Bay Campus), Fabian Way, SA1 8EN Swansea, United Kingdom}
\affiliation{School of Mathematical Sciences, Queen Mary University of London, Mile End Road,
London, E1 4NS, United Kingdom}

\author{Maurizio Piai\,\orcidlink{\orcidauthorPIAI}}
\email{m.piai@swansea.ac.uk}
\affiliation{Department of Physics, Faculty  of Science and Engineering, Swansea University, Singleton Park, SA2 8PP, Swansea, United Kingdom}
\affiliation{Centre for Quantum Fields and Gravity, Faculty  of Science and Engineering, Swansea University, Singleton Park, SA2 8PP, Swansea, United Kingdom}

\author{Gianmarco Simonetti\,\orcidlink{\orcidauthorSIMONETTI}}
\email{s2820577@ed.ac.uk}
\affiliation{Department of Physics, Faculty  of Science and Engineering, Swansea University, Singleton Park, SA2 8PP, Swansea, United Kingdom}
\affiliation{Centre for Quantum Fields and Gravity, Faculty  of Science and Engineering, Swansea University, Singleton Park, SA2 8PP, Swansea, United Kingdom}
\affiliation{School of Physics and Astronomy, The University of Edinburgh, Peter Guthrie Tait Road, Edinburgh EH9 3FD, United Kingdom}

\author{Davide Vadacchino\,\orcidlink{\orcidauthorVADACCHINO}}
\email{davide.vadacchino@plymouth.ac.uk}
\affiliation{Centre for Mathematical Sciences, University of Plymouth, Plymouth, PL4 8AA, United Kingdom}

\author{Alexis Verney-Provatas\,\orcidlink{\orcidauthorPROVATAS}}
\email{2414441@swansea.ac.uk}
\affiliation{Department of Physics, Faculty  of Science and Engineering, Swansea University, Singleton Park, SA2 8PP, Swansea, United Kingdom}
\affiliation{Centre for Quantum Fields and Gravity, Faculty  of Science and Engineering, Swansea University, Singleton Park, SA2 8PP, Swansea, United Kingdom}
\affiliation{School of Physics and Astronomy, The University of Edinburgh, Peter Guthrie Tait Road, Edinburgh EH9 3FD, United Kingdom}

\collaboration{(on behalf of the TELOS collaboration) \vspace{4pt}\\ 
\href{https://telos-collaboration.github.io}{ \includegraphics[height=1cm]{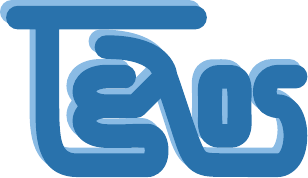} } }
\noaffiliation

\title{Symplectic lattice gauge theories in the Grid framework: domain wall fermions and continuum extrapolations}
\date{\today}

\begin{abstract}

We report the results of the first numerical lattice study using domain-wall fermions in the $Sp(4)$ gauge theory coupled to two flavours of (Dirac) fermions, transforming in the fundamental representation of the gauge group.  This theory plays a prominent role in the literature on extensions of the Standard Model with composite dynamics. It provides a short-distance completion for a class of composite Higgs models, or,  alternatively,  of dark matter models based on  the strongly interacting massive particle paradigm.   We adopt the M\"{o}bius formulation of domain-wall fermions (MDWF), implemented within the Grid software environment. We report the  results of extensive tests of the algorithm implementation, and of the optimisation of the choices of  algorithmic parameters appearing in the MDWF action. We then measure masses and decay constants of the lightest flavoured mesons in ensembles with moderately large fermion masses and several choices of lattice coupling, and perform an extrapolation to the continuum. We compare our results for the physical observables to published measurements obtained in the same field theory, but derived on the lattice by employing Wilson fermions. We demonstrate that, with the deployment of moderate computational resources,  the MDWF formulation can yield order-of-magnitude gains in the approach to the continuum limit, in regions of physical parameter space relevant to phenomenological applications of this theory.

\end{abstract}
\preprint{CTPU-PTC-26-20}
\preprint{PNUTP-26/A02}
\maketitle

\tableofcontents

\section{Introduction } 
\label{Sec:introduction}

The domain-wall formulations of fermions provide powerful tools for the numerical lattice study of non-perturbative properties of strongly coupled gauge theories. In particular, M\"{o}bius domain-wall fermions (MDWFs)~\cite{Brower:2004xi,Brower:2005qw}, supplemented by Pauli-Villars fields~\cite{Vranas:1997da,Vranas:2000tz,Berruto:2001ty}, are tantamount to a unified framework that can be implemented algorithmically~\cite{Brower:2012vk}. By tuning a set of algorithmic parameters, one can access the small-mass regime of the lattice theory\footnote{The Nielsen-Ninomiya theorem~\cite{Nielsen:1980rz,Nielsen:1981xu}  is circumvented by extending the lattice space-time with an additional dimension, and by introducing a non-trivial domain-wall operator~\cite{Kaplan:1992bt},  to exploit the existence of exactly massless Weyl fermions localised on the domain wall in the continuum theory (see also Ref.~\cite{Callan:1984sa}). Hence, one can  realise approximate chiral symmetry in finite lattice theories, without fermion doublers, by arranging for the two Weyl components of a light Dirac field to be localised near distinct slices of space, along the fifth dimension, which suppresses their overlap and the resulting residual Dirac mass. The additive renormalisation of the mass of Wilson fermions  in the extra dimension vanishes in the limit in which said dimension has infinitely many sites,  $N_5\rightarrow +\infty$~\cite{Aoki:1997mu}. The size of  symmetry breaking effects introduced by the finiteness of the extra dimension is measured by the residual mass,  $a m_{\rm res}$, and is independent of the mass of the physical fermions, in four dimensions (although estimates of the latter on the lattice are affected by the former). The underlying mechanism relies on the existence of the Ginsparg-Wilson relation~\cite{Ginsparg:1981bj} (see  Refs.~\cite{Hasenfratz:1997ft,Bietenholz:1995cy,Bietenholz:1996pf,Luscher:1998pqa} and the pedagogical discussion in Ref.~\cite{Luscher:2000hn}), which is also at the basis of the proof of the equivalence of several different proposals in the literature, such as those in Refs.~\cite{Shamir:1993zy,Furman:1994ky} and~\cite{Narayanan:1993zzh,Narayanan:1993sk,Narayanan:1993ss,Narayanan:1994gw,Neuberger:1997bg,Neuberger:1997fp,Borici:1999zw,Chiu:2002ir}. }
and approach the continuum limit.\footnote{Appropriate choices of the Pauli-Villars action remove heavy modes and enlarged symmetries introduced by the extra dimension, reinstating the short-distance behaviour and chiral symmetry pattern of the continuum theory of interest. Furthermore, dimension-5 operators (e.g., the Sheikholeslami-Wohlert term~\cite{Sheikholeslami:1985ij}) are forbidden in the $N_5\rightarrow +\infty$ limit (see, e.g., the discussion in Section III of Ref.~\cite{Bietenholz:1998ut}). Hence, residual lattice artefacts scale as ${\cal O}(a^2)$ with the lattice spacing, $a$, providing an automatic improvement of the fermion part of the lattice action. } The optimisation of the aforementioned tuning of parameters is critical to explore interesting regions of parameter space, as it allows to maximise the physics gain with available computational resources.\footnote{As discussed in Ref.~\cite{Brower:2012vk}, other types of domain-wall fermions have been applied to the study of finite-temperature quantum chromodynamics (QCD)~\cite{Cheng:2009be} and supersymmetry~\cite{Giedt:2008xm}. These studies required expensively large extra dimensions, with $N_5\sim {\cal O}(48)$, to achieve the desired precision.}

This lattice strategy has become increasingly relevant in the study of QCD~\cite{RBC:2014ntl,Bhattacharya:2014ara}. For example, it has been deployed to compute hadronic contributions to the anomalous magnetic moment of the muon~\cite{RBC:2018dos,Blum:2019ugy,RBC:2024fic} and to the hadronic decays of the kaons~\cite{RBC:2015gro,RBC:2020kdj}. A number of collaborations have adopted MDWFs to perform high-precision lattice-QCD  studies---see for instance Refs.~\cite{Colquhoun:2022atw,Zhang:2025vns,Gavai:2024mcj}, as well as the hybrid approach described in Refs.~\cite{Meyer:2021vfq,Chang:2017oll}.
The literature on domain-wall fermions in the context of extensions of the Standard Model (SM) of particle physics based on new strongly coupled dynamics is more limited  (see, e.g., the adoption of Shamir's domain wall treatment in Ref.~\cite{LSD:2014nmn}, and of MDWFs in Refs.~\cite{Hasenfratz:2017qyr,Hasenfratz:2019dpr,Hasenfratz:2020ess,Hasenfratz:2022zsa,Park:2025ggx,Ayyar:2026wht}). By exploiting recent algorithm, software, and hardware developments, we exemplify the practical benefits of the adoption of domain-wall fermions, and in particular their MDWF implementation, for theories of new physics, by applying it to the $Sp(4)$ gauge theory coupled to $N_{\rm f}=2$ Dirac fermions transforming as the fundamental representation of the group. We devote the rest of this introduction to motivating lattice studies of new strong dynamics, and to contextualising the choice of theory we made for this paper.

Despite the spectacular success of its predictions, including the discovery of the Higgs boson~\cite{ATLAS:2012yve,CMS:2012qbp}, it is generally accepted that the Standard Model of particle physics cannot be a complete theory of nature, valid at arbitrarily short distances,  for a combination of reasons, of both theoretical and phenomenological origin. Among the former, it does not include a theory of quantum gravity, and many of the SM couplings are not asymptotically safe at short length scales (see, e.g., Refs.~\cite{Aizenman:2019yuo,Luscher:1987ay,Luscher:1987ek, Luscher:1988uq,Bulava:2012rb, Molgaard:2014mqa, Chu:2018ldw}). Hence, it is natural to think of it as an effective field theory (EFT), valid only for energies up to a physical cut-off scale, $\Lambda$. Among the latter, the SM does not provide a compelling description for any of the three main components of the observable universe today, according to the Standard Model of Cosmology: dark energy, dark matter, and the baryon-antibaryon asymmetry.  In fact, the cosmological constant problem, the absence of a viable cold dark matter candidate, and the failure to satisfy Sakharov's conditions~\cite{Sakharov:1967dj} for electroweak baryogenesis,\footnote{The SM electroweak phase transition is predicted to be too weak to meet the out-of-equilibrium condition~\cite{Kajantie:1996mn, Laine:2012jy}.} all point to the existence of new physics beyond the Standard Model (BSM), which may arise above its cut-off scale, $\Lambda$.

Composite particles originating as bound states in new strongly coupled theories are ubiquitous in proposals for  BSM physics. They provide the necessary setting for SM completions that preserve its established long-distance successes, while also addressing its short-distance foibles and phenomenological  shortcomings. Models exploiting compositeness have been proposed as the origin of the physics of electroweak symmetry breaking (EWSB), of the Higgs boson, of the top quark, of dark matter, and of primordial phase transitions that might be detectable by gravitational wave experiments. The theory considered in this paper can be viewed as a minimal benchmark that may play a role in all these far-reaching proposals, and is hence an important target for numerical lattice studies, as we shall explain shortly.  We will later outline tangible aims and objectives for the study performed for this publication.

In a broad class of Composite Higgs Models (CHMs), the Higgs fields responsible for EWSB arise as Pseudo-Nambu-Goldstone-Bosons (PNGBs)~\cite{Kaplan:1983fs,Georgi:1984af,Dugan:1984hq} in a new, strongly coupled, physics sector. The compositeness scale, $\Lambda$,  arises dynamically,  slightly above the electroweak scale, and naturally cuts off the sensitivity of long-distance observables to short-distance details. The idea and its implementations, in particular the triggering of EWSB via vacuum misalignment (a variation upon vacuum alignment~\cite{Das:1967it,Peskin:1980gc,Preskill:1980mz}),  are reviewed in Refs.~\cite{Panico:2015jxa,Witzel:2019jbe,Cacciapaglia:2020kgq,Bennett:2023wjw}---useful guidance is also found in the summary tables in Refs.~\cite{Ferretti:2013kya,Ferretti:2016upr,Cacciapaglia:2019bqz}. In an independent development, the large mass of the top quark can be explained with Top Partial Compositeness (TPC), as first proposed in Ref.~\cite{Kaplan:1991dc}, and also discussed in Refs.~\cite{Grossman:1999ra,Gherghetta:2000qt,Chacko:2012sy}.  The extensive literature on implementations of these two ideas includes  phenomenological studies~\cite{Katz:2005au,Barbieri:2007bh,Lodone:2008yy,Gripaios:2009pe,Mrazek:2011iu,Marzocca:2012zn,Grojean:2013qca,Cacciapaglia:2014uja,Ferretti:2014qta,Arbey:2015exa,Cacciapaglia:2015eqa,Vecchi:2015fma,Ma:2015gra,Feruglio:2016zvt,DeGrand:2016pgq,Fichet:2016xvs,Galloway:2016fuo,Agugliaro:2016clv,Belyaev:2016ftv,Csaki:2017cep,Chala:2017sjk,Golterman:2017vdj,Csaki:2017jby,Alanne:2017rrs,Alanne:2017ymh,Sannino:2017utc,Alanne:2018wtp,Bizot:2018tds,Cai:2018tet,Agugliaro:2018vsu,Cacciapaglia:2018avr,BuarqueFranzosi:2018eaj,Gertov:2019yqo,Ayyar:2019exp,Cacciapaglia:2019ixa,BuarqueFranzosi:2019eee,Cacciapaglia:2019dsq,Cacciapaglia:2020vyf,Appelquist:2020bqj,Dong:2020eqy,Cacciapaglia:2021uqh,Banerjee:2022izw,Ferretti:2022mpy, Cai:2022zqu,Appelquist:2022qgl, Cacciapaglia:2024wdn,Banerjee:2024zvg,Caliri:2024jdk,Cacciapaglia:2026jlv}, bottom-up holographic models~\cite{Contino:2003ve,Agashe:2004rs,Agashe:2005dk,Agashe:2006at,Contino:2006qr,Falkowski:2008fz,Contino:2010rs,Contino:2011np,Erdmenger:2020lvq,Erdmenger:2020flu,Elander:2020nyd,Elander:2021bmt,Elander:2023aow,Erdmenger:2023hkl,Elander:2024lir,Erdmenger:2024dxf,Alfano:2024aek}, and  top-down holographic theories~\cite{Imoto:2009bf,Elander:2021kxk}.

 The minimal theory that  is amenable to lattice studies, while implementing both CHM and TPC mechanisms, was put forward in Ref.~\cite{Barnard:2013zea}---see also Refs.~\cite{Ferretti:2013kya,Ferretti:2016upr,Cacciapaglia:2019bqz}. The Higgs-doublet fields arise as four of the five PNGBs along the $SU(4)/Sp(4)$ coset that describes the pattern of global symmetries acting on the $N_{\rm f}=2$ hyperquarks in a gauge theory with $Sp(4)$ gauge group. These (Dirac) fermions transform in the fundamental representation of the gauge group. The further  addition of  $N_{\rm as}=3$ (Dirac) fermions transforming in the 2-index antisymmetric representation serves to implement the TPC mechanism. The admixture of fermions in different representations results in new fermion bound states, dubbed chimera baryons, which have the right quantum numbers to act as partners of the top quarks. While it also affects the underlying dynamics, making its numerical studies demanding in terms of resources (see, e.g., Refs.~\cite{Bennett:2024tex,TELOS:2025ash}), it leaves the CHM mechanism, to large extent, unaffected. The study of the lattice theory with $N_{\rm f}=2$ and $N_{\rm as}=0$ is hence the natural starting point towards understanding the dynamics of this whole class of  phenomenologically promising models.

As mentioned above, despite extensive experimental searches, a dark sector  consisting of fields carrying new strongly coupled interactions, but only feeble couplings to the SM fields, would have escaped experimental detection~\cite{Strassler:2006im,Cheung:2007ut,Hambye:2008bq,Feng:2009mn,Cohen:2010kn,Foot:2014uba,Bertone:2016nfn}. In conjunction with the  evidence for dark matter (see, e.g., the review in Ref.~\cite{Cirelli:2024ssz}), this idea has been given a plethora of possible realisations, which, depending on the details of the underling dynamics, are referred to as composite dark matter models~\cite{DelNobile:2011je, Hietanen:2013fya,Cline:2016nab,Cacciapaglia:2020kgq,Dondi:2019olm, Ge:2019voa,Beylin:2019gtw,Yamanaka:2019aeq,Yamanaka:2019yek,Cai:2020njb}, or strongly interacting massive particle (SIMP) models~\cite{Hochberg:2014dra,Hochberg:2014kqa,Hochberg:2015vrg,Bernal:2017mqb,Berlin:2018tvf, Bernal:2019uqr,Tsai:2020vpi,Kondo:2022lgg,Chu:2024rrv}---see also Refs.~\cite{Maas:2021gbf,Zierler:2021cfa,Kulkarni:2022bvh} and~\cite{Pomper:2024otb,Appelquist:2024koa}. Remarkably, the minimal SIMP dark-matter model proposal in Ref.~\cite{Hochberg:2014kqa} is based on the same $Sp(4)$, $N_{\rm f}=2$ and $N_{\rm as}=0$ theory that we discuss in this paper, hence providing additional motivation for this lattice study.

New dark sectors are also interesting for a complementary reason: if such a physical sector underwent a  first-order phase transition in the early universe, it might have left behind a  stochastic background of  relic  gravitational waves (GW)~\cite{Witten:1984rs,Kamionkowski:1993fg,Allen:1996vm,Schwaller:2015tja, Croon:2018erz,Christensen:2018iqi}, which would be testable in future experiments~\cite{Seto:2001qf, Kawamura:2006up,Crowder:2005nr,Corbin:2005ny,Harry:2006fi, Hild:2010id,Yagi:2011wg,Sathyaprakash:2012jk,Thrane:2013oya, Caprini:2015zlo, LISA:2017pwj, LIGOScientific:2016wof,Isoyama:2018rjb,Baker:2019nia, Brdar:2018num,Reitze:2019iox,Caprini:2019egz, Maggiore:2019uih}---see also Ref.~\cite{afzal2023nanograv}. Computing the GW power spectrum in strongly coupled theories is a challenging task, as it requires detailed knowledge of the thermodynamics in proximity of the phase transition, where phase coexistence and metastability  appear. Such studies are currently being pursued via several  complementary investigation strategies---see the discussions in Ref.~\cite{Huang:2020crf}, and in Refs.~\cite{Halverson:2020xpg,Kang:2021epo,Reichert:2021cvs,Reichert:2022naa,Pasechnik:2023hwv}. They rely on  supplementing effective field theory (EFT) approaches, such as Polyakov-loop~\cite{Pisarski:2000eq, Pisarski:2001pe,Pisarski:2002ji,Sannino:2002wb, Ratti:2005jh,Fukushima:2013rx,Fukushima:2017csk,Lo:2013hla,Hansen:2019lnf} or matrix  models~\cite{ Meisinger:2001cq,Dumitru:2010mj, Dumitru:2012fw, Kondo:2015noa,Pisarski:2016ixt,Nishimura:2017crr,Guo:2018scp, KorthalsAltes:2020ryu,Hidaka:2020vna}, with input from numerical lattice results. It is currently not known whether the $Sp(4)$, $N_{\rm f}=2$ gauge theory undergoes a  phase transition, and whether it might be strong enough to have generated a detectable GW signal.

Lattice field theory is the instrument of choice to predict the dynamical behaviour of strongly coupled gauge theories, such as those advocated by the aforementioned  composite physics  proposals. A number of dedicated investigations of candidate completions of CHMs exist, for gauge theories with $SU(2)$ group and  matter fields transforming in the fundamental representation~\cite{Hietanen:2014xca,Detmold:2014kba,Arthur:2016dir,Pica:2016zst,Lee:2017uvl,Drach:2017btk,Drach:2020wux,Drach:2021uhl,Bowes:2023ihh}, with $SU(4)$ group and  fermion content in an admixture of fundamental and 2-index antisymmetric representation~\cite{DeGrand:2015lna,DeGrand:2016htl,Ayyar:2017qdf,Ayyar:2018zuk,Ayyar:2018ppa,Ayyar:2018glg, Cossu:2019hse, Lupo:2021nzv,DelDebbio:2022qgu,Hasenfratz:2023sqa}, with $SU(2)$ group and fermions transforming as an admixture of fundamental and adjoint representation~\cite{Bergner:2020mwl, Bergner:2021ivi}. Lattice studies of the flavor-singlet sector relevant to dark matter models also exist, both for $SU(2)$~\cite{Arthur:2016ozw} and $Sp(4)$ theories with matter fields~\cite{Bennett:2023rsl,Bennett:2024wda}.

The characterisation of phase transitions in strongly coupled theories is a central arena for the application and testing of  lattice techniques. The finite-temperature behaviour of $SU(3)$ gauge theories is summarised  in Ref.~\cite{borsanyi:2022xml} and in  the review~\cite{Aarts:2023vsf}---see also Refs.~\cite{Svetitsky:1982gs,Yaffe:1982qf} and~\cite{Kajantie:1981wh,Celik:1983wz,Kogut:1983mn,Svetitsky:1983bq,Gottlieb:1985ug,Brown:1988qe,Fukugita:1989yb,Bacilieri:1989ir,Alves:1990pn,Boyd:1995zg,Boyd:1996bx,Borsanyi:2012ve,Shirogane:2016zbf} for pure gauge, and Refs.~\cite{Saito:2011fs,Ejiri:2019csa,Kiyohara:2021smr,Fromm:2011qi,Cuteri:2020yke,Borsanyi:2021yoz} for heavy quarks. For other gauge theories, see for instance Refs.~\cite{Lucini:2002ku,Lucini:2003zr,Lucini:2005vg,Panero:2009tv,Datta:2010sq,Lucini:2012wq} for $SU(N_c)$, Ref.~\cite{Rummukainen:2026etj} for $SU(8)$, Ref.~\cite{Pepe:2005sz,Pepe:2006er,Cossu:2007dk,Bruno:2014rxa} for $G_2$, Ref.~\cite{Appelquist:2015yfa,Appelquist:2015zfa,LatticeStrongDynamics:2020jwi,LatticeStrongDynamicsLSD:2023vtk,Ayyar:2026wht} for stealth dark matter with $SU(4)$ gauge group, Ref.~\cite{Holland:2003kg} for pioneering work on $Sp(2N)$ Yang-Mills theories, and Ref.~\cite{Bruno:2024dha} for the thermodynamics of the $Sp(4)$ Yang-Mills theory. Most recently, inspired by flat histogram~\cite{Berg:1991cf} and density of states~\cite{Wang:2000fzi} methods, the Logarithmic Linear Relaxation (LLR) algorithm~\cite{Langfeld:2012ah,Langfeld:2013xbf,Langfeld:2015fua,Cossu:2021bgn} has been developed, providing a new numerical strategy aimed at  the high precision characterisation of first-order transitions near criticality. Finite-temperature LLR studies of the $SU(3)$ and $Sp(4)$ Yang-Mills theories have produced the first physical results, published in Refs.~\cite{Mason:2022trc,Mason:2022aka,Lucini:2023irm,Bennett:2024bhy,Bennett:2025whm}. Finite-temperature studies using the LLR algorithm exist also for $SU(4)$~\cite{Springer:2021liy, Springer:2022qos} and $SU(N_c)$~\cite{Springer:2023wok,Springer:2023hcc}. A finite-temperature study of the $Sp(4)$ theory with $N_f=2$ is currently underway, and the results will be presented in a forthcoming publication. Establishing its zero-temperature properties is a necessary first step to such future endeavour.

In recent years, the TELOS Collaboration has been pursuing an extensive programme of lattice studies of the $Sp(2N)$ gauge theories, coupled to fermion matter fields transforming according to different combinations  of irreducible  representations of the group~\cite{Bennett:2017kga,Lee:2018ztv,  Bennett:2019jzz, Bennett:2019cxd, Bennett:2020hqd, Bennett:2020qtj,  Bennett:2022yfa, Bennett:2022gdz, Bennett:2022ftz, Bennett:2023wjw,  Bennett:2023gbe, Bennett:2023mhh, Bennett:2023qwx, Bennett:2024cqv,  Bennett:2024wda, Bennett:2024tex,   TELOS:2025ash,Bennett:2025whm,TELOS:2026alk}. The main targets of this programme are observable quantities that are   relevant for model-building and phenomenological purposes, in the context of BSM composite models. These include  measurements of  masses, decay constants and matrix elements of bound states, topological observables, spectral densities and PNGB scattering amplitudes---see also Refs.~\cite{Hong:2017suj,Kulkarni:2022bvh,Bennett:2023rsl, Dengler:2024maq,Bennett:2024bhy}. Matter fields have been introduced in the (unimproved) action as Wilson fermions. The relative simplicity of this first-principles approach has made it possible to collect useful preliminary information for a broad class of theories and across relevant portions of their parameter space. Yet, the adoption of a different action is necessary to yield a high-precision characterisation of the interesting theories, over the entirety of their parameter space, particularly in extrapolating toward the continuum and massless limits. In view of the prominent role that the $Sp(4)$, $N_{\rm f}=2$ theory plays in the context of BSM physics, and of its minimality within such context,  we use it as a testing ground for one special choice of such action: the MDWF formulation.

For this work, we perform extensive tests and optimisation of our algorithms, by tuning the MDWF parameters. We exploit the Grid software suite~\cite{Boyle:2015tjk,Boyle:2016lbp,Yamaguchi:2022feu,Grid-repo}, supplemented by our own implementation of the $Sp(4)$ group~\cite{Bennett:2023gbe}, on supercomputing machines with GPU-based architecture. We measure masses and decay constants of the lightest flavoured mesons, using the Hadrons analysis framework~\cite{antonin_portelli_2023_8023716}. We compare our results with those published in Ref.~\cite{TELOS:2026alk} (see also Ref.~\cite{Bennett:2019jzz}), which have been obtained with Wilson fermions, with the HiRep code~\cite{HiRepSpN,HiRepSUN,pica_github}, and our earlier adaptations for $Sp(2N)$ theories~\cite{Bennett:2017kga, sa2c}, running on CPU-based facilities. In approaching the continuum limit, the MDWF formulation adopted in this study outperforms the simpler unimproved Wilson-fermion formulation used in the earlier study, as expected because of the favourable scaling of lattice artefacts. As part of this assessment,  we also provide an estimate of the computational demands needed to produce our ensembles, setting the stage for future large-scale investigations.

The paper is organised as follows.  We define the continuum and lattice theories of interest in Sect.~\ref{Sec:lattice}. We provide a pedagogical discussion of our implementation of the MDWF formulation, and describe in detail all the notational conventions and numerical choices used  in the optimisation process, in Sect.~\ref{sec:dwf_par_optimization}. While a significant portion of the material presented in this early part of the paper can be found in the literature on the subject,  we intend to make the presentation self-contained, and set the reference framework, notation and conventions, for the use of domain-wall fermions in the future research programme of the TELOS collaboration. In Sect.~\ref{Sec:ensemble_characterization}, we characterise the resulting ensembles used for this work,  discuss our scale setting procedure and our treatment of the topology. To these purposes, we make use of the gradient flow, which we implement along the guidelines outlined in Ref.~\cite{Bennett:2022ftz}. Section~\ref{Sec:spectr_decayconst} contains the main physics analysis, presenting our new measurements of masses and decay constants of the lightest meson particles in the theory. We summarise our results, and outline the future research programme of the TELOS collaboration, in Sect.~\ref{Sec:outlook}. The paper is complemented by three Appendices, providing additional technical details about the numerical tests we conducted using the Grid software (Appendix~\ref{Sec:gauge_invariance}), 
our assessment of finite volume effects (Appendix~\ref{Sec:finite_size_effects}), 
and our treatment of the fitting procedure used for the residual mass and the continuum extrapolation (Appendix~\ref{Sec:fitting_details}). 

\section{The gauge theory}
\label{Sec:lattice}

In this section, we present the action of both the continuum and lattice field theories of interest. We start in Sect.~\ref{Sec:continuum}, by showing the Lagrangian density for the field theory, and discussing the symmetry properties that are used to classify meson bound states that play a central role for applications in the CHM and SIMP contexts. We refer to the Appendix of Ref.~\cite{Bennett:2019cxd} for notational details (see also Ref.~\cite{
,Bennett:2023rsl}), in particular pertaining our choices of conventions in relating  2-component and 4-component spinors. We then provide a pedagogical description of the full lattice action and the MDWF implementation, in Sect~\ref{Sec:dwf_discretization}, following  Ref.~\cite{Brower:2012vk} and references therein.

\subsection{Field theory, continuum formulation}
\label{Sec:continuum}

The $Sp(4)$ gauge theory is coupled to $N_{\rm f} = 2$ fundamental Dirac fermions, $Q^{i\,a}$, that carry hypercolour index, $a=1,\,\ldots,\,4$, referring to the $Sp(4)$  gauge group, and flavour index, $i=1,\,\,2$, labelling the fermion species.  Spinor indices are understood. The continuum Lagrangian density is the following:
\begin{equation}
\mathcal{L}= -\frac{1}{2} \Tr \ G_{\mu\nu} G^{\mu\nu}
\,+\,\frac{1}{2}\left(i\overline{Q^{i}}_a \gamma^{\mu}\left(D_{\mu} Q^i\right)^a
\,-\,i\overline{D_{\mu}Q^{i}}_a \gamma^{\mu}Q^{i\,a}\right)\,-\,m^{\rm f}\overline{Q^i}_a Q^{i\,a}\,,
\label{eq:lagrangian}
\end{equation}
where the bar on the spinor denotes the Dirac adjoint, $\bar Q\equiv Q^{\dagger}\gamma_0$, and the field-strength tensor is defined as
\begin{equation}
G_{\mu\nu} \equiv \partial_\mu A_\nu - \partial_\nu A_\mu + i g [A_\mu, A_\nu]\,,
\end{equation}
with $g$ the gauge coupling, $A_\mu \equiv \sum_A A_\mu^A T^A$ the gauge fields, and $T^A$, for $A=1,\,\cdots,\,10$,  the $4\times 4$ Hermitian generators of $Sp(4)$, normalised so that $\Tr\,T^AT^B=\frac{1}{2}\delta^{AB}$.
Suppressing hypercolour indices, the covariant derivative acting on the fermions, $D_\mu Q$, reads as follows:
\beqs
D_\mu Q^i &\equiv& \partial_\mu Q^i + i g A_\mu Q^i\,,
\eeqs
and  transforms  under a local gauge transformation, $U(x) \in Sp(4)$, in the same way as the fermion fields,
\begin{equation}
Q^i(x) \rightarrow U(x) Q^i(x), \qquad D_\mu Q^i(x) \rightarrow U(x) D_\mu   Q^i (x)\,.
\end{equation}
In this paper we assume that the Dirac mass term is diagonal and degenerate,  and denote the mass as $m^{\rm f}$.

In the massless limit, $m^{\rm f}=0$, the pseudo-real nature of the fundamental representation of  the $Sp(4)$ gauge group leads to an enhancement of  the natural $U(2)_L\times U(2)_R$ global symmetry acting on the chiral components of the $N_{\rm f}=2$ Dirac fermions. To see this, one can follow the notational choices in Ref.~\cite{Bennett:2019cxd}, and define the charge-conjugate spinors,  $Q^{i \,a}_C \equiv - \gamma_5 \Omega^{ab} C \left(\overline{Q^i}^T\right)_b$,  where $C$ is the charge-conjugation matrix, $\gamma_5$ the fifth gamma matrix, and $\Omega$ the symplectic matrix---the $4\times 4$ antisymmetric matrix characterising the $Sp(4)$ gauge group. The two conjugate spinors, $Q_C^{i\,a}$,  transform as the $4$ of the $Sp(4)$ gauge group. One then finds that the same Lagrangian density, as in Eq.~(\ref{eq:lagrangian}), can be rewritten  in terms of 2-component spinors, $q^{j\,a}$, with $j=1,\,\cdots,\,4$,  transforming as a $4$ of the enhanced $SU(4)$ global symmetry, with  $q^j = (P_L Q^1, P_L Q^2, \, P_L Q_C^1, \, P_L Q^2_C)$---omitting here the hypercolour indices, for simplicity---where $P_L\equiv\frac{1}{2}\left(\mathbb{1}+\gamma_5\right)$ is the left-handed chiral projector, in spinor space. The $SU(4)$ symmetry contains the global $SU(2)_L\times SU(2)_R \times U(1)_B$ subgroup, in which the Abelian factor is the non-anomalous combination of the $U(1)$ factors acting on the fermions. The other combination is the anomalous $U(1)_A$.
  
  In the presence of a (small)  non-vanishing mass, $m^{\rm f}\neq 0$, one can show explicitly that the last term in Eq.~(\ref{eq:lagrangian}), written in the space of 2-component fermions, is proportional to the symplectic matrix, $\Omega$, in flavour space, which implies that the global $SU(4)$ symmetry is broken explicitly to its $Sp(4)$ subgroup. The enlarged, approximate, $U(4)=U(1)_A\times SU(4) \sim U(1)_A\times SO(6)$ global symmetry is also spontaneously broken  by the dynamical formation of a fermion condensate, to the $Sp(4)\sim SO(5) \subset SU(4)\sim SO(6)$   subgroup, resulting in the emergence of five PNGBs. In the absence of other symmetry-breaking effects, the vacuum aligns with the mass term inside the $SU(4)$ space~\cite{Peskin:1980gc}, so that explicit and spontaneous breaking leave the $Sp(4)$ global subgroup intact, while  the five PNGBs spanning the $SU(4)/Sp(4)$ coset acquire a (degenerate) mass.

The enhanced global symmetry acting on the matter field content makes  this theory especially interesting in the CHM context.  The construction of CHMs starts with the identification of the global $SO(4)$ symmetry acting on the SM classical Higgs potential as a subgroup of the aforementioned, unbroken, global symmetry, with $SO(4)\subset Sp(4)\sim SO(5)$.  The five PNGBs describing the $SU(4)/Sp(4)\sim SO(6)/SO(5)$ coset acquire a mass (and a potential) controlled by $m^{\rm f}$, and can be decomposed as $5=1 \oplus 4$ under the unbroken $SO(4)$ subgroup. The $4$ is identified with the SM Higgs doublet, because of the well-known property that $SO(4)\sim SU(2)\times SU(2)$. The $SO(4)$ singlet provides an additional (composite) scalar field. The embedding of  SM gauge  and Yukawa interactions further modifies the potential, destabilising the vacuum,  and leading  to EWSB---we refer the reader to Ref.~\cite{Barnard:2013zea} for a discussion of the coupling to the SM fields. For the purposes of this paper, we consider the $Sp(4)$, $N_{\rm f} = 2$ theory in isolation, without coupling it to external fields, and  focus our attention on the lightest mesons.

We classify operators and the states (particles) they source in terms of representations of the unbroken, global  $Sp(4)\sim SO(5)$ symmetry. It is important to highlight that the equivalent of the baryon $U(1)_B$ symmetry of the Standard Model (or, better, the anomaly-free $U(1)_{B-L}$ combination)  is identified with an Abelian subgroup of the $Sp(4)$ global symmetry, hence it does not provide a new quantum number. In fact, di-quark bound states in this theory are part of the meson multiplets, ultimately as a consequence of the fact that the fermion fields, $Q$, and their conjugates, $Q_C$, both transform as $4$'s of the gauged $Sp(4)$.  Conversely, the anomalous Abelian factor, $U(1)_A\sim SO(2)$ is also broken by the mass term, and is not manifest at low energies. This may be important in the dark-matter context~\cite{Cacciapaglia:2019bqz,Arthur:2016ozw,Drach:2021uhl,Bennett:2023rsl,Bennett:2024wda}, and it can provide another useful quantum number, as some of the mesons arrange naturally in (split) $SO(2)$ doublets. As its study involves some of the heavier bound states in the theory, we do not further discuss it here. We refer to Ref.~\cite{Bennett:2019cxd} for a complete discussion and classification of heavier, flavoured and unflavoured, states. The mesons of interest for this publication are hence the following: the PNGBs, that transform as a $5$ of $Sp(4)$ and we denote as ${\rm PS}$ states, the vector states, denoted ${\rm V}$, transforming as a $10$ of $Sp(4)$, and the axial-vector states, labelled ${\rm AV}$, transforming as $5$ of $Sp(4)$.

We conclude by emphasising that, because  the global symmetries in the theory with pseudo-real representation are an extension of those of a theory with complex representations, then all the symmetry arguments that would apply to the latter apply also to the former. The $U(2)_L\times U(2)_R$ global symmetry  is a subgroup of  $U(4)$, and the unbroken $U(1)_B\times SU(2)_V$ subgroup of the former is itself contained in the unbroken $Sp(4)$ subgroup of the latter. It is hence at times convenient to decompose the irreducible representations of $Sp(4)$ into representations of this $U(1)_B\times SU(2)_V$ subgroup. For the pseudoscalar and axial-vector mesons, one finds the decomposition $5= 3_0 \oplus 1_{2}$, yielding a triplet with no baryon number and a (complex) singlet with baryon number, $B=2$ (a di-baryon). For the vectors, one finds that $10=3_0 \oplus 3_2 \oplus 1_0$, one triplet and a singlet being real, and corresponding to the $\rho$ and $\omega$ mesons of a QCD-like theory, supplemented by a complex triplet of di-quarks. In the absence of explicit symmetry-breaking interactions, these multiplets are degenerate, as they combine to form $Sp(4)$ representations.

\subsection{Lattice discretisation}
\label{Sec:dwf_discretization}

The  five-dimensional, Euclidean hypercubic lattice has spacing  in the physical four dimensions, $a$, that differs from that  along the fifth dimension,  $a_5$. The time direction has extent $T = N_t a$, the three physical space directions have isotropic length, $L = N_s a$, and the fifth dimension has length $L_5 = N_5 a_5$. The lattice  action is written as follows:
\begin{equation}
S \equiv S_g + S_f + S_{\rm PV}\,,
\label{eq:lattice_action}
\end{equation}
where $S_g$ stands for the gauge action, $S_f$ for the fermion action, and $S_{\rm PV}$ for the action of the  Pauli-Villars regulators.

The gauge action for the $Sp(2N)$ theory is given in the standard Wilson plaquette form:
\begin{equation}
S_g \equiv \beta \sum_x \sum_{\mu<\nu} \left(1 - \frac{1}{2N} \, \text{Re} \, \Tr \, \mathcal{P}_{\mu\nu}(x) \right)\,,
\end{equation}
where the inverse coupling is defined as $\beta \equiv \frac{4N}{g_0^2}$, with $g_0$ the bare coupling. The plaquette operator is:
\begin{equation}
\label{eq:elementary_plaquette}
\mathcal{P}_{\mu\nu}(x) \equiv U_\mu(x) U_\nu(x + \hat{\mu}) U_\mu^\dagger(x + \hat{\nu}) U_\nu^\dagger(x)\,,
\end{equation}
where $U_{\mu}(x)\in Sp(4)$  is a gauge link at position $x$, in the direction $\mu$, and $\hat{\mu},\,\hat{\nu}$ have unit length.

The fermion action, living  in five dimensions, is defined as follows:
\begin{equation}
\label{eq:fermionaction}
S_f \equiv a^4 a_5 \sum_{j=1}^{N_{\rm f}} \sum_{x, y} \sum_{s, r=0}^{N_5-1} \overline{\Psi}^j(x, s) D^{\mathrm{M}}(x, s, y, r, m_j) \Psi^j(y, r)\,,
\end{equation}
where $\Psi^j$ are five-dimensional fields with flavour index $j$, and $m_j$ denotes the mass of the $j$-th flavoured fermion.  In the present case, the masses are degenerate: $m_1 = m_2 \equiv m_0$. We denote as  $D^{\mathrm{M}}$ the M\"obius domain-wall kernel, which we shall define shortly, following the discussion in Ref.~\cite{Brower:2012vk}.

The purpose of introducing the fifth dimension is to separate the two chiral components of the fermions, localise them near the two five-dimensional  boundaries at the end of space (domain walls),  and suppress their overlap, as well as the resulting (Dirac) mass of the four-dimensional fermion.  Along the  fifth dimension, which we label by the coordinate $s$, a constant mass term, $m_5$, is present, and  open boundary conditions are used at the end points, $s=0$ and $s=N_5-1$, so as to create a domain wall at $s = 0$ and an anti-domain wall at $s = N_5-1$---see the discussion in Ref.~\cite{Vranas:1997da} and references therein, in particular Refs.~\cite{Boyanovsky:1986dt,Shamir:1993zy, Furman:1994ky}. The zero modes of chiral fermions (solutions to the Dirac equation with zero eigenvalue) localise exponentially (the fermion wave-functions decay exponentially away from the domain walls). The fifth-dimensional mass acts as a potential wall in the Dirac equation, and $m_5$ is called the domain wall height. In the  limit $N_5 \to \infty$, the fermions localised at the boundaries represent the physical four-dimensional chiral fermions, and decompose as follows:
\begin{equation}
\psi(x) = P_R \Psi(x, 0) + P_L \Psi(x, N_5 - 1)\,, \qquad {\rm where} \qquad P_{L/R} = \frac{1}{2}(1 \pm \gamma_5)\,.
\label{eq:dwf_physical_fields}
\end{equation}

The M\"{o}bius  kernel, which appears in the MDWF formulation in Eq.~(\ref{eq:fermionaction}), is defined in the block-tridiagonal form~\cite{Brower:2005qw}:
\begin{equation}
D^{\mathrm{M}} \equiv \left[
\begin{array}{ccccc}
D_L & D_R P_R & 0 & \cdots & -m_0 D_R P_L \\
D_R P_L & D_L & D_R P_R & \cdots & 0 \\
0 & D_R P_L & D_L & \cdots & 0 \\
\vdots & \vdots & \vdots & \ddots & \vdots \\
-m_0 D_R P_R & 0 & 0 & \cdots & D_L \\
\end{array}
\right]\,,
\end{equation}
with
\begin{equation}
D_L = b D_{\rm W} + 1\,, \qquad D_R = c D_{\rm W} - 1\,,
\end{equation}
where $b,c > 0$ are real  parameters. 
The dynamics is contained inside of $D_W$, which is the four-dimensional Wilson-Dirac operator, and is defined as follows:
\beqs
D_{\rm W} \psi(x) &\equiv& (4/a-m_5) \psi(x)\label{eq:DiracF} \\
&&-\frac{1}{2a}\sum_\mu \nonumber
\left\{(1-\gamma_\mu)U_\mu(x)\psi(x+\hat{\mu})
+(1+\gamma_\mu)U^{\dagger}_\mu(x-\hat{\mu})\psi(x-\hat{\mu})\frac{}{}\right\}\,,
\eeqs
where the link variables, $U_\mu(x)$ and $U^{\dagger}_\mu(x-\hat{\mu})$, are the same that enter the gauge action, via Eq.~(\ref{eq:elementary_plaquette}).
This kernel is a generalization of the Shamir formulation~\cite{Shamir:1993zy, Furman:1994ky},  defined as:
\begin{equation}
\label{eq:dwf_kernel}
D_{\mathrm{S}}(x, s, y, r)\equiv\delta_{s, r} D_{\rm W}(x,y)+\delta_{x, y} D_5^{\mathrm{S}}(s,r)\,,
\end{equation}
where
\begin{equation}
\label{eq:dwf_kernel_def_5d}
D_5^{\mathrm{S}}(s, r) \equiv\delta_{s, r}-\left(1-\delta_{s, N_s-1}\right) P_{R} \delta_{s+1, r}-\left(1-\delta_{s, 0}\right) P_{L} \delta_{s-1, r}  +m_0\left(P_{R} \delta_{s, N_s-1} \delta_{0, r}+P_{L} \delta_{s, 0} \delta_{N_s-1, r}\right)\,.
\end{equation}
Shamir's operator, $D_{\rm S}$, can be rewritten using identical copies of the Wilson-Dirac operators along a fifth-dimension, to take the form~\cite{Brower:2012vk}:
\begin{equation}
D_{\mathrm{S}} = \frac{a_5 D_{\rm W}}{2 + a_5 D_{\rm W}}\,,
\end{equation}
provided one makes the identification $a_5 \equiv (b - c)a > 0$.
 This operator relates to the MDWF  formulation, through the rescaling
 by the real positive scale factor, $\alpha$:
\begin{equation}
D^{\mathrm{M}} = \alpha D_{\mathrm{S}}\,, \qquad {\rm where} \qquad \alpha = \frac{(b + c)a}{a_5}\,.
\end{equation}

The MDWF formulation contains one additional tunable parameter, $\alpha$, with respect to Shamir's formulation, which can be recovered for $\alpha=1$ (or $c=0$)~\cite{Brower:2005qw,Brower:2012vk}. When varying  $\alpha$, the eigenvectors of the M\"{o}bius domain-wall kernel, $D^{\mathrm{M}}$,  are unchanged, and coincide with those of $D_{\rm S}$. Hence, the MDWF formulation can be viewed as an algorithmic generalisation of the same underlying lattice Dirac action. By tuning $\alpha$, one can reach a given target level of global symmetry and precision with fewer points along the fifth dimension, $N_5$, and hence optimising computational costs~\cite{Brower:2005qw,Brower:2012vk}. Furthermore, the precise choice of value of $\alpha$ optimised to tuning the algorithm may also change as a function of $N_5$. While more detailed discussions of these properties of the MDWF formulation can be found in the literature, we will explicitly test that they remain valid for our theory of interest.

The third and last contribution to the action, in Eq.~(\ref{eq:lattice_action}), is the Pauli-Villars term, ${S}_{\rm PV}$. Pauli-Villars bosonic fields, $\Phi$ and $\bar{\Phi}$, are introduced to cancel the contribution of unphysical, heavy fermion modes that appear for finite values of  $N_5$~\cite{Berruto:2001ty,Vranas:2000tz,Narayanan:1994gw}. Formally,  $S_{\rm PV}$ is analogous to the fermion action:
\begin{equation}
S_{\rm PV}[\Phi, \bar{\Phi}, U] = a^4 a_5 \sum_{x,y} \sum_{s,r=0}^{N_5 - 1} \bar{\Phi}(x,s) D^{\mathrm{M}}_{\rm PV}(x,s,y,r, m_{\rm PV}) \Phi(y,r),
\end{equation}
where $D^{\mathrm{M}}_{\rm PV}$ is constructed similarly to $D^{\mathrm{M}}$, but with a new mass parameter, $am_{\rm PV}$, which is customary to set to $am_{\rm PV} = 1$, so that the cancellation is optimised. The subtraction implemented by this term ultimately allows the extrapolation to the continuum limit of the four-dimensional gauge theory, as it removes the short distance contributions due to heavy modes that complete it to a higher dimensional one.

The partition function of the lattice theory is then the following:
\begin{equation}
Z \equiv \int \mathcal{D}U \, \mathcal{D}\Psi \, \mathcal{D}\bar{\Psi} \, \mathcal{D}\Phi \, \mathcal{D}\bar{\Phi} \,
e^{-S_g[U] - S_f[\Psi, \bar{\Psi}, U] - S_{\rm PV}[\Phi, \bar{\Phi}, U]}
\label{eq:dwf_path_integral}\,,
\end{equation}
where we schematically denoted as $\mathcal{D}\cdot$ the different factors entering the measure. Because
$S_f$ and  $S_{\rm PV}$ are bilinear in the fermions and PV fields, respectively,  their  contributions take the form of  determinants. They factorise into overlap-like and PV parts, which can be written as
follows~\cite{Brower:2012vk,Neuberger:1997bg,Boyle:2014hxa}:
\begin{equation}
\det\!\left[D^{\rm M}(am_0)\right]
=
\det\!\left[D^{\rm M}_{\mathrm{ov}}(am_0)\right]
\det\!\left[D^{\mathrm{M}}_{\rm PV}(am_{\rm PV}=1)\right],
\end{equation}
where the effective four-dimensional M\"obius overlap operator is
\begin{equation}
D^{\mathrm{M}}_{\mathrm{ov}}(am_0)
=
\frac{1+am_0}{2}
+
\frac{1-am_0}{2}\gamma_5\,
\epsilon_{N_5}\!\left(H_{\mathrm{M}}\right)\,,
\qquad
H_{\mathrm{M}}
\equiv
\gamma_5
\frac{(b+c)D_{\mathrm{W}}(-am_5)}
{2+(b-c)D_{\mathrm{W}}(-am_5)}\,.
\end{equation}
Here, \(\epsilon_{N_5}(H_{\mathrm{M}})\) denotes the finite-\(N_5\)
rational approximation to the sign function, which satisfies
\(\epsilon_{N_5}(H_{\mathrm{M}})\rightarrow
\operatorname{sgn}(H_{\mathrm{M}})\) as
\(N_5\rightarrow\infty\). In this limit, the massless operator
\(D_{\mathrm{ov}}^{\rm M}(0)\) satisfies the Ginsparg--Wilson relation:
\begin{equation}
\label{eq:GW_equation}
\gamma_5 D_{\mathrm{ov}}^{\rm M}(0)
+
D_{\mathrm{ov}}^{\rm M}(0)\gamma_5
=
2D_{\mathrm{ov}}^{\rm M}(0)
\gamma_5D_{\mathrm{ov}}^{\rm M}(0).
\end{equation}

Observables computed in the five-dimensional domain wall theory match those obtained in the effective four-dimensional overlap theory,
\begin{equation}
\langle O[\Psi, \bar{\Psi}, U] \rangle_{5d} = \langle O[\psi, \bar{\psi}, U] \rangle_{4d, \mathrm{ov}}\,,
\end{equation}
which demonstrates the equivalence of the two formulations.
A significant, practical  advantage of using the MDWF formulation is its natural compatibility with numerical strategies developed for Wilson fermions. Since the M\"{o}bius kernel is constructed from the Wilson-Dirac operator, it allows for the direct application of mature linear solvers and preconditioning techniques~\cite{Brower:2005qw, Brower:2012vk}, developed in the context of Wilson fermion simulations. This observation greatly simplifies the integration of the MDWF formulation into established lattice gauge theory software. For our work, we elected to use the Grid software suite~\cite{Boyle:2015tjk, Boyle:2016lbp}, to generate gauge configurations with fully dynamical MDWFs,  in the context of gauge theories with symplectic group. Furthermore, the MDWF formalism introduces extra tunable parameters, the presence of which can be exploited to minimise the condition number of the fermion matrix, thereby accelerating solver convergence~\cite{Brower:2012vk}. The results of consistency checks on our $Sp(4)$ implementation and gauge covariance are detailed in Appendix~\ref{Sec:gauge_invariance}.

\section{MDWF optimisation}
\label{sec:dwf_par_optimization}

In this section, we discuss our choices of parameters in the MDWF formulation, which are driven by a compromise between computational efficiency and precision goal for the observables of interest: the  masses of the lightest flavoured mesons in the theory.  The starting point of this analysis is the definition of fermion currents on the lattice, for which we follow Ref.~\cite{Furman:1994ky}, or equivalently, in the MDWF formulation, Ref.~\cite{Boyle:2014hxa}. The axial-vector currents of the continuum theory are recovered from the M\"{o}bius domain wall fermion formulation by exploiting the symmetry properties  of the lattice action.

We first define, in four dimensions, the five components of the axial-vector (AV) current, associated with the five generators of the $SU(4)/Sp(4)$ coset, that we label with the index  $A=1,\,\cdots 5$. Following  the conventions of Ref.~\cite{Bennett:2019cxd}, they are given by the following operators:
\begin{equation}
	\begin{aligned} 
		\mathcal{O}^{\mathrm{AV}, \, 1}_{\mu} & =\overline{Q^{1 a}} \gamma_\mu \gamma_5 Q^{2 a}+\overline{Q^{2 a}} \gamma_\mu \gamma_5 Q^{1 a}\,, \\ 
		\mathcal{O}^{\mathrm{AV}, \, 2}_{\mu} & = -i \overline{Q^{1 a}} \gamma_\mu \gamma_5 Q^{2 a}+i \overline{Q^{2 a}} \gamma_\mu \gamma_5 Q^{1 a}\,, \\ 
		\mathcal{O}^{\mathrm{AV}, \, 3}_{\mu} & =\overline{Q^{1 a}} \gamma_\mu \gamma_5 Q^{1 a}-\overline{Q^{2 a}} \gamma_\mu \gamma_5 Q^{2 a}\,, \\ 
		\mathcal{O}^{\mathrm{AV}, \, 4}_{\mu} & =-i \overline{Q^{1 a}} \gamma_\mu Q_C^{2 a}+i \overline{Q_C^{2 a}} \gamma_\mu Q^{1 a}\,, \\ 
		\mathcal{O}^{\mathrm{AV}, \, 5}_{\mu} & =\overline{Q^{1 a}} \gamma_\mu Q_C^{2 a}+\overline{Q_C^{2 a}} \gamma_\mu Q^{1 a}\,,
	\end{aligned}
\end{equation}
where, as anticipated,  $Q^{i\, a}_C \equiv - \gamma_5 \Omega^{ab} C (\overline{Q^i}^T)_b$ is the charge-conjugated spinor of $Q^{i\,a}$. We suppressed the spinor indices, which are summed over. 
The components of the corresponding pseudoscalar (PS) density current, that are associated with the PNGBs, are given by
\begin{equation}
	\begin{aligned}
		\mathcal{O}^1_{\mathrm{PS}} &= \overline{Q^{1 a}} \gamma_5 Q^{2 a}+\overline{Q^{2 a}} \gamma_5 Q^{1 a}\,, \\
		\mathcal{O}^2_{\mathrm{PS}} &= -i \overline{Q^{1 a}} \gamma_5 Q^{2 a}+i \overline{Q^{2 a}} \gamma_5 Q^{1 a}\,, \\
		\mathcal{O}^3_{\mathrm{PS}} &= \overline{Q^{1 a}} \gamma_5 Q^{1 a}-\overline{Q^{2 a}} \gamma_5 Q^{2 a}\,, \\
		\mathcal{O}^4_{\mathrm{PS}} &= -i\left(\overline{Q^{1 a}} Q_C^{2 a}+\overline{Q_C^{2 a}} Q^{1 a}\right)\,, \\
		\mathcal{O}^5_{\mathrm{PS}} &= i\left(-i \overline{Q^{1 a}} Q_C^{2 a}+i \overline{Q_C^{2 a}} Q^{1 a}\right)\,,
	\end{aligned}
\end{equation}
in which one recognises the three generators of the $SU(2)_L\times SU(2)_R/SU(2)_V\sim SO(4)/SO(3)$ coset, that can be identified with the first three such operators. The two additional di-quark operators complete the multiplet of the unbroken global $Sp(4)\sim SO(5)$. Because of the symmetry properties of the system, quantities involving these generators are degenerate, hence we focus our attention on one generator only, for each multiplet, which we choose to be $\mathcal{O}^{\mathrm{AV}, \, 1}_{\mu}$ 
and $\mathcal{O}^1_{\mathrm{PS}}$, respectively, without loss of generality.

In order to define the corresponding lattice current in terms of the five-dimensional fermion fields, $\Psi(x, s)$, by borrowing from Ref.~\cite{Furman:1994ky}, we introduce the current:
\begin{equation}
	j_\mu^1(x, s) =  \dfrac{1}{\sqrt{2}} \left[\bar{\Psi}(x+\hat{\mu}, s)(1+\gamma_\mu) U^{\dagger}_{\mu} (x) T^1 \Psi(x, s) - \bar{\Psi}(x, s)(1-\gamma_\mu) U_{\mu} (x) T^1 \Psi(x+\hat{\mu}, s) \right] ,
\end{equation}
where $T^1$ is the first generator of the $SU(4)/Sp(4)$ coset, chosen to match the conventions for $\mathcal{O}^{\mathrm{AV}, \, 1}_{\mu} $ and $\mathcal{O}^1_{\mathrm{PS}}$ (see, e.g.,  Eq.~(B4) of Ref.~\cite{Lee:2017uvl}). Here,  $\mu = 1,\, \dots,\, 4$ denotes the four Euclidean directions, and $\hat{\mu}$ denotes a unit displacement in the $\mu$-direction. The four-dimensional axial-vector current on the lattice is obtained by summing over, with the sum anti-symmetrised with respect to the midpoint of the fifth dimension, as follows:
\begin{equation}
	\mathcal{A}_\mu^1(x) \equiv \sum_{s=0}^{N_5-1} \operatorname{sgn}\left(s-\frac{N_5-1}{2}\right) j_\mu^1(x, s)\,.
\end{equation}
In the limit of infinite size for the fifth dimension, $N_5\rightarrow +\infty$, this current is related to the local axial-vector operator, $\mathcal{O}_\mu^{\mathrm{AV}, \, 1}$, by multiplicative renormalisation:
\begin{equation}
	\mathcal{A}_\mu^1\,=\,Z_A \mathcal{O}_\mu^{\mathrm{AV}, \, 1}\,.
\end{equation}

\begin{figure}[t]
    \centering
    \begin{tabular}{c}
    \includegraphics{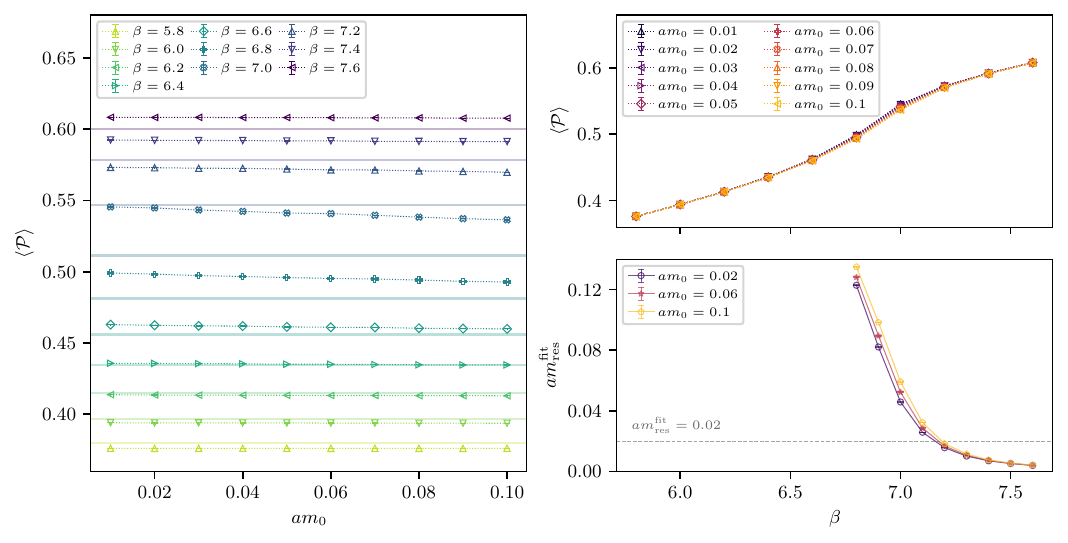} 
    \end{tabular}
    \caption{\label{fig:bulkphase_study} 
        Average plaquette and residual mass, computed  in the MDWF lattice formulation, in the limiting case in which it reproduces the Shamir action, for different values of lattice coupling,  $\beta$, and bare fermion mass, $am_0$. The algorithmic parameters are fixed to $\alpha = 1.0$ (or, equivalently, $b=1$ and $c=0$), so that the MDWF treatment reproduces the Shamir case, as well as $am_5 = 1.8$,  $a_5/a = 1.0$, and fifth-dimension extent $N_5=8$. In the left panel, the average plaquette, $\left\langle {\cal P} \right\rangle$,  is plotted as a function of the bare mass, $a m_0$. The horizontal solid lines indicate the corresponding pure Yang–Mills values, for comparison. Different data sets are distinguished by the value of the lattice coupling, $\beta$. We also show the average plaquette as a function of $\beta$ (top-right panel), and the residual mass (bottom-right panel) fitted from plateaux according to Eq.~\eqref{eq:residual_mass}. For the average plaquette analysis we use a lattice volume of $\tilde{V}=8\times 8^3$, while for the residual mass study we use a larger lattice volume, $\tilde{V}=32\times16^3$.}
\end{figure}

       \begin{figure}[t]
       \begin{center}
       \begin{tabular}{cc}
        \includegraphics{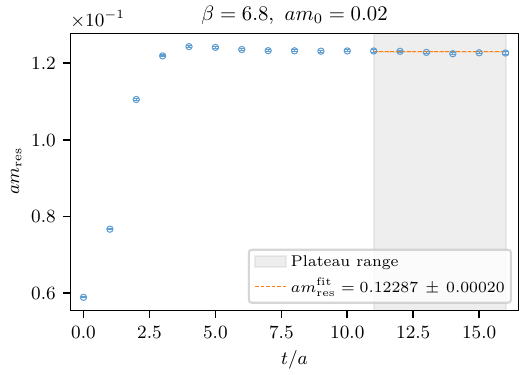} &
          \includegraphics{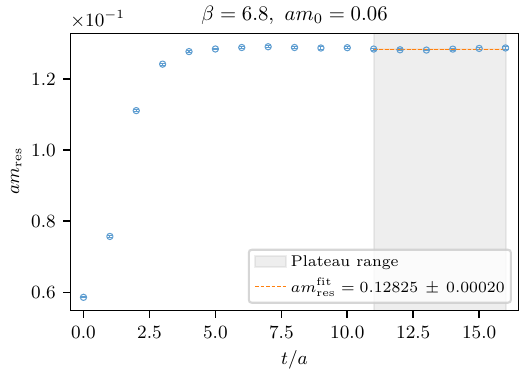} \\
          \includegraphics{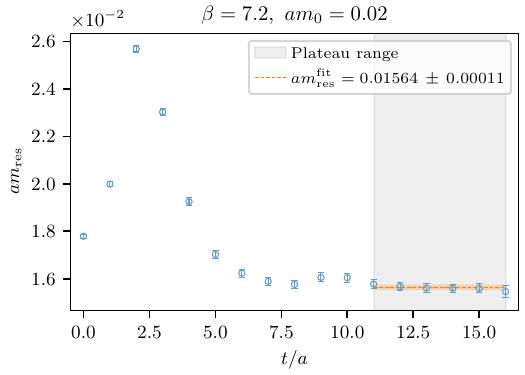} &
          \includegraphics{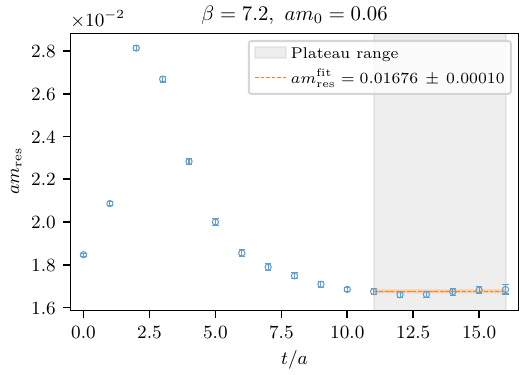}\\  
        \end{tabular}
         \caption{ \label{fig:mres_plateau} 
         Representative examples of (folded) plateaux in the residual mass, $a m _{\rm res}$, and their corresponding correlated fits, $a m _{\rm res}^{\rm fit} (t)$, in the $Sp(4)$ gauge theory with $N_{\rm f}=2$, as a function of the time separation, $t/a$ (expressed in lattice units), between sink and source, obtained from the (volume and ensemble averaged) two-point correlation functions involving the pseudoscalar meson operator, ${\cal O}^1_{\rm PS}$, evaluated at the midpoint and end-point of the fifth dimensions. The lattice has sizes are $N_5=8$, $N_t=16$ and $N_t=32$. The values of the M\"obius parameters are chosen so as to reproduce the Shamir formulation, namely $\alpha=1.0$, $a_5/a=1.0$, $am_5=1.8$, and $am_{\rm PV}=1.0$, while the corresponding values of $\beta$ and $am_0$ are displayed in the titles of the individual panels.}
         \end{center}
          \end{figure}

For  finite values of the number of sites in the fifth dimension, $N_5$, and  nonzero (Dirac) masses for the fermions, the global symmetry is only approximate. The divergence of the partially-conserved axial-vector current (PCAC), $\mathcal{A}_\mu^1$, receives a non-vanishing contribution which, following Ref.~\cite{Furman:1994ky}, can be written as the sum of two terms:
\begin{equation}
	\label{eq:pcac_dw}
	\Delta_\mu \mathcal{A}^1_{\mu}(x) = 2 a m_0 \mathcal{O}^1_{\rm PS}(x) + 2 \mathcal{O}^1_{\text{PS}, \,  q}(x)\,,
\end{equation}
where $\Delta_\mu f(x) \equiv f(x) - f(x - \hat{\mu})$ stands for the left-discretised derivative computed on the lattice. Using the continuity equation for the fifth dimension, one finds that the two pseudoscalar densities  appearing on the right-hand side of the PCAC relation, Eq.~(\ref{eq:pcac_dw}), can be written in terms of the five-dimensional fields evaluated at the boundaries and at the middle points of the fifth dimension, as shown in Ref.~\cite{Furman:1994ky}. The first term is connected to the bare fermion mass, hence it enters the physics of the continuum system of interest. It depends on the fields evaluated at the boundaries of the fifth dimension,  and can be written in the following form:
\begin{equation}
	\mathcal{O}^1_{\rm PS}(x) 
	= \sqrt{2} \left[\bar{\Psi}(x, 0) P_L T^1 \Psi(x, N_5-1) - \bar{\Psi}(x, N_5-1) P_R T^1 \Psi(x, 0) \right]\,.
\end{equation}
The second term in Eq.~(\ref{eq:pcac_dw})  comes from the  midpoint contribution, it is an unphysical bi-product of the finiteness of the fifth dimension, and it is given by the following expression:
\begin{equation}
	\mathcal{O}^1_{\text{PS}, \, q}(x) = \sqrt{2}\left[ \bar{\Psi}(x, N_5/2) P_L T^1 \Psi(x, N_5/2-1) -\bar{\Psi}(x, N_5/2-1) P_R T^1 \Psi(x, N_5/2) \right]\,.
\end{equation}

This last term introduces in the system a residual mass, $a m_{\rm res}$, that  quantifies the extent of (unphysical)  breaking of the continuous global symmetry due to the finiteness of $N_5$. At low energies, the midpoint term satisfies the approximate identity~\cite{Blum:2000kn, CP-PACS:2000fmi, Aoki:2002vt}:
\begin{equation}\label{eq:Opsq_approx}
	\mathcal{O}_{\text{PS}, \, q}^1 \simeq a m_{\mathrm{res}} \mathcal{O}_{\rm PS}^1 + \mathcal{O}(a^2)\,,
\end{equation}
which is exploited in order  to extract estimates of the residual mass  from appropriate combinations of correlation functions. Operationally, the effective residual mass can be defined from  the ratio of pseudoscalar two-point functions evaluated at the midpoints and endpoints of the fifth dimension:
\begin{equation}
	\label{eq:residual_mass}
	a m_{\rm res}(t) \equiv \frac{\sum_{\vec{x},\vec{y}} \langle \mathcal{O}^1_{\text{PS}, \, q} (\vec{y},t) \mathcal{O}^1_{\rm PS} (\vec{x},0) \rangle}{\sum_{\vec{x},\vec{y}} \langle \mathcal{O}^1_{\rm PS} (\vec{y},t) \mathcal{O}^1_{\rm PS} (\vec{x},0) \rangle}\,.
\end{equation}
The estimation of the effective mass,  $a m_{\rm res}$, proceeds as for other spectroscopy  measurements, and requires the identification of a plateau (at asymptotically large $t$) in this ratio.

The Ginsparg-Wilson relation~\cite{Ginsparg:1981bj}, in Eq.~\eqref{eq:GW_equation}, encodes an exact lattice realisation of the continuous  global symmetry of the theory. For domain wall fermions, with finite extent of the fifth dimension, $N_5$, the effective four-dimensional Dirac operator, \( D_{\text{eff}} \), replaces $D_{\mathrm{ov}}$, and does not  satisfy exactly Eq.~\eqref{eq:GW_equation}, with
\begin{equation}
\Delta_{\text{GW}}(x, y) \equiv \gamma_5 D_{\text{eff}}(x, y) + D_{\text{eff}}(x, y) \gamma_5 - a D_{\text{eff}}
 \gamma_5 D_{\text{eff}}(x, y)\,\neq \,0\,.
\end{equation}
In the limit $N_5 \to \infty$, this spurious symmetry-breaking effect vanishes, $\Delta_{\text{GW}}(x, y)\rightarrow 0$,  restoring the exact Ginsparg-Wilson relation. For finite $N_5$, the residual mass represents an additive term to the quark mass, that arises as part of the low-momentum term of the symmetry-breaking operator, $\Delta_{\text{GW}}$~\cite{Brower:2012vk}.
Minimising $a m_{\rm res}$ is hence crucial for maintaining good symmetry properties in domain-wall fermion simulations performed at finite $N_5$~\cite{Brower:2012vk}, as the symmetries of the continuum theory are recovered only in the limit  $a m_{\rm res} \to 0$.

\begin{figure}[t]
    \centering
    \includegraphics{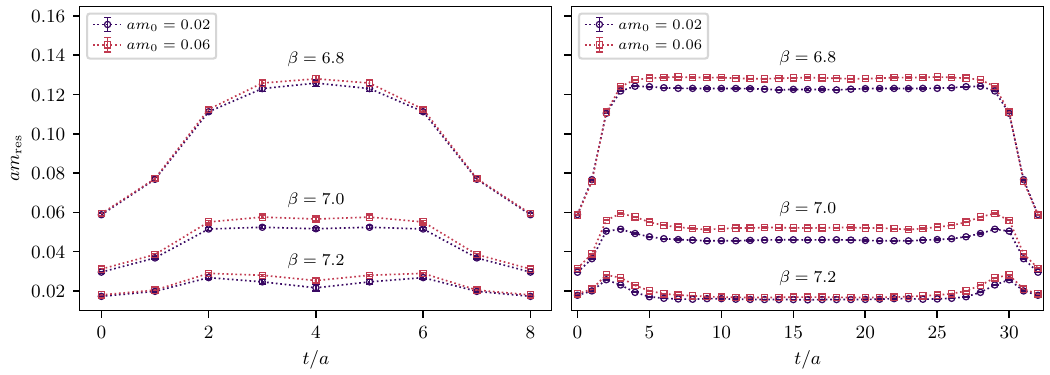}    
    \caption{\label{fig:mres_Shamir_beta} 
        Examples of the residual mass $am_{\rm res}$ as a function of $t/a$, computed with the MDWF formulation, but in the limit it reproduces the Shamir action, showing the increase of the residual mass for coarse lattice spacings. We used representative values of the lattice coupling, $\beta = 6.8,\, 7.0,\, 7.2$, and the fermion bare mass, $am_0 = 0.02,\, 0.06$. The algorithmic parameters are fixed to $\alpha = 1.0$, $am_5 = 1.8$, and $a_5/a = 1.0$. All simulations use a fifth-dimension extent of $N_5=8$, while the lattice volumes are $\tilde{V}=8\times 8^3$ in the left panel and $\tilde{V}=32\times16^3$ in the right panel.  }
\end{figure}

\begin{figure}[t]
    \centering
    \begin{tabular}{c}
    \includegraphics{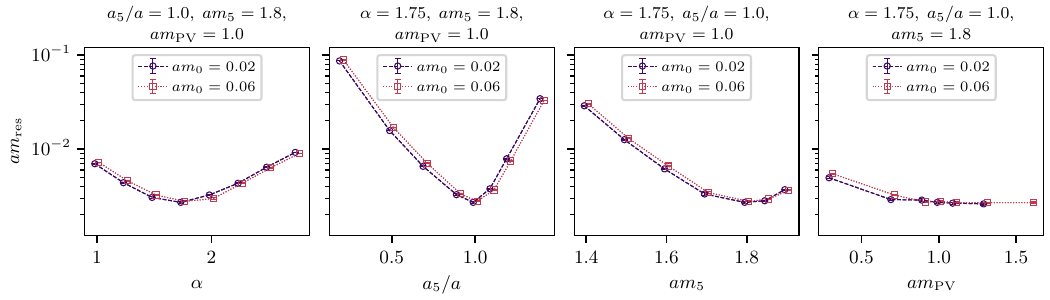} \\ 
    \includegraphics{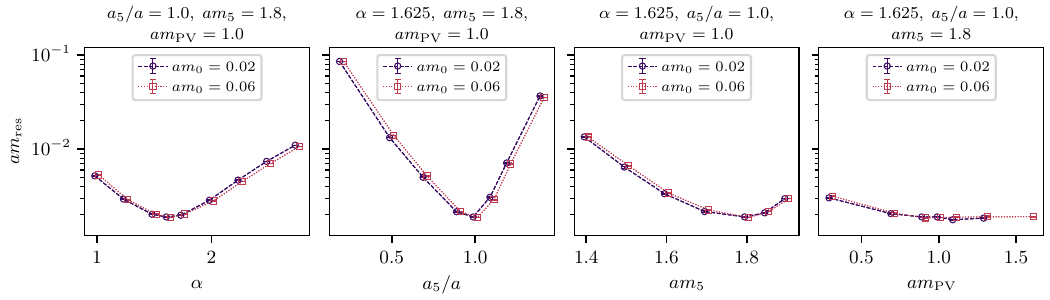} \\ 
    \includegraphics{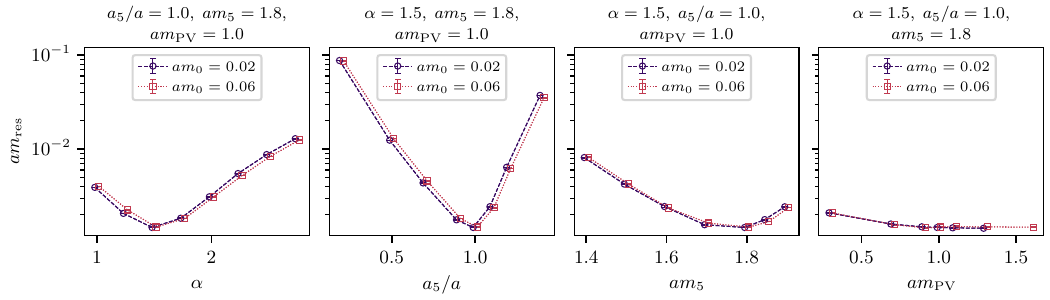} \\
    \end{tabular}
    \caption{ \label{fig:par_scan_1} 
      Examples of parameter scans performed in order to optimise the MDWF formulation for the $Sp(4)$ theory with $N_{\rm f}=2$, demonstrating how the residual mass, $a m_{\rm res}$, depends on the MDWF parameters, $\alpha$, $a_5/a$, $am_5$, and $a m_{\rm PV}$ (left to right),  by varying them one at the time. All calculations use lattices with extensions $N_5=8$, $N_s=16$ and $N_t=32$. Plots in the first row have lattice coupling $\beta=7.4$, in the second row $\beta=7.5$, and in the third row $\beta=7.6$. We show two representative values of fermion bare mass, $a m_0=0.02$ and $a m_0=0.06$, chosen at the two extrema of the region of parameter space of interest in our spectroscopy study.}
\end{figure}

In view of the intrinsically five-dimensional nature of domain-wall fermions, and the associated growth of the computational cost with respect to four-dimensional formulations, a preliminary optimisation of the MDWF formulation by tuning its parameters is essential to ensure suppression of systematic effects (in particular the violation of the Ginsparg-Wilson relation) with manageable computational resources. The benefit of the M\"{o}bius domain wall fermion formulation is that it yields an  improved control over the unphysical breaking of the continuous global symmetry, reflected in a reduced residual mass, obtained by introducing additional freedom compared to other physically equivalent approaches, such as Shamir's formulation of domain-wall fermions.

Nevertheless, Shamir's  formulation of domain wall fermions, which corresponds to the hyper-surface $\alpha=1.0$ in the MDWF parameter space, plays an important role in setting up our numerical process, as it provides the natural baseline for  further investigations of the M\"{o}bius theory. We hence devote some time to analyse its behaviour. In this process, we fix $a_5/a=1.0$ and a domain wall height of $am_5=1.8$~\cite{PhysRevD.77.014509,PhysRevD.83.074508}. This first study is guided by two main considerations. First, we aim to avoid regions of parameter space affected by large lattice artefacts, by investigating the possible presence of a bulk phase transition as the gauge coupling is varied. Such lattice phases significantly break the infrared physics and must be avoided in order to obtain numerical results that can be extrapolated towards the continuum theory of interest. Second, we seek to identify a range of $\beta$ in which the residual mass indicates that the LH and RH fermion modes are well localised on the domain walls, with small overlap.
Indeed, at sufficiently strong coupling, domain wall fermions may fail to describe the continuum properties of the fermion fields, as the localisation is lost and the residual mass becomes large \cite{Berruto:2001ty,Brower_2000}. Therefore, studying the behaviour of \(am_{\rm res}\) as a function of $\beta$ provides a first useful diagnostic for identifying the region of parameter space where the domain wall mechanism is operating as intended. Once we have established a safe region for the coupling,  $\beta$, and the fermion bare mass, $am_0$, we will  then proceed to tune the algorithmic parameters of the M\"{o}bius formulation.

To address the first point, we examine the behaviour of the average plaquette, $\left\langle {\cal P} \right\rangle$, for different values of the lattice coupling, $\beta$, and  fermion bare mass, $am_0$. In the left panel of Fig.~\ref{fig:bulkphase_study}, we show $\left\langle {\cal P} \right\rangle$ as a function of the bare fermion mass, $am_0$, for values of the gauge coupling, $\beta$, ranging from $\beta = 5.8$ to $\beta = 7.6$, together with the corresponding pure Yang-Mills results. In the right panel, instead, $\left\langle {\cal P} \right\rangle$ is plotted as function of $\beta$. The data exhibit a smooth and mild dependence on $a m_0$ at fixed $\beta$, with no indication of discontinuities and hence no evidence of a bulk phase transition. This behaviour is consistent with previous findings, for example in the $SU(3)$ theories studied in Ref.~\cite{Hasenfratz:2022zsa,Hasenfratz_2021}, for small number of fermions, $N_{\rm f}=2$.

\begin{figure}[t]
    \centering
    \begin{tabular}{c}
    \includegraphics{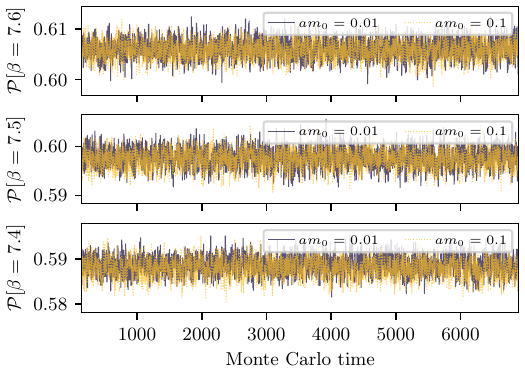} 
    \includegraphics{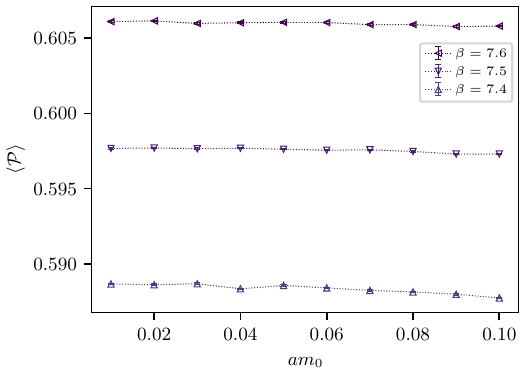} \\
    \end{tabular}
        \caption{ \label{fig:plaq_bulk_tuned} 
      Left panels: Monte Carlo time history of the plaquette, ${\cal P}$,  for the tuned MDWF action (parameters listed in Table~\ref{tab:mres_tuned_mobius}), for  $\beta = 7.4,\, 7.5,\, 7.6$ (top to bottom), considering two extreme choices of fermion  bare mass, $am_0 = 0.01$ and $0.1$. Right panel: corresponding average plaquette, $\left\langle {\cal P} \right\rangle$, as a function of the bare mass, for the three chosen values of $\beta$. The lattice volume is $\tilde{V}=8\times 8^3$, with $N_5=8$.  }
\end{figure}

For the second point, in the bottom-right panel of Fig.~\ref{fig:bulkphase_study} we also show the residual mass, \(am_{\rm res}^{\rm fit}\), fitted from plateaux emerging at large lattice time, $t/a$, as a function of the lattice coupling, \(\beta\),  for three values of the fermion bare mass, \(am_0=0.02,\,0.06, \, 0.1\). Some representative examples of these plataux and the corresponding fits are shown in Fig.~\ref{fig:mres_plateau}. We observe that as $\beta$ increases, the residual mass decreases, indicating a behaviour closer to the continuum for what pertains to global symmetries. Since \(am_{\rm res}\) quantifies the extent of residual spurious symmetry breaking, we require that it is kept at least one order of magnitude smaller than the smallest input fermion mass considered in our calculations, namely \(am_0 = 0.02\). The motivation for this requirement can be seen by plugging Eq.~\eqref{eq:Opsq_approx} in the axial Ward identity for domain wall fermions, Eq.~\eqref{eq:pcac_dw},
\begin{equation}\label{eq:pcac_dw_mres}
	\Delta_\mu \mathcal{A}^1_{\mu}(x)\simeq 2 (a m_0 +am_{\rm res})\mathcal{O}^1_{\rm PS}(x)
\end{equation}
which shows that the residual mass enters additively to the bare mass, effectively shifting the physical fermion mass. Our numerical results show that \(am_{\rm res} \sim 0.02\) at \(\beta \approx 7.2\), which is comparable to our lightest choice of fermion mass and therefore not sufficiently suppressed. To illustrate the numerical evidence of this behaviour, in Fig.~\ref{fig:mres_Shamir_beta} we present the full time-slice dependence of the residual mass for two different volumes, showing that at coarser lattice spacing (i.e. smaller values of $\beta$) the domain-wall mechanism for recovering continuous global symmetries is no longer guaranteed to work. In this regime, the residual mass increases significantly, signalling a severe violation of the symmetry~\cite{qcboyle2015charmphysicsphysicallight}.

In view of these results, we choose to restrict our attention to large lattice coupling, \(\beta \geq 7.4\), for which we are confident  \(am_{\rm res}\) can be effectively suppressed as shown in Fig.~\ref{fig:par_scan_1} and Tab~\ref{tab:mres_tuned_mobius}. I. Within this safer region of parameter space, we then proceed with the tuning of the M\"{o}bius formulation parameters, that should allow for further suppression of the residual mass.

\begin{table}[t]
\caption{Results for the tuning of the M\"obius formulation in the $Sp(4)$ theory with $N_{\rm f}=2$ dynamical domain-wall fermions. All calculations are performed in the MDWF formulation on lattices with $N_5=8$, $N_s=16$, and $N_t=32$. The table lists, for each ensemble, the lattice coupling, $\beta$, the fermion bare mass, $am_0$, the lattice extents, $N_t$, $N_s$, and $N_5$,  the tuned values of the  M\"obius parameters, $\alpha$, $a_5/a$, $am_5$, and $am_{\rm PV}$. We also report the integrated autocorrelation times of the plaquette, $\tau_{\rm int}^{\rm plaq}$, and of the residual mass, $\tau_{\rm int}^{m_{\rm res}}$, the latter computed from configurations separated by five molecular-dynamics trajectories. Our estimate of the resulting residual mass, $am_{\rm res}$, is listed together with  the extent of the fit window $\left[ t_{\rm start}^{\rm res}/a,\, t_{\rm end}^{\rm res}/a\right]$, and the reduced $\chi$ squared, $\chi^2/N_{\rm d.o.f.}$. \\}
\centering
\begin{tabular}{|l|c|c|c|c|c|c|c|c|c|c|c|c|c|c|c|}
\hline\hline
Ensemble & $\beta$ & $am_0$ & $N_t$ & $N_s$ & $N_5$ & $\alpha$ & $a_5/a$ & $am_5$ & $am_{\rm PV}$ & $\tau_{\rm int}^{\rm plaq}$ & $\tau_{\rm int}^{m_{\rm res}}$ & $am_{\rm res}$ & $t^{m_{\rm res}}_{\rm start}/a$ & $t^{m_{\rm res}}_{\rm end}/a$ & $\chi^2/N_{\rm d.o.f.}$ \\
\hline
MB74M6Sc3 & 7.4 & 0.06 & 32 & 16 & 8 & 1.75 & 1 & 1.8 & 1 & 0.64(19) & 0.60(15) & 0.00278(4) & 12 & 16 & 1.056 \\
MB74M2Sc3 & 7.4 & 0.02 & 32 & 16 & 8 & 1.75 & 1 & 1.8 & 1 & 1.0(3) & 0.39(8) & 0.00271(5) & 11 & 16 & 1.107 \\
MB75M6Sc3 & 7.5 & 0.06 & 32 & 16 & 8 & 1.625 & 1 & 1.8 & 1 & — & 0.60(12) & 0.001867(19) & 12 & 16 & 1.474 \\
MB75M2Sc3 & 7.5 & 0.02 & 32 & 16 & 8 & 1.625 & 1 & 1.8 & 1 & 0.49(8) & 0.47(5) & 0.001896(27) & 10 & 16 & 1.360 \\
MB76M6Sc2 & 7.6 & 0.06 & 32 & 16 & 8 & 1.5 & 1 & 1.8 & 1 & 0.59(28) & 0.55(12) & 0.001476(14) & 13 & 16 & 1.326 \\
MB76M2Sc2 & 7.6 & 0.02 & 32 & 16 & 8 & 1.5 & 1 & 1.8 & 1 & — & 0.50(4) & 0.001462(17) & 11 & 16 & 1.093 \\
\hline\hline
\end{tabular}

\label{tab:mres_tuned_mobius}
\end{table}

We then explore the space of the MDWF parameters---the M\"{o}bius kernel parameter, $\alpha$, the M\"obius scale, $a_5/a$, the domain-wall height, $am_5$, and the Pauli-Villars mass $am_{\rm PV}$---for representative choices of the lattice parameters, the extent of the fifth dimension, $N_5$, the bare fermion mass, $am_0$, and the lattice coupling, $\beta$. For completeness, we also investigate the dependence on the Pauli-Villars mass $am_{\rm PV}$, although the choice $am_{\rm PV}=1.0$ is generally required by intrinsic cancellation properties of the formulation. This part of the study is carried out through a comprehensive set of scans of the parameter space, measuring the residual mass, $am_{\rm res}$, on a lattice of size $N_t=32$ and $N_s=16$, while keeping the fifth-dimensional extent fixed to $N_5=8$. The results are summarised in Fig.~\ref{fig:par_scan_1}, where we show the dependence of $am_{\rm res}$ on each tuning parameter, with the others held fixed. The scans are performed at three values of the lattice coupling, $\beta = 7.4$, $7.5$, and $7.6$, and for two representative choices of  fermion bare mass, $am_0 = 0.02$ and $am_0 = 0.06$.

Starting from the scans in $\alpha$ (leftmost column in Fig.~\ref{fig:par_scan_1}), we observe a clear and well-defined minimum of the residual mass, $am_{\mathrm{res}}$, for all three values of $\beta$. The optimal choice of $\alpha$ shifts to smaller values when increasing  the lattice coupling, taking values $\alpha \simeq 1.75$, $1.625$, and $1.5$, for $\beta = 7.4$, $7.5$, and $7.6$, respectively. Since $\alpha$ controls the rescaling of the kernel spectrum entering the approximation $\varepsilon_{N_5}(H)\simeq \operatorname{sign}(H)$, the presence of a sharp minimum reflects an optimal choice of $\alpha$ for approximating the sign function over the relevant spectral range. Deviations from these optimal values lead to a noticeable increase in $am_{\rm res}$, confirming the expected  sensitivity to this parameter of  the global symmetries and their breaking pattern.

The scans over the M\"{o}bius scale parameter, $a_5/a$ (second column in Fig.~\ref{fig:par_scan_1}), and the domain-wall height, $am_5$ (third column in Fig.~\ref{fig:par_scan_1}), exhibit the expected behaviour known to hold also in the Shamir formulation. In particular, the residual mass shows a pronounced minimum at $a_5/a = 1.0$, for all values of $\beta$, confirming this choice as optimal and consistent with the literature on the Shamir formulation. Deviations from this value lead to a rapid increase in $am_{\rm res}$, highlighting its importance in controlling the symmetry properties of the theory. Similarly, the dependence on the domain-wall height, $am_5$, which governs the domain-wall potential in the fifth dimension, displays a clear minimum at $am_5 = 1.8$ for all choices of lattice coupling of interest for this study. This value represents a common choice in the literature for both Shamir and M\"{o}bius domain-wall fermions. By contrast, the rightmost column in Fig.~\ref{fig:par_scan_1} shows only a mild dependence of $am_{\rm res}$ on the Pauli-Villars mass, $am_{\mathrm{PV}}$, which we set to $am_{\rm PV}=1.0$.

The comparison between the two choices of  fermion bare mass, $am_0 = 0.02$ and $am_0 = 0.06$, shows no evidence of a significant dependence of the residual mass on the input bare mass, even when the MDWF parameters, $\alpha$, $a_5/a$, and $am_5$, are not tuned  to their optimal values. This behaviour is consistent with expectations, indicating that the M\"{o}bius formulation maintains good symmetry properties over a range of fermion masses that covers those of interest for this study---see also Ref.~\cite{PhysRevD.72.114505}.

\begin{figure}[t]
    \centering
    \begin{tabular}{c}
    \includegraphics{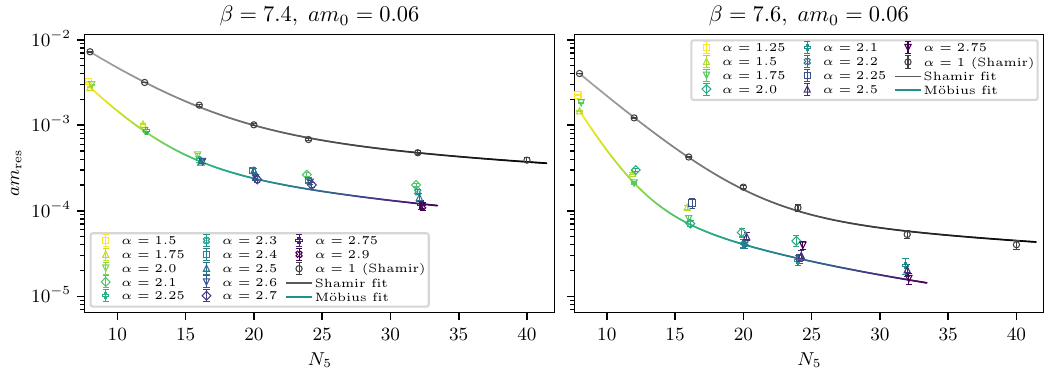} \\
    \end{tabular}
   \caption{ \label{fig:mres_Ls}
   The residual mass, $am_{\rm res}$,  in the MDWF formulation of the $Sp(4)$ theory with $N_{\rm f}=2$, as a function of the extent of the fifth dimension, $N_5$, for fixed values of the MDWF parameters, $a_5/a=1.0$, $am_5=1.8$, and $am_{\rm PV}=1.0$, but for different values of $\alpha$. The choice $\alpha=1$ corresponds to the Shamir formulation. All calculations are performed on lattices with temporal extent $N_t=32$ and spatial extent $N_s=16$.  The solid lines show fits of the minima of $am_{\rm res}$ as a function of $\alpha$ performed using Eq.~\eqref{eq:mres_Ls} for the general MDWF formulation.  For the Shamir formulation, we fix $\nu = 1$, while for the more general MDWF case, $\nu$ is treated as a  fit parameter. The left panel is obtained with lattice coupling $\beta=7.4$, and the right panel with $\beta=7.6$, while in both cases the fermion bare mass is $am_0 = 0.06$---other choices of $\beta$ and $a m_0$ in the range of interest to this study yield similar results. We notice that at small \(N_5\), the residual mass displays an approximately linear behaviour on the logarithmic scale, as expected when the exponential term in Eq.~\eqref{eq:mres_Ls} dominates, while at larger \(N_5\) the power-law contribution becomes increasingly important, leading to visible deviations from the exponential behaviour.}
\end{figure}

Having optimised the choice of parameters of the MDWF formulation, we also verify that our ensembles are not affected by lattice artefacts due to bulk phases. To this end, we perform a scan of the average plaquette for the selected values of $\beta$, using the optimised MDWF parameters which are summarised in Table~\ref{tab:mres_tuned_mobius}. The full scan of the average plaquette is shown in Fig.~\ref{fig:plaq_bulk_tuned}. As in the Shamir case, we observe a smooth, mild  dependence of the average plaquette on the range of bare quark masses considered, for each value of $\beta$. This behaviour indicates that our parameter choices lie well away from any bulk phase transitions.

It is useful to investigate the behaviour of the residual mass, $am_{\rm res}$, with the extent of the fifth dimension, $N_5$, for optimised choices of MDWF parameters,  for at least two reasons. First, it provides a direct check that chiral symmetry is progressively improved as $N_5$ increases, since the overlap between left- and right-handed modes localised on the domain walls is exponentially suppressed with their separation; this behaviour is expected to hold for both Shamir and MDWF formulations. Second, it allows us to quantify the improvement obtained with M\"obius domain-wall fermions.  To this end, in Fig. \ref{fig:mres_Ls} we show the results obtained for two representative lattice gauge couplings, $\beta=7.4$ and $7.6$, for the bare mass $am_0=0.06$, by explore a range of fifth-dimensional extents, bot for the Shamir and MDWF cases. For the MDWF measurements, we estimated the minima of the residual mass obtained by tuning $\alpha$, while keeping $am_5 = 1.8$, $a_5/a = 1.0$, and $am_{\rm PV}=1$ fixed.  By directly comparing $a m_{\rm res}$ for the Shamir and MDWF formulations, we observe a systematical suppression of the values of $a m_{\rm res}$, indicating an improved realisation of the global symmetry.

Furthermore, studying the scaling of the residual mass as a function of $N_5$ allows us to quantify the mobility edge~\cite{Golterman:2004cy, PhysRevD.77.014509}. To this end, we fit our numerical results for the values of $a m_{\rm res}$ obtained with optimised MDWF parameters with the functional form
\begin{equation}\label{eq:mres_Ls}
   a m_{\rm res} \simeq c_1 e^{-\lambda_c N_5} + \frac{c_2}{N_5^\nu}\,.
\end{equation}
 The first term in Eq.~\eqref{eq:mres_Ls} arises from extended modes with eigenvalues close to the mobility edge, $\lambda_c$, which separates the localised and extended regions of the spectrum of the Hermitian Wilson kernel. This contribution is exponentially suppressed with increasing $N_5$, with the rate of suppression controlled by $\lambda_c$. The second term originates from low-lying localised modes and decreases only as a power of $N_5$. Its coefficient, $c_2$, is related to the density of near-zero localised modes, whose contribution to the residual  global symmetry breaking is therefore more difficult to suppress by simply increasing $N_5$. A nonzero and sufficiently large $\lambda_c$ is therefore crucial to ensure good chiral properties and locality. Its value is directly related to the phase structure of the Wilson kernel: the vanishing of $\lambda_c$ signals the onset of the Aoki phase, where near-zero modes proliferate and no clear separation between localised and extended modes exists~\cite{Golterman_2005}. In this regime, the suppression of $m_{\rm res}$ with $N_5$ becomes ineffective. Conversely, a finite mobility edge indicates that the system lies outside this phase, where domain wall fermions provide a local and well-defined formulation of the fermions.

 \begin{table}[t]
\caption{Results of the fit of the residual mass, $am_{\rm res}$, as a function of the extent of the fifth dimension, $N_5$, using Eq.~\eqref{eq:mres_Ls} as fitting ansatz. For the Shamir formulation, we fix $\nu=1$, while for the more general MDWF formulation, $\nu$ is treated as a free fit parameter. We show the best fit values of the parameters, with their statistical uncertainties,  the number of degrees of freedom of the fit, $N_{\rm d.o.f.}$, and the value of its reduced $\chi$ squared, $\chi^2/N_{\rm d.o.f.}$  \\}
\centering
\begin{tabular}{|l|c|c|c|c|c|c|c|c|c|}
\hline\hline
Fit & $\beta$ & $am_0$ & $(N_5,\alpha)$ used & $c_1$ & $\lambda_c$ & $c_2$ & $\nu$ & $N_{\rm d.o.f.}$ & $\chi^2/N_{\rm d.o.f.}$ \\
\hline
\multirow{2}{*}{Shamir} & \multirow{2}{*}{7.4} & \multirow{2}{*}{0.06} & $(8, 1)$, $(12, 1)$, $(16, 1)$, $(20, 1)$ & \multirow{2}{*}{0.0406(18)} & \multirow{2}{*}{0.2532(57)} & \multirow{2}{*}{0.01486(48)} & \multirow{2}{*}{1} & \multirow{2}{*}{4} & \multirow{2}{*}{1.309} \\
& & & $(24, 1)$, $(32, 1)$, $(40, 1)$ & & & & & &  \\
\multirow{2}{*}{Möbius} & \multirow{2}{*}{7.4} & \multirow{2}{*}{0.06} & $(8, 1.75)$, $(12, 2.25)$, $(16, 2.5)$ & \multirow{2}{*}{0.0487(97)} & \multirow{2}{*}{0.395(37)} & \multirow{2}{*}{0.0100(98)} & \multirow{2}{*}{1.27(30)} & \multirow{2}{*}{2} & \multirow{2}{*}{1.570} \\
& & & $(20, 2.7)$, $(24, 2.7)$, $(32, 2.9)$ & & & & & &  \\
\multirow{2}{*}{Shamir} & \multirow{2}{*}{7.6} & \multirow{2}{*}{0.06} & $(8, 1)$, $(12, 1)$, $(16, 1)$, $(20, 1)$ & \multirow{2}{*}{0.0467(11)} & \multirow{2}{*}{0.3137(30)} & \multirow{2}{*}{0.00178(11)} & \multirow{2}{*}{1} & \multirow{2}{*}{4} & \multirow{2}{*}{1.457} \\
& & & $(24, 1)$, $(32, 1)$, $(40, 1)$ & & & & & &  \\
\multirow{2}{*}{Möbius} & \multirow{2}{*}{7.6} & \multirow{2}{*}{0.06} & $(8, 1.5)$, $(12, 1.75)$, $(16, 2)$ & \multirow{2}{*}{0.180(53)} & \multirow{2}{*}{0.623(44)} & \multirow{2}{*}{0.015(15)} & \multirow{2}{*}{1.98(31)} & \multirow{2}{*}{2} & \multirow{2}{*}{0.133} \\
& & & $(20, 2.2)$, $(24, 2.1)$, $(32, 2.75)$ & & & & & &  \\
\hline\hline
\end{tabular}

\label{tab:mres_ls_fit_results}
\end{table}
 
 The Shamir formulation predicts that $\nu=1$, while for M\"obius domain-wall fermions  we treat $\nu$ as a free parameter of the fit, though for an ideally tuned M\"obius formulation one expects $\nu=2$~\cite{PhysRevD.77.014509,Brower:2012vk,Golterman_2005}. The outcome of this study is shown in Fig.~\ref{fig:mres_Ls}, and  the fit results are reported in Table~\ref{tab:mres_ls_fit_results}. Overall, Eq.~\eqref{eq:mres_Ls} provides a good description of the dependence on $N_5$  of the residual mass for the Shamir formulation at both $\beta=7.4$ and $\beta=7.6$. For the MDWF formulation, the quality of the fit and the value of $\nu$ that we were able to extract are also satisfactory and consistent with predictions for lattice coupling $\beta=7.6$. Conversely, when $\beta=7.4$ we obtain a larger reduced $\chi^2_{\rm red}\simeq \ChiSmallestBeta $ and the result for  the exponent in Eq.~\eqref{eq:mres_Ls},   $\nu\simeq \NuSmallestBeta $, is significantly below the asymptotic expectation, $\nu=2$---see Table \ref{tab:mres_ls_fit_results}. This behaviour is consistent with the increased difficulty of tuning the MDWF parameter, $\alpha$, on the coarser lattices. This is also reflected in the fact that, for some values of $N_5$, we were not able to clearly identify the minimum of $am_{\rm res}$ from the scan over $\alpha$, as shown in Fig.~\ref{fig:mres_Ls}. Indeed, at smaller $\beta$, corresponding to a coarser lattice spacing, gauge-field dislocations are less strongly suppressed and the density of low-lying localised modes of the Hermitian Wilson kernel is expected to increase. Correspondingly, the mobility edge $\lambda_c$ becomes smaller and the separation between localised and extended modes is less pronounced. This has two related consequences. First, the larger contribution of low-lying localised modes enhances the residual effects of global symmetry breaking, while the smaller value of $\lambda_c$ makes the contribution from extended modes, $c_1e^{-\lambda_c N_5}$, less efficiently suppressed with increasing $N_5$. As a result, reducing $m_{\rm res}$ at finite $N_5$ becomes more challenging on the ensembles with coarser lattices.

\begin{figure}[t]
    \centering
    \begin{tabular}{c}
    \includegraphics{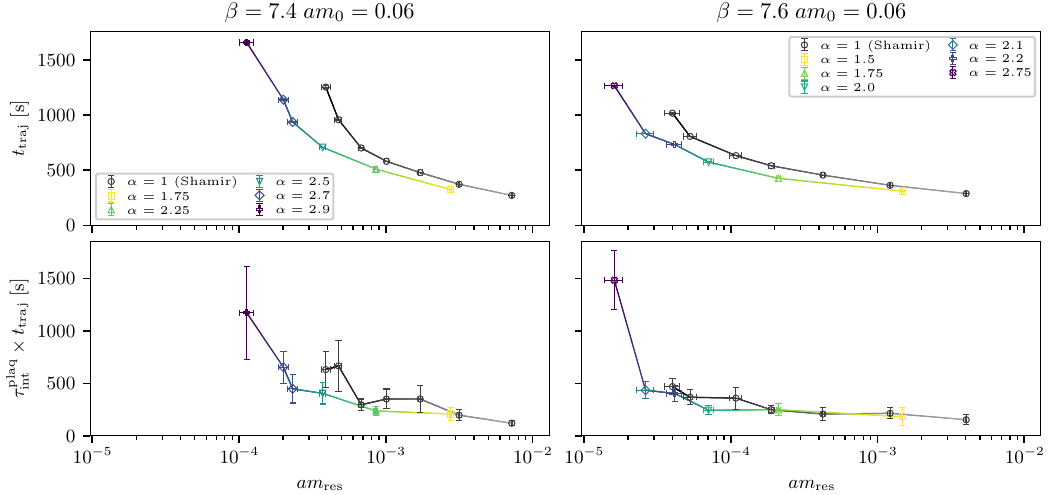}
    \end{tabular}
   \caption{  \label{fig:mres_costs} Comparison between the computational cost of the Shamir and MDWF formulations for the ensembles listed in Table~\ref{tab:mres_ls_fit_results}. In the top panels, we show the elapsed time per HMC trajectory as a function of the residual mass, $am_{\rm res}$, for $\beta=7.4$ (left panel) and $\beta=7.6$ (right), in both cases for $am_0=0.06$. From right to left, the data points correspond to increasing values of $N_5$. All simulations were performed using a single NVIDIA A100-PCIE-40GB GPU and 20 molecular-dynamics steps per HMC trajectory. In the bottom panels, we show the same elapsed time per trajectory multiplied by the integrated autocorrelation time of the plaquette, providing an estimate of the computational cost associated with generating statistically independent configurations.}
\end{figure}

A second implication of the use of coarser lattices is that the presence of small eigenvalues also makes the numerical simulation more expensive. Small eigenvalues lead to a poorer conditioning of the fermion linear systems that must be solved repeatedly during the HMC evolution. In particular, the evaluation of the pseudofermion contribution and of the corresponding fermion force requires repeated inversions of the domain wall Dirac operator. A poorer conditioned system therefore requires a larger number of solver iterations, increasing the computational cost and the elapsed time of each HMC trajectory. This effect is illustrated in Fig.~\ref{fig:mres_costs}, whose top panel shows the elapsed time per HMC trajectory for the Shamir and MDWF formulations at $\beta=7.4$ and $7.6$, with $am_0=0.06$, while the bottom panel shows the same quantity multiplied by the integrated autocorrelation time of the plaquette, providing an estimate of the computational cost of generating statistically independent configurations. For each value of $N_5$, the MDWF data correspond to the value of $\alpha$ that minimises the residual mass.

Finally, we notice that the tuning of the M\"obius parameter introduces a further source of computational cost. As already discussed, increasing the scaling parameter $\alpha$ rescales the eigenvalue spectrum entering the finite-$N_5$ approximation to the sign function, allowing the relevant low-lying part of the spectrum to be approximated more accurately and therefore reducing $a m_{\rm res}$ at fixed $N_5$~\cite{Brower:2012vk}. However, this improvement comes at an additional numerical cost. Although the MDWF formulation does not increase the number of Wilson--Dirac applications required for a single conjugate-gradient iteration, by increasing $\alpha$ one may worsen the condition number of the five-dimensional operator and hence increase the overall cost of the inversion. This effect was already described in Ref.~\cite{Brower:2012vk}, where an additional computational overhead for increasing values of $\alpha$ at fixed $N_5$ was observed due to the increased condition number of the operator.

\begin{figure}[t]
    \centering
    \begin{tabular}{cc}

    \includegraphics{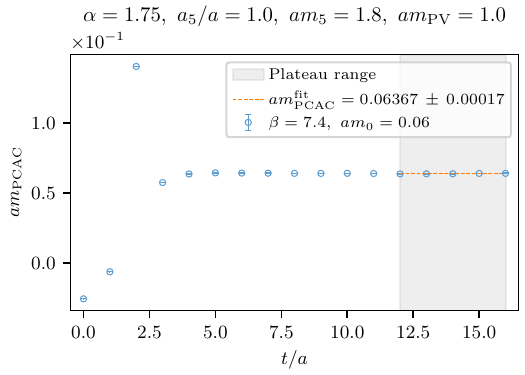} &
    \includegraphics{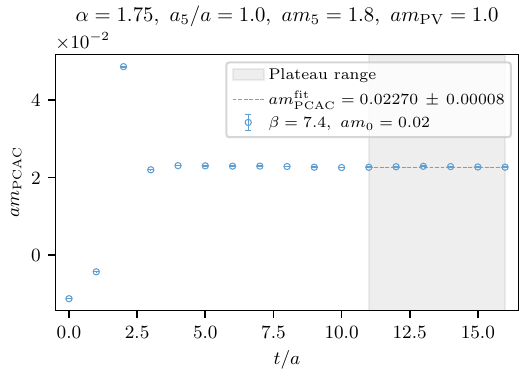} \\
    \includegraphics{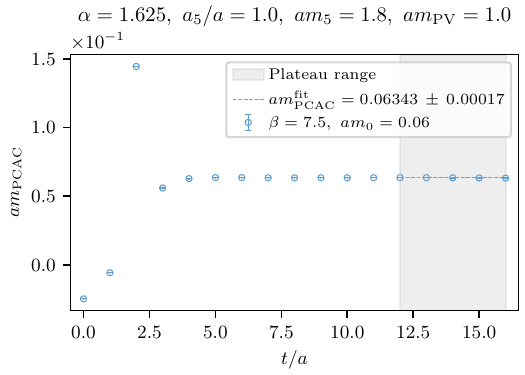} &
    \includegraphics{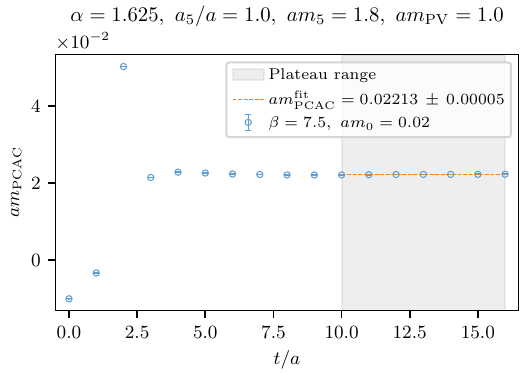} \\
    \includegraphics{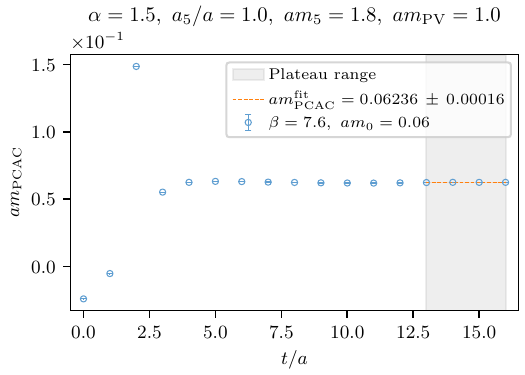} &
    \includegraphics{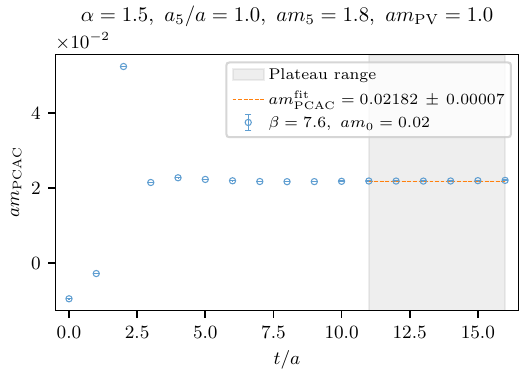} 
    
    \end{tabular}
    \caption{ \label{fig:mpcac_tuned} 
      Representative examples of (folded) plateaux in the PCAC mass, $a m _{\rm PCAC}$, and their corresponding correlated fits, $a m _{\rm PCAC}^{\rm fit}$, in the $Sp(4)$ gauge theory with $N_{\rm f}=2$, as a function of the time separation, $t/a$ (expressed in lattice units), between sink and source, obtained from the (volume and ensemble averaged) two-point correlation functions involving the pseudoscalar meson operator, ${\cal O}^1_{\rm PS}$, evaluated at the midpoint and end-point of the fifth dimensions. The lattice has sizes are $N_5=8$, $N_s=16$, and $N_t=32$. The values of the other parameters, for the examples reported, are displayed on top of the corresponding individual panels and in the legends.}
\end{figure}

As a consistency check of the restoration of the global symmetries of the continuum theory, obtained by suppressing the residual mass with the tuned MDWF parameters listed in Table~\ref{tab:mres_tuned_mobius}, we also determine the partially conserved axial current (PCAC) mass for representative ensembles,  computed following Ref.~\cite{Bursa:2011ru}:
\begin{equation}
am_{\rm PCAC}(t)
\equiv
\frac{\partial_0\, C_{\mathcal{A}\mathcal{O}_{\rm PS}}(t)}
     {2\,C_{\mathcal{O}_{\rm PS}\mathcal{O}_{\rm PS}}(t)},
\end{equation}
where
\begin{equation}
C_{\mathcal{A}\mathcal{O}_{\rm PS}}(t)
=
\sum_{\mathbf{x}}
\left\langle
\mathcal{A}_0^1(\mathbf{x},t)\mathcal{O}^1_{\rm PS}(0)
\right\rangle,
\qquad
C_{\mathcal{O}_{\rm PS}\mathcal{O}_{\rm PS}}(t)
=
\sum_{\mathbf{x}}
\left\langle
\mathcal{O}^1_{\rm PS}(\mathbf{x},t)\mathcal{O}^1_{\rm PS}(0)
\right\rangle,
\end{equation}
and $\partial_0$ denotes the lattice temporal derivative. As already discussed in Eq.~\eqref{eq:pcac_dw_mres}, at sufficiently large Euclidean distances the domain-wall axial Ward identity takes into account the effect of the finite extent of the fifth dimension, at leading order, through the residual mass, $am_{\rm res}$. Comparing Eq.~\eqref{eq:pcac_dw_mres} with the defining PCAC relation,
\begin{equation}
\Delta_\mu \mathcal{A}_\mu^1(x)
=
2am_{\rm PCAC}\,
\mathcal{O}_{\rm PS}^1(x),
\end{equation}
one therefore expects
\begin{equation}
am_{\rm PCAC}
\simeq
am_0+am_{\rm res}.
\label{eq:mpcac_mres}
\end{equation}
Hence, a direct comparison between the PCAC mass and the sum of the input bare mass and the residual mass provides a useful consistency check.

\begin{table}[t]
\centering
\caption{Results for the PCAC mass, $am_{\rm PCAC}$, for the tuned MDWF formulation, in the $Sp(4)$ theory with $N_{\rm f}=2$ dynamical domain-wall fermions. All calculations are performed on lattices with $N_5=8$, $N_s=16$, and $N_t=32$. The table lists, for each ensemble, the lattice coupling, $\beta$, the fermion bare mass, $am_0$, the lattice extents, $N_t$, $N_s$, and $N_5$,  the tuned values of the  MDWF parameters, $\alpha$, $a_5/a$, $am_5$, and $am_{\rm PV}$. Our estimate of the resulting PCAC mass, $am_{\rm PCAC}$, is listed together with the sum of the residual mass and the bare mass, $am_0+am_{\rm res}$ for  direct comparison, the extent of the fit window $\left[ t_{\rm start}^{\rm res}/a,\, t_{\rm end}^{\rm res}/a\right]$, and the reduced $\chi$ squared, $\chi^2/N_{\rm d.o.f.}$. \\}
\begin{tabular}{|l|c|c|c|c|c|c|c|c|c|c|c|c|c|c|}
\hline\hline
Ensemble & $\beta$ & $am_0$ & $N_t$ & $N_s$ & $N_5$ & $\alpha$ & $a_5/a$ & $am_5$ & $am_{\rm PV}$ & $am_{\rm PCAC}$ & $am_0 + am_{\rm res}$ & $t^{m_{\rm PCAC}}_{\rm start}/a$ & $t^{m_{\rm PCAC}}_{\rm end}/a$ & $\chi^2/N_{\rm d.o.f.}$ \\
\hline
MB74M6Sc3 & 7.4 & 0.06 & 32 & 16 & 8 & 1.75 & 1 & 1.8 & 1 & 0.06367(17) & 0.06278(4) & 12 & 16 & 1.239 \\
MB74M2Sc3 & 7.4 & 0.02 & 32 & 16 & 8 & 1.75 & 1 & 1.8 & 1 & 0.02270(8) & 0.02271(5) & 11 & 16 & 1.405 \\
MB75M6Sc3 & 7.5 & 0.06 & 32 & 16 & 8 & 1.625 & 1 & 1.8 & 1 & 0.06343(17) & 0.061867(19) & 12 & 16 & 1.669 \\
MB75M2Sc3 & 7.5 & 0.02 & 32 & 16 & 8 & 1.625 & 1 & 1.8 & 1 & 0.02213(5) & 0.021896(27) & 10 & 16 & 0.525 \\
MB76M6Sc2 & 7.6 & 0.06 & 32 & 16 & 8 & 1.5 & 1 & 1.8 & 1 & 0.06236(16) & 0.061476(14) & 13 & 16 & 1.781 \\
MB76M2Sc2 & 7.6 & 0.02 & 32 & 16 & 8 & 1.5 & 1 & 1.8 & 1 & 0.02182(7) & 0.021462(17) & 11 & 16 & 0.986 \\
\hline\hline
\end{tabular}

\label{tab:pcac_tuned_mobius}
\end{table}

We show in Fig.~\ref{fig:mpcac_tuned} the PCAC mass ratios and fit intervals for representative choices of the ensembles. The fitted values are reported in Table~\ref{tab:pcac_tuned_mobius}, together with the combination $am_0+am_{\rm res}$, to facilitate  direct comparison. We find that the two determinations agree at the percent level, although in several cases their difference is larger than the quoted statistical uncertainties. Such deviations are not unexpected, since the relation $am_{\rm PCAC}\simeq am_0+am_{\rm res}$ is only approximate at finite $N_5$. In particular, the mid-point contribution $\mathcal{O}^1_{\text{PS}, \, q}$ is not necessarily described exactly by a single additive residual-mass term, and additional deviations may arise from finite-lattice-spacing effects, the discretisation of the temporal derivative, residual excited-state contamination, and finite-volume or finite-time effects. Nevertheless, the close numerical agreement provides a useful diagnostic test confirming  that, once $am_{\rm res}$ is suppressed, the PCAC mass entering the axial Ward identity is predominantly determined by the input bare mass of the fermions, with only a small contribution from residual lattice artefacts.

\section{Scale setting and topology}
\label{Sec:ensemble_characterization}

In this section, we briefly describe our treatment of the gradient flow, which we use both as a scale setting procedure, and as a smoothening device to compute (and monitor) the  topology of our ensembles.
The discussion is based on the seminal works, such as those in Refs.~\cite{Luscher:2010iy,Luscher:2013vga,BMW:2012hcm} and ~\cite{Luscher:1981zq,Campostrini:1989dh,Luscher:2011kk,Alexandrou:2017hqw}, but we follow the notation and conventions outlined in Refs.~\cite{Bennett:2017kga,Bennett:2022ftz}.

\subsection{Scale setting}
\label{Sec:scale_setting}

\begin{figure}[t]
    \centering
    \begin{tabular}{cc}

    \includegraphics{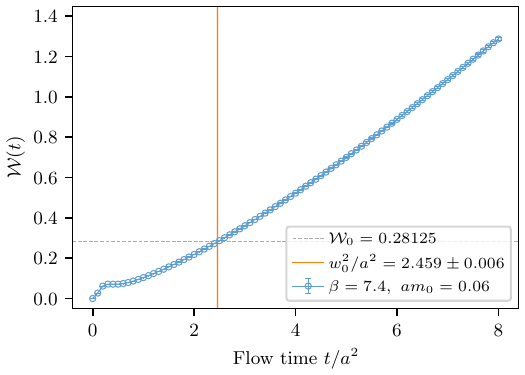} &
    \includegraphics{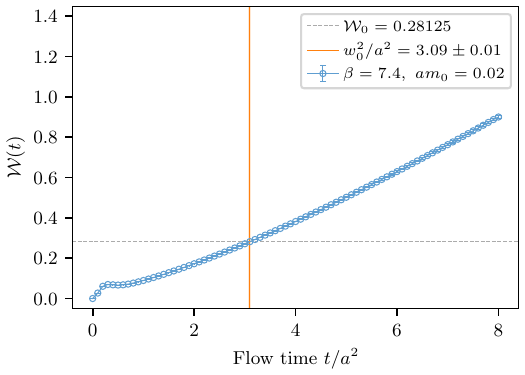} \\
    \includegraphics{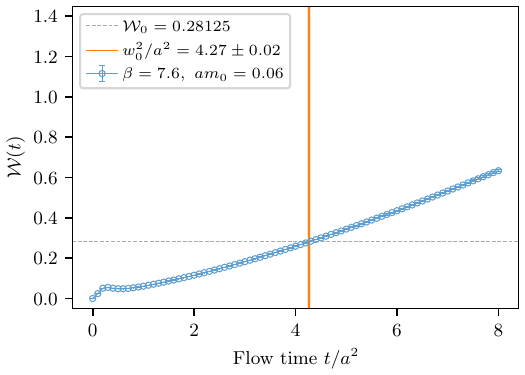} &
    \includegraphics{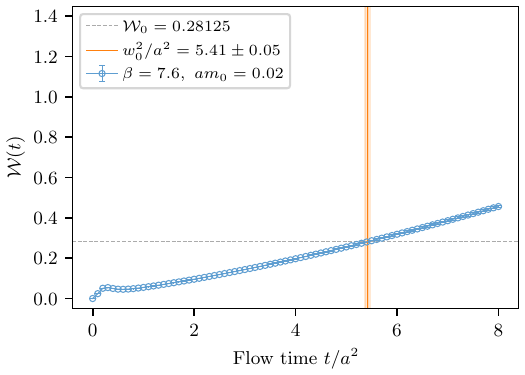} 
    
    \end{tabular}
    \caption{ \label{fig:w0_extraction} 
      Representative examples of the extraction of the Wilson flow scale, $w_0$, from $W(t)$, along the configurations (sampled by skipping five trajectories between adjacent configurations), in the $Sp(4)$ theory coupled to $N_{\rm f} = 2$ Dirac fermions transforming in the fundamental representation, treated with the MDWF formulation. The scale $w_0$ is extracted by solving numerically the constraint $\mathcal{W}(t) \big|_{t = w_0^2} = 0.28125$. The MDWF parameters are $\alpha = 1.75$, $a_5/a = 1.0$, $am_5 = 1.8$, $\beta = 7.4$ (two top panels), $\alpha = 1.5$, $a_5/a = 1.0$, $am_5 = 1.8$, $\beta = 7.6$ (two bottom panels) and $am_0=0.06$ (two left panels), $am_0=0.02$ (two right panels). The horizontal lines correspond to the reference value ${\cal W}_0=0.28125$.}
\end{figure}

Having tuned the MDWF parameters to suppress the residual mass, as discussed in Sect.~\ref{sec:dwf_par_optimization}, we proceed to generate gauge configurations using the Grid software package~\cite{Boyle:2015tjk, Boyle:2016lbp}. The gauge links are, by construction, independent of the fifth dimension; they are thus identical across each four-dimensional slice of the lattice. This property allows us to directly apply the standard scale setting analysis developed for four-dimensional lattice gauge theories.

We elect to employ the gradient flow~\cite{Luscher:2010iy} and its discretised counterpart, the Wilson flow~\cite{Luscher:2013vga}, which serve two complementary purposes. First, the Wilson flow provides a robust, universal approach to setting the lattice scale---independent of the particular theory under study or any model-dependent assumptions. Second, it smoothens out short-distance fluctuations in the gauge configurations, reducing numerical noise and discretisation artefacts, and improving the measurement of observables that are sensitive to all energy scales, such as the topological charge, \( Q \). This dual functionality is especially important in studies of theories for which no known experimental benchmarks exist, and hence having a reliable and model-independent scale is essential for comparing results across different regions of parameter space. 

 The gradient flow is defined by first introducing the flow time, \( t \), as an auxiliary fifth coordinate---not to be confused with the one appearing in the MDWF formalism. The evolution of the gauge fields under the flow is governed by the differential equation
\begin{equation}
\label{eq:Wilsonflowdiffeq}
\frac{\mathrm{d} B_\mu(x,\,t)}{\mathrm{d}t} = D_\nu G_{\nu\mu}(x,\,t)\,,
\end{equation}
with the initial condition \( B_\mu(x,\,0) = A_\mu(x) \), where \( A_\mu(x) \) denotes the original gauge fields. The covariant derivative of the adjoint fields is defined as \( D_\mu \equiv \partial_\mu + [B_\mu,\,\cdot\,] \), and the field strength tensor as \( G_{\mu\nu}(t) = [D_\mu, D_\nu] \).  

\begin{table}[t]
\centering
\caption{\label{tab:ensembles_DWF} 
Characterisation of the ensembles generated to study the dynamics of $Sp(4)$ with $N_{\rm f} = 2$ Dirac fermions transforming in the fundamental representation, realised with the MDWF formalism. The optimised choices of MDWF parameters employed are $am_5 = 1.8, \, a_5 / a= 1.0,  \, am_{\rm PV} = 1.0$ for every ensemble and $\alpha=1.75, \, 1.625, \, 1.5$, for ensembles with $\beta=7.4, \, 7.5, \, 7.6$, respectively. For each ensemble, we tabulate here the lattice coupling, $\beta$, the number of sites in the lattice directions, $N_5$, $N_t$, and $N_s$, the bare mass, $a m_0$, the Wilson flow scale, $w_0/a$ and the ensemble average of the topological charge, $\langle Q_L(w^2_0) \rangle$, the integrated autocorrelation times computed from measurements separated by five HMC trajectories, using the topological charge, \(\tau_{\rm int}^{Q}\), and the Wilson flow scale, \(\tau_{\rm int}^{w_0}\). We also report the number of approximately independent configurations, \(N_{\rm cfg}\), which is obtained by skipping \(2\times(\tau_{\rm int}^{w_0}\times5)\) HMC trajectories between consecutive configurations.\\}
\begin{tabular}{|l|c|c|c|c|c|c|c|c|c|c|c|}
\hline\hline
Ensemble & $\beta$ & $N_5$ & $N_t$ & $N_s$ & $am_0$ & $w_0/a$ & $\langle Q_L(w_0^2) \rangle$ & $\tau_{\rm int}^{Q}$ & $\tau_{\rm int}^{w_0}$ & $am_{\rm res}$ & $N_{\rm cfg}$ \\
\hline
MB74M2Sp & 7.4 & 8 & 48 & 24 & 0.02 & 1.758(3) & 0.1(6) & 12.10(3.75) & 6.94(1.78) & 0.002682(9) & 72 \\
MB74M3Sp & 7.4 & 8 & 48 & 24 & 0.03 & 1.7023(25) & -0.2(7) & 16.16(5.22) & 11.03(3.14) & 0.002688(13) & 89 \\
MB74M4Sp & 7.4 & 8 & 48 & 24 & 0.04 & 1.6397(22) & 0.2(9) & 20.96(6.82) & 7.65(1.75) & 0.002711(11) & 112 \\
MB74M5Sp & 7.4 & 8 & 48 & 24 & 0.05 & 1.6018(18) & 0.0(8) & 18.77(5.37) & 8.82(1.93) & 0.002732(8) & 146 \\
MB74M6Sp & 7.4 & 8 & 48 & 24 & 0.06 & 1.5682(20) & -0.7(6) & 12.61(3.09) & 10.62(2.45) & 0.002742(8) & 126 \\
MB75M2Sp & 7.5 & 8 & 48 & 24 & 0.02 & 2.033(7) & 0.3(8) & 31.06(13.32) & 9.60(2.93) & 0.001895(12) & 55 \\
MB75M3Sp & 7.5 & 8 & 48 & 24 & 0.03 & 1.957(4) & -0.2(7) & 29.44(10.43) & 10.16(2.49) & 0.001885(8) & 111 \\
MB75M4Sp & 7.5 & 8 & 48 & 24 & 0.04 & 1.898(4) & -0.1(1.0) & 56.83(23.80) & 7.36(1.58) & 0.001890(7) & 128 \\
MB75M5Sp & 7.5 & 8 & 48 & 24 & 0.05 & 1.8499(27) & -0.5(6) & 26.83(8.25) & 8.34(1.68) & 0.001902(6) & 170 \\
MB75M6Sp & 7.5 & 8 & 48 & 24 & 0.06 & 1.8073(24) & -0.4(7) & 25.53(7.34) & 9.83(2.02) & 0.001919(5) & 185 \\
MB76M2Sp & 7.6 & 8 & 48 & 24 & 0.02 & 2.326(10) & 0.5(1.4) & 111.64(55.07) & 16.69(5.99) & 0.001470(8) & 37 \\
MB76M3Sp & 7.6 & 8 & 48 & 24 & 0.03 & 2.260(8) & 0.7(6) & 45.22(18.27) & 22.63(7.43) & 0.001469(4) & 59 \\
MB76M4Sp & 7.6 & 8 & 48 & 24 & 0.04 & 2.174(7) & 0.1(9) & 49.57(20.80) & 12.27(3.35) & 0.001482(5) & 56 \\
MB76M5Sp & 7.6 & 8 & 48 & 24 & 0.05 & 2.114(7) & -0.2(7) & 48.96(18.14) & 32.07(10.42) & 0.001485(4) & 86 \\
MB76M6Sp & 7.6 & 8 & 48 & 24 & 0.06 & 2.065(5) & 0.0(7) & 57.71(21.49) & 19.59(5.12) & 0.001495(3) & 100 \\
\hline\hline
\end{tabular}

\label{tab:spectrum_ensembles}
\end{table}

To extract a physical scale from this process, one defines the following quantity~\cite{Luscher:2013vga}:
\beqs
\label{eq:GF_coupling}
\mathcal{E}(t) &\equiv \frac{t^2}{2} \left\langle \mathrm{Tr}\left[ G_{\mu\nu}(t) G_{\mu\nu}(t) \right] \right\rangle\,, 
\eeqs
and the closely related one~\cite{BMW:2012hcm}
\beqs
\label{eq:flowW}
\mathcal{W}(t) &\equiv t \frac{\mathrm{d}}{\mathrm{d}t} \mathcal{E}(t)\,.
\eeqs
One then chooses a reference value for either $\mathcal{E}$ or $\mathcal{W}$, to define a characteristic flow time. Two popular choices in the literature are the flow scales \( t_0 \) and \( w_0 \), defined by the conditions~\cite{BMW:2012hcm}
\beqs
\label{eq:scale_t0}
\mathcal{E}(t_0) &= \mathcal{E}_0\,, 
\eeqs
or, alternatively, 
\beqs
\label{eq:scale_w0}
\mathcal{W}(w_0^2) &= \mathcal{W}_0\,.
\eeqs
The reference values, \( \mathcal{E}_0 \) and \( \mathcal{W}_0 \), are chosen on the basis of theoretical considerations (and convenience). We adopt the prescription in Ref.~\cite{Bennett:2024tex, Bennett:2024cqv} for \( w_0 \), according to which \( \mathcal{W}(t) \big|_{t = w_0^2} = 0.28125 \). This choice descends from the expected scaling of the flow with the quadratic Casimir of the group, and is made to facilitate comparison with similar theories based on other gauge groups. Representative examples of $\mathcal{W}(t)$ and of the extraction of $w_0$ are shown in Fig.~\ref{fig:w0_extraction}, for choices of the lattice parameters, $\beta$ and $a m_0$, at the extrema of the ranges of interest for this study.

The Wilson flow is obtained by replacing the field theory quantities with their lattice counterparts. The Wilson flow acts as a Gaussian smearing of the gauge configurations, effectively averaging over a radius \( \sqrt{8t} \), hence suppressing ultraviolet noise in lattice observables. The continuum gauge field, \( A_\mu(x) \), is replaced by the link variables, \( U_\mu(x) \). Accordingly,  the flow equation is rewritten in terms of flowed fields denoted \( V_\mu(x,t) \), satisfying the initial condition \( V_\mu(x,0) = U_\mu(x) \). The discretised version of the field strength tensor, \( G_{\mu\nu} \), may be provided by the elementary plaquette, \( \mathcal{P}_{\mu\nu} \), introduced in Eq.~\eqref{eq:elementary_plaquette}, or, alternatively, by the clover-leaf operator, \( \mathcal{C}_{\mu\nu} \), which provides a simple form of improvement. Given a generic configuration of link variables, \( U_\mu(x) \), the clover-leaf operator is constructed as a weighted average over the four elementary plaquettes that share a common corner, and is known to reduce discretisation effects for the gauge action and the flow equation employed in this work, particularly for observables like the topological susceptibility. The explicit form we use is the following~\cite{Sheikholeslami:1985ij}:
\beqs
\label{Eq:clover-leaf}
\mathcal{C}_{\mu\nu}(x) &\equiv& 
\frac{1}{8}
\left\{ \frac{}{}U_\mu(x) U_\nu(x+\hat{\mu}) U^\dag_\mu(x+\hat{\nu})U_\nu^\dag(x)       +\right.\nonumber\\
        &&\left. 
  + U_\nu(x) U^\dag_\mu(x+\hat{\nu}-\hat{\mu}) U^\dag_\nu(x-\hat{\mu})U_\mu(x-\hat{\mu})\,+\right.\\
        &&\left.
        + \,U^\dag_\mu(x-\hat{\mu}) U^\dag_\nu(x-\hat{\nu}-\hat{\mu}) U_\mu(x-\hat{\nu}-\hat{\mu})
        U_\nu(x-\hat{\nu})  +\right.\nonumber\\
        &&\left.
              + U^\dag_\nu(x-\hat{\nu}) U_\mu(x-\hat{\nu}) U_\nu(x-\hat{\nu}+\hat{\mu})U_\mu^\dag(x)
        -{\rm h.c.}\frac{}{}\right\}\,.\nonumber
        \eeqs
Table~\ref{tab:ensembles_DWF} lists the ensembles we generated in order to study the dynamics of the $Sp(4)$ gauge theory with $N_{\rm f} = 2$ MDWF, transforming in the  fundamental representation of the group. Together with the choice of bare parameters characterising the ensemble, we report also the flow scale, $w_0 / a$, its integrated autocorrelation time, the residual mass $a m_{\rm res}$, and topological quantities that we introduce in Sect.~\ref{Sec:Q_and_autocorrQ}.

\begin{figure}[t]
    \centering
    \begin{tabular}{cc}
        \includegraphics{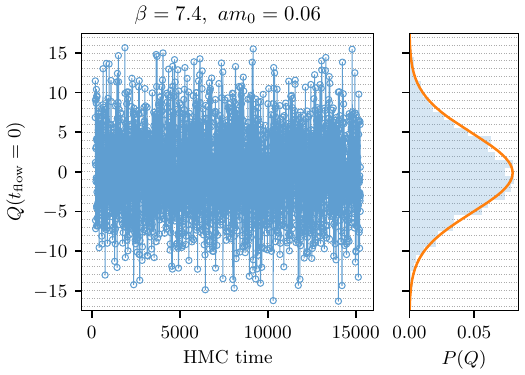}& 
        \includegraphics{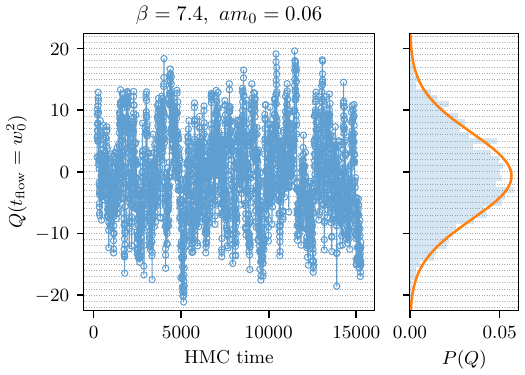}\\
        \includegraphics{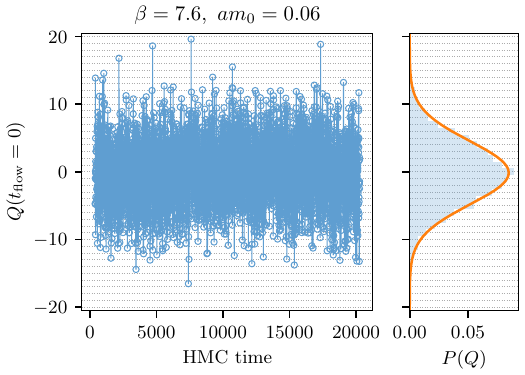} & 
        \includegraphics{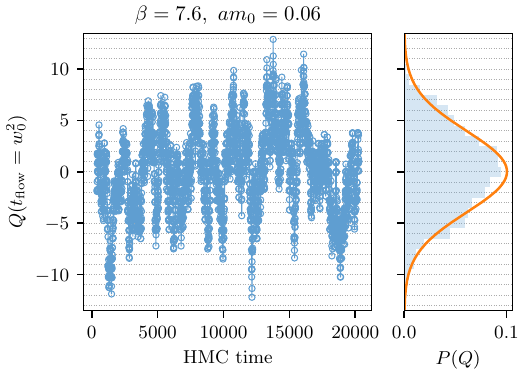}
    \end{tabular}
    \caption{ \label{fig:topological_Q} Representative examples of the evolution of the topological charge, $Q_L$,  along the configurations (sampled from configurations separated by five HMC trajectories), in the Sp(4) theory coupled to $N_{\rm f}=2$ Dirac fermions transforming in the fundamental representation, treated with the MDWF formulation. The topological charge is evaluated at flow times $t=0$ and $t = w_0^2$, and the MDWF parameters are  $\alpha = 1.75$, $a_5 /a = 1.0$, $am_5 = 1.8$ and $\beta=7.4$ (two top panels), $\alpha = 1.5$, $a_5 /a = 1.0$, $am_5 = 1.8$ and $\beta=7.6$ (two bottom panels) and $am_0=0.06$ for all the plots.}
\end{figure}

\subsection{Topological charge and autocorrelation}
\label{Sec:Q_and_autocorrQ}

In the continuum theory, the topological charge, \( Q \),  is defined as follows::
\begin{align}
    Q \equiv \frac{1}{32\pi^2} \int \mathrm{d}^4x\, \epsilon^{\mu \nu \rho \sigma} \, \mathrm{Tr} \left[ G_{\mu\nu}(x)\, G_{\rho\sigma}(x) \right]\,,
\end{align}
where \( G_{\mu\nu} \) is the field strength tensor. On the lattice, we define the flowed topological charge density as
\begin{align}
    Q_L(x,t) \equiv \frac{1}{32 \pi^2} \varepsilon^{\mu\nu\rho\sigma} \, \mathrm{Tr}\left[ \mathcal{C}_{\mu\nu}(x,t)\, \mathcal{C}_{\rho\sigma}(x,t) \right]\,,
\end{align}
where \( \mathcal{C}_{\mu\nu}(x,t) \) denotes the clover-leaf definition of the field-strength tensor, as in Eq.~(\eqref{Eq:clover-leaf}), evolved with the Wilson flow equation, and evaluated at flow time \( t \), to remove ultraviolet fluctuations, as stated in Sect.~\ref{Sec:scale_setting}. In order to monitor the behaviour of the topology in our ensembles, we compute the ensemble average of the flowed topological charge, \( \langle Q_L(t = w_0^2) \rangle \), with \( w_0 \) the Wilson flow scale.

To assess the size of statistical uncertainties in our measurements, we estimate the autocorrelation times of relevant observables. For a generic observable, \( X \), we denote the individual measurements as \( X_i \), where \(i=1,\ldots,N\) labels the configurations in Monte Carlo time, and the sample mean as \( \bar{X} \). We assume that the time series, \( X_i \), is fully thermalised (if otherwise, we proceed to excluding portions of the ensembles, in the early stages of the Monte Carlo generation, until we retain only thermalised configurations). The Madras--Sokal integrated autocorrelation time~\cite{Madras:1988ei,Wolff:2003sm,Luscher:2004pav}
is then defined as
\begin{align}
    \tau^{X}_{\mathrm{int}}(k)
    = \frac{1}{2} + \sum_{t=1}^{k} \rho^{X}(t)\,,
\end{align}
where the normalised autocorrelation function is
\begin{align}
    \rho^{X}(t)
    = \frac{\Gamma^{X}(t)}{\Gamma^{X}(0)}\,,
\end{align}
with
\begin{align}
    \Gamma^{X}(t) = \frac{1}{N - t}
    \sum_{i=1}^{N - t}
    \left( X_i - \bar{X} \right)
    \left( X_{i + t} - \bar{X} \right)\,.
\end{align}

In Table~\ref{tab:ensembles_DWF}, we report the average topological charge, \( \langle Q_L(t = w_0^2) \rangle \), and the integrated autocorrelation time, \( \tau_{\mathrm{int}}^Q \), associated with the topological charge. Representative examples of our monitoring of the topological charge are shown in Fig.~\ref{fig:topological_Q}. The value of \( \langle Q_L(t = w_0^2) \rangle \) for all ensembles is compatible with zero, at the $2\sigma$ level. The distribution of \( Q \) is approximately Gaussian and centred around zero, as expected in the limit of infinite simulation time. These observations indicate that, in the range of parameter space explored in this study, the simulations do not show evidence of topological freezing. 

In summary, we find that the \( Sp(4) \) gauge theory with \( N_{\mathrm{f}} = 2 \) Dirac fermions transforming in the fundamental representation, treated with the MDWF formulation, and using the optimised choices of parameters, is well-suited for large-scale numerical simulations. The topological sector is effectively sampled with no evidence of topological freezing, and the residual mass \( am_{\mathrm{res}} \) remains small compared to the bare quark masses \( am_0 \), in the ensembles we consider, ensuring good global symmetry properties for the fermions.

\section{Meson masses and decay constants}
\label{Sec:spectr_decayconst}

This section presents the main results of the paper. We discuss our measurements of the spectral quantities extracted from meson two-point correlation functions---masses and decay constants of the ground state composite particles. We  analyse the lattice measurements in terms of a generalisation of chiral perturbation theory ($\chi$PT) that allows to perform a continuum (and massless) extrapolation. We compare our results to existing ones, published in the literature, for the same theory~\cite{TELOS:2026alk},  but obtained using  Wilson fermions. We find agreement with previous measurements, confirming both the validity of our implementation of the MDWF formulation, and the effectiveness of our extrapolations. But we also find that the measurements obtained with MDWF are much closer to the continuum limit than what available with Wilson fermions,  at least for the region of moderate fermion masses relevant to this paper. This observation confirms that the automatic improvement that is built into the MDWF formulation of the lattice theory allows us to achieve high-precision measurements in theories in this class, by suppressing discretisation artefacts below the statistical uncertainties.

\subsection{Two-point correlation functions, spectroscopy and decay constants}
\label{Sec:spectra_decay_theo}

\begin{table}[t]
    \centering
    \caption{Interpolating operators used to study flavoured mesons in the $Sp(4)$ theory with $N_{\rm f}=2$, Dirac fermions in the fundamental representation,  $Q^i$. We omit hypercolour and spinor indices for simplicity, while we specify the choice of combination of flavors used to represent the non-trivial multiplets of the unbroken global $Sp(4)$ symmetry. For each operator, we tabulate their spin and parity quantum numbers, $J^P$, and the transformation properties under the global, unbroken $Sp(4)$ group. The naming conventions follow Ref.~\cite{Bennett:2024cqv}.\\}
    \begin{tabular}{ |c|c|c|c|c| }
        \hline\hline
        $~~~~$Label  $~~~~$ &   $~~~~$Operator $\mathcal{O}$  $~~~~$ & $J^P$ & $Sp(4)$  \\
        \hline
        PS & $ \overline{Q^1} \gamma_5 Q^2$ & $0^{-}$ & $5$  \\
        V  & $ \overline{Q^1} \gamma_\mu Q^2$ & $1^{-}$ & $10$  \\
        AV & $ \overline{Q^1}\gamma_5 \gamma_\mu Q^2$ & $1^{+}$ & $5$  \\ 
        \hline\hline
    \end{tabular}
    \label{tab:meson_ops}
\end{table}

We summarise in this subsection the methodology adopted to extract spectral quantities and matrix elements from numerical lattice data. We focus on the  pseudoscalar (PS), vector (V), and axial-vector (AV), flavored mesons. In order to avoid contamination with singlets,  all  fermion-bilinear interpolating operators take the form:
\begin{equation}
    \mathcal{O} = \overline{Q^1}(x) \Gamma Q^2(x),
    \label{eq:meson_ops}
\end{equation}
where $\Gamma$ denotes a suitable combination of Dirac gamma matrices, chosen to source particles with the quantum numbers of interest, such as the spin and parity,  $J^P$, and the multiplet of the unbroken $Sp(4)$ global symmetry they belong to. The meson channels investigated in this work are summarised in Table~\ref{tab:meson_ops}.

The starting points of the physics analysis are the Euclidean two-point correlation function. The extraction of the mass and decay constant of the PS states requires to perform the lattice measurement of the following ensemble averages: 
\begin{align}
\label{eq:corr_1}
    C_{\rm PS, PS}(t) &= \sum_{\vec{x}} \langle O_{\rm PS}(\vec{x}, t) \, O_{\rm PS}^\dagger(\vec{0}, 0) \rangle\,, \\
\label{eq:corr_2}
    C_{\rm AV, PS}(t) &= \sum_{\vec{x}} \langle O_{\rm AV}(\vec{x}, t) \, O_{\rm PS}^\dagger(\vec{0},0) \rangle\,,
\end{align}
where the fermions are evaluated at the four-dimensional boundaries of the fifth dimension, $s = 0$ and $s = N_5 - 1$, as defined in Eq.~\eqref{eq:dwf_physical_fields}.  We are also interested in the V mesons, their mass being extracted from the correlation function
\beqs
C_{\rm V, V}(t) &= \sum_{\vec{x}} \langle O_{\rm V}(\vec{x}, t) \, O_{\rm V}^\dagger(\vec{0},0) \rangle\,.
\eeqs
In the region of  large Euclidean time separations, $t$, and with anti-periodic boundary conditions for the fermions, these correlation functions are expected to decay exponentially,
\beqs
\label{eq:large_t_meson_PS}
    C(t) & \approx A \left( e^{-m t} \pm e^{-m(T - t)} \right)\,,
\eeqs
where \(m\) stands for the mass of the lightest meson in the channel under consideration, with the plus (minus) sign applying to correlation functions that are symmetric (antisymmetric) under time reflection about the temporal midpoint, \(t\rightarrow T-t\).

Operationally, the mass of the meson of interest, $a m_{\rm PS}$ or $a m_{\rm V}$, is measured by studying the effective mass
\begin{equation}
    am_{\rm eff}(t) \equiv \cosh^{-1} \left[ \frac{C(t+1) + C(t-1)}{2C(t)} \right]\,,
\end{equation}
and identifying plateaux in $am_{\rm eff}(t)$, in an interval of  asymptotically large values of $t$. This process defines  fitting windows in $t$, and ground-state isolation is implemented through multi-exponential fits, by minimising the correlated chi-square,
\begin{equation}
\label{eq:correlated_chisq}
    \chi^2 \equiv \chi^2_{\rm prior}+\sum_{t, t'} \left[ h^{(k)}(t) - C(t) \right] \text{Cov}^{-1}_{tt'} \left[ h^{(k)}(t') - C(t') \right]\,,
\end{equation}
in which we take the model functions of the form $h^{(k)} (t) = \sum_{n = 1}^{k} \mathcal{B}_n \big(\, e^{-tE_n}\pm e^{-(T-t)E_n}\big)$. The Bayesian prior, $\chi^2_{\rm prior}$,  is included to better isolate the energy states. Following to Ref.~\cite{Lepage:2001ym}, this contribution explicitly reads as follows:
\begin{equation}
\label{eq:chi_priors}
\chi^2_{\rm prior}
= \sum_{n=1}^{k}
\left( \frac{\mathcal{B}_n - \mathcal{B}_n^{\rm prior}}
{\sigma^{\rm prior}_{\mathcal{B}_n}} \right)^2
+ \sum_{n=1}^{k}
\left( \frac{E_n - E_n^{\rm prior}}
{\sigma^{\rm prior}_{E_n}} \right)^2 \,.
\end{equation}
For $k > 1$, the prior is obtained from a fit with $k - 1$ exponential terms, and the resulting values for $E_n$ and $\mathcal{B}_n$ used as priors in Eq.~\eqref{eq:chi_priors}. For $k = 1$, the prior is set to $\chi^2_{\rm prior} = 0$.

The uncertainties on the fit parameters, generically denoted as ${\theta_i}$, are obtained from the covariance matrix computed at the minimum of the $\chi^2$, defined in Eq.~(\ref{eq:correlated_chisq}). In the Gaussian approximation around the best-fit point, \(\theta_i=\theta_i^\star\), the covariance matrix around for the parameters is given by the inverse of the Hessian,
\begin{equation}
\mathrm{Cov}_{ij}  \approx 2
\left( \frac{\partial^2 \chi^2}{\partial \theta_i \, \partial \theta_j} \right)^{-1}_{\theta = \theta^\star} \,.
\end{equation}

\begin{table}[t]
\centering
\caption{Numerical results of the lattice measurements of $am_{\rm PS}$, $af_{\rm PS}$, $am_{\rm V}$, and $Z_A$, obtained by using the optimised MDWF formulation to study the $Sp(4)$ theory coupled to $N_{\rm f}=2$ Dirac fermions transforming in the fundamental representation. The ensembles are characterised in Tab.~\ref{tab:ensembles_DWF}.  For each ensemble, we report the inferred measurement of the mass of the lightest flavored pseudoscalar particle, $a m_{\rm PS}$, and the corresponding reduced chi square, $\chi^2_{\rm PS}/N_{\rm d.o.f.}$, the result of the combined extraction of the pseudoscalar mass and (unrenormalised) decay constant, $a m_{\rm PS}^{\rm comb}$,  and $af_{\rm PS}$, with the associated reduced chi square, $\chi^2_{\rm comb}/N_{\rm d.o.f.}$, the mass of the lightest  flavored vector,  $a m_{\rm V}$, and the associated reduced chi square, $\chi_{\rm V}^2/N_{\rm d.o.f.}$, and the measurement of the axial renormalisation constant, $Z_A$, with its associated reduced chi square, $\chi^2_{\rm Z}/N_{\rm d.o.f.}$. In parentheses next to each reduced chi square we also report the corresponding number of degrees of freedom, $N_{\rm d.o.f.}$. Statistical uncertainties are returned by the chi-square analysis described in the main text.
\\}
\label{tab:spectrum_results}
\begin{tabular}{|l|c|c|c|c|c|c|c|c|c|}
\hline\hline
Ensemble & $am_{\rm PS}$ & $\chi^2_{\rm PS}/N_{\rm d.o.f.}$ & $am_{\rm PS}^{\rm comb}$ & $af_{\rm PS}$ & $\chi^2_{\rm comb}/N_{\rm d.o.f.}$ & $am_{\rm V}$ & $\chi^2_{\rm V}/N_{\rm d.o.f.}$ & $Z_A$ & $\chi^2_{Z}/N_{\rm d.o.f.}$ \\
\hline
MB74M2Sp & 0.2717(13) & 0.49 (13) & 0.2702(17) & 0.0817(14) & 0.44 (12) & 0.435(9) & 0.22 (11) & 0.726(7) & 1.09 (10) \\
MB74M3Sp & 0.3283(12) & 0.80 (6) & 0.3278(12) & 0.0869(13) & 0.66 (12) & 0.470(7) & 1.37 (7) & 0.747(6) & 0.72 (6) \\
MB74M4Sp & 0.3816(9) & 0.74 (8) & 0.3823(10) & 0.0965(10) & 1.04 (12) & 0.522(3) & 0.76 (12) & 0.742(4) & 0.68 (9) \\
MB74M5Sp & 0.4289(7) & 0.77 (8) & 0.4288(7) & 0.1026(8) & 0.47 (14) & 0.5576(29) & 1.31 (7) & 0.7487(25) & 0.81 (12) \\
MB74M6Sp & 0.4733(6) & 0.45 (10) & 0.4731(7) & 0.1080(7) & 0.54 (14) & 0.593(3) & 1.30 (9) & 0.7536(22) & 0.54 (14) \\
MB75M2Sp & 0.2482(26) & 1.10 (6) & 0.2464(22) & 0.0689(12) & 1.19 (16) & 0.386(13) & 0.35 (10) & 0.764(5) & 0.75 (9) \\
MB75M3Sp & 0.3031(14) & 0.66 (6) & 0.3022(15) & 0.0767(12) & 0.86 (8) & 0.431(5) & 0.49 (9) & 0.762(5) & 0.61 (9) \\
MB75M4Sp & 0.3529(13) & 1.03 (6) & 0.3535(13) & 0.0849(10) & 0.63 (12) & 0.469(4) & 0.49 (6) & 0.7553(26) & 0.43 (12) \\
MB75M5Sp & 0.3975(8) & 1.21 (7) & 0.3978(8) & 0.0894(7) & 0.62 (12) & 0.5049(27) & 0.91 (10) & 0.7647(26) & 0.74 (12) \\
MB75M6Sp & 0.4412(8) & 1.01 (6) & 0.4409(8) & 0.0948(7) & 0.74 (8) & 0.5403(22) & 0.93 (8) & 0.7591(22) & 0.55 (12) \\
MB76M2Sp & 0.233(3) & 1.01 (7) & 0.234(5) & 0.059(3) & 1.50 (18) & 0.349(11) & 0.96 (6) & 0.779(8) & 1.48 (15) \\
MB76M3Sp & 0.2852(16) & 1.11 (7) & 0.2846(14) & 0.0680(11) & 0.73 (12) & 0.386(9) & 0.61 (6) & 0.777(7) & 0.94 (8) \\
MB76M4Sp & 0.3327(19) & 0.96 (7) & 0.3324(19) & 0.0744(17) & 1.14 (14) & 0.432(5) & 0.96 (10) & 0.778(6) & 1.07 (9) \\
MB76M5Sp & 0.3744(13) & 0.62 (6) & 0.3739(14) & 0.0815(10) & 1.33 (8) & 0.468(4) & 0.71 (6) & 0.768(3) & 1.14 (12) \\
MB76M6Sp & 0.4150(9) & 0.62 (8) & 0.4148(9) & 0.0845(9) & 0.91 (14) & 0.4957(29) & 0.75 (6) & 0.7760(27) & 0.39 (14) \\
\hline\hline
\end{tabular}

\end{table}

Besides the masses of the lightest states, we measure also hadron-to-vacuum matrix elements (and decay constants) for the mesons. They are extracted from the pre-factors in the correlation functions. These constants are encoded in the spectral decomposition of the correlation function matrix:
\begin{equation}
\label{eq:correlation_matrix}
    C_{ij}(t) \equiv \langle O_i(t) O_j^\dagger(0) \rangle = \sum_n \frac{1}{2E_n} \langle 0 | O_i | n \rangle \langle n | O_j^\dagger | 0 \rangle \left( e^{-E_n t} \pm e^{-E_n (T - t)} \right)\,,
\end{equation}
with appropriate choice of the sign in the terms inside the bracket as detailed in Eqs.~\eqref{eq:c_psps},~\eqref{eq:c_avps} below.
The decay constant, $ f_{\rm PS}$, of the pseudoscalar meson is defined by the following relations:\footnote{These normalisation  are such that applying the same conventions to two-flavour QCD yields $f_{\rm PS}=93$ MeV.}
\beqs
    \label{eq:fpi} 
    \langle 0 | \overline{Q^1} \gamma_5 \gamma_\mu Q^2 | {\rm PS} \rangle &=&{ \sqrt{2}} f_{\rm PS} p_\mu\,,\\
     \langle 0 | \overline{Q^1}  \gamma_\mu Q^2 | {\rm V} \rangle &=& { \sqrt{2}}  f_{\rm V} m_{\rm V} \epsilon_\mu\,,\\
     \langle 0 | \overline{Q^1} \gamma_5 \gamma_\mu Q^2 | {\rm AV} \rangle &=& { \sqrt{2}}  f_{\rm AV} m_{AV} \epsilon_\mu\,,
\eeqs
with the polarisation vector, $\epsilon_\mu$, satisfying $\epsilon^*_\mu \epsilon^\mu = 1$ and orthogonal to $p_\mu$, $\epsilon_{\mu}p^{\mu}=0$. For large values of $t$, the correlation functions relevant to the measurement of $f_{\rm PS}$  behave as follows:
\begin{align}\label{eq:c_psps}
    C_{\rm PS, PS}(t) &\xrightarrow[t \to \infty]{} \frac{|\langle 0 | O_{\rm PS} | {\rm PS} \rangle|^2}{2 m_{\rm PS}} \left( e^{-m_{\rm PS} t} + e^{-m_{\rm PS}(T - t)} \right), \\ \label{eq:c_avps}
    C_{\rm AV, PS}(t) &\xrightarrow[t \to \infty]{} \frac{f_{\rm PS} \langle 0 | O_{\rm PS} | {\rm PS} \rangle^*}{\sqrt{2}} \left( e^{-m_{\rm PS} t} - e^{-m_{\rm PS}(T - t)} \right),
\end{align}
allowing for the simultaneous fit of both correlation functions, and  the extraction of  $f_{\rm PS}$ from their combination.

The decay constants are multiplicatively renormalised, with the renormalised quantities defined as follows:
\begin{equation}
    f_{\rm PS}^{\rm ren} \equiv Z_A f_{\rm PS}\,,
    \label{eq:ren_const_meson}
\end{equation}
where $Z_A$ is the axial current renormalisation constant. In the context of domain-wall fermions, this  renormalisation is related to the PCAC relation, and to the renormalisation of $\mathcal{A}_\mu^A$, as detailed in Eq.~\eqref{eq:pcac_dw}. The coefficient $Z_A$ is computed non-perturbatively, by defining the following ratio~\cite{Aoki:2002vt}:
\begin{equation}
\label{eq:ren_const_dwfs}
    Z_A (t) \equiv \frac{1}{2} \left[ \frac{R(t+a/2) + R(t-a/2)}{2L(t)} + \frac{2R(t+a/2)}{L(t)+L(t+a)} \right]\,,
\end{equation}
where
\begin{equation}
R\left(t+\tfrac{a}{2} \right)\equiv \sum_{\vec{x}, \vec{y}} \langle \mathcal{A}_0^{A}(\vec{x}, t) \mathcal{O}^{\rm A}_{\rm PS}(\vec{y},0)\rangle\,, \qquad {\rm and} \qquad 
L(t) \equiv  \sum_{\vec{x}} \langle \mathcal{O}^{\rm AV,\,A}_0(\vec{x},t) \mathcal{O}^{\rm A}_{\rm PS}(\vec{y},0)\rangle\,.
\end{equation} 
One then looks for a plateau at large $t$. The appearance of shifts by $a/2$  accounts for the fact that $\mathcal{A}_\mu^A$ is defined between adjacent lattice sites. The structure  of Eq.~\eqref{eq:ren_const_dwfs} ensures that scaling violations arise only at $\mathcal{O}(a^2)$, not  $\mathcal{O}(a)$.

     \begin{figure}[t]
     \begin{center}
    \includegraphics{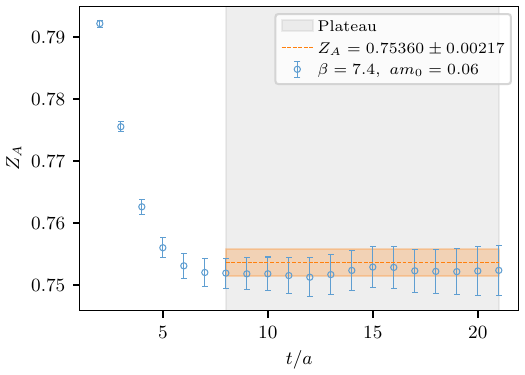}
    \includegraphics{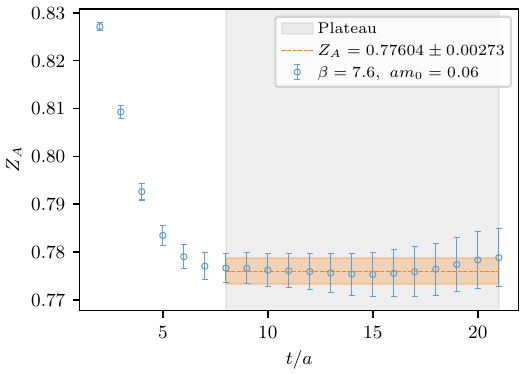}
    \caption{Representative examples of plateaux appearing in the non-perturbative calculation of the renormalisation factor, $Z_A$, for the two ensembles MB76M6Sp and MB74M6Sp listed in Table~\ref{tab:ensembles_DWF}.
    \label{fig:ZA_fit_plateau}}
    \end{center}
     \end{figure}

Lattice calculations are affected by finite-size effects, which introduce systematic effects. We performed a systematic study of these effects, to ensure they are  subdominant in our analysis, and report some illustrative examples of the results in Appendix~\ref{Sec:finite_size_effects}. For domain-wall fermions, additional artefacts arise due to the finite extent, $N_5$, in the fifth dimension, as discussed in Sect.~\ref{sec:dwf_par_optimization}. We extract the ${\rm PS}$ meson mass, $a m_{\rm PS}$, and unrenormalised decay constant, $a f_{\rm PS}$, in the ensembles listed in Table~\ref{tab:ensembles_DWF}, for which both systematic effects due to finite volume and finite $N_5$ are negligibly small, compared to statistical uncertainties.  For each ensemble, we report in Table~\ref{tab:spectrum_results} our estimates of $am_{\rm PS}$, $am_{\rm V}$, the decay constant $af_{\rm PS}$, and the renormalisation factor $Z_A$, together with the corresponding reduced chi-square values and numbers of degrees of freedom. We note that the largest and smallest reduced chi-square values occur for the lightest ensembles. In particular, the largest value, $\chi^2/{\rm d.o.f.}=1.50$, is found for the combined $f_{\rm PS}$ fit at $\beta=7.6$, $am_0=0.02$. This ensemble has the smallest number of independent configurations, $N_{\rm cfg}=37$, and the largest autocorrelations, which makes the fit more challenging. The smallest value, $\chi^2/{\rm d.o.f.}=0.22$, is found in the vector channel at $\beta=7.4$, $am_0=0.02$, where the plateau uncertainties are comparatively large. Discretisation effects are expected to be largest for this ensemble, since it corresponds to the coarsest lattice spacing in the spectrum analysis, and it also exhibits the largest residual mass fraction, $m_{\rm res}/m_0 \simeq 0.134$. For all the spectrum measurements, we analyse the two-point functions using the Hadrons framework~\cite{antonin_portelli_2023_8023716}. The meson masses are extracted using the interpolating operators listed in Table~\ref{tab:meson_ops}, and $a f_{\rm PS}^{\rm ren}$ is determined from Eq.~\eqref{eq:ren_const_meson}. We also show two examples of plateaux used to extract $Z_A$, in Fig.~\ref{fig:ZA_fit_plateau}.

\subsection{Continuum extrapolation and comparison to Wilson fermions}
\label{Sec:chiPT_masses_fpi}

\begin{figure}
        \centering
        \includegraphics{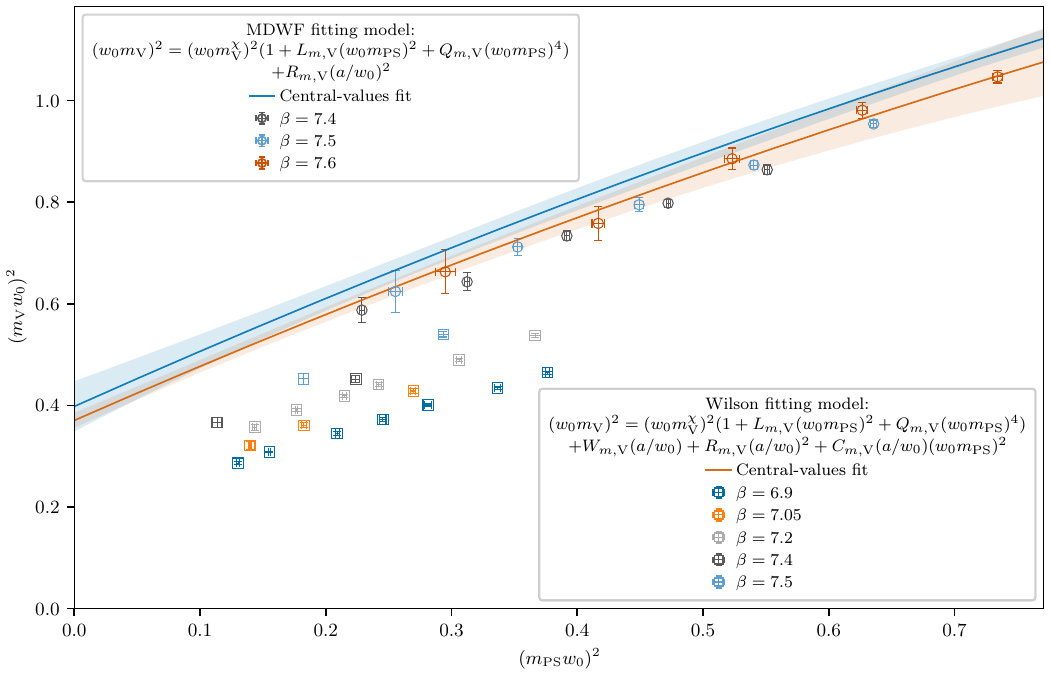}
    \caption{ \label{fig:mv_mps} Numerical results for the measurement of the mass of the lightest, flavored ${\rm  V}$ meson,  $w_0^2m_{\rm V}^2$, expressed in units of the Wilson scale, $w_0$, extracted with the  reference value $\mathcal{W}_0=0.28125$, as a function of the mass of the lightest ${\rm PS}$ meson, $w_0^2m_{\rm PS}^2$.  The ensembles have been generated with the optimised choices of  MDWF parameters listed in Table~\ref{tab:spectrum_ensembles}. Circle markers of different color denote the three values of the lattice coupling, $\beta = 7.4, \, 7.5, \, 7.6$, as explained in the legend. We also display, for comparison, the results of the measurements of the same observables from lattice calculations using Wilson fermions, taken from Ref.~\cite{TELOS:2026alk}, represented by squared markers of different color, representing lattice couplings  $\beta = 6.9, \, 7.05, \, 7.2, \, 7.4, \, 7.5$, as quoted in the legend. The plot shows also the two continuum extrapolations, represented by the continuum lines, with the shaded bands illustrating the uncertainties of our extrapolations.}
\end{figure}

\begin{figure}[t]
\includegraphics{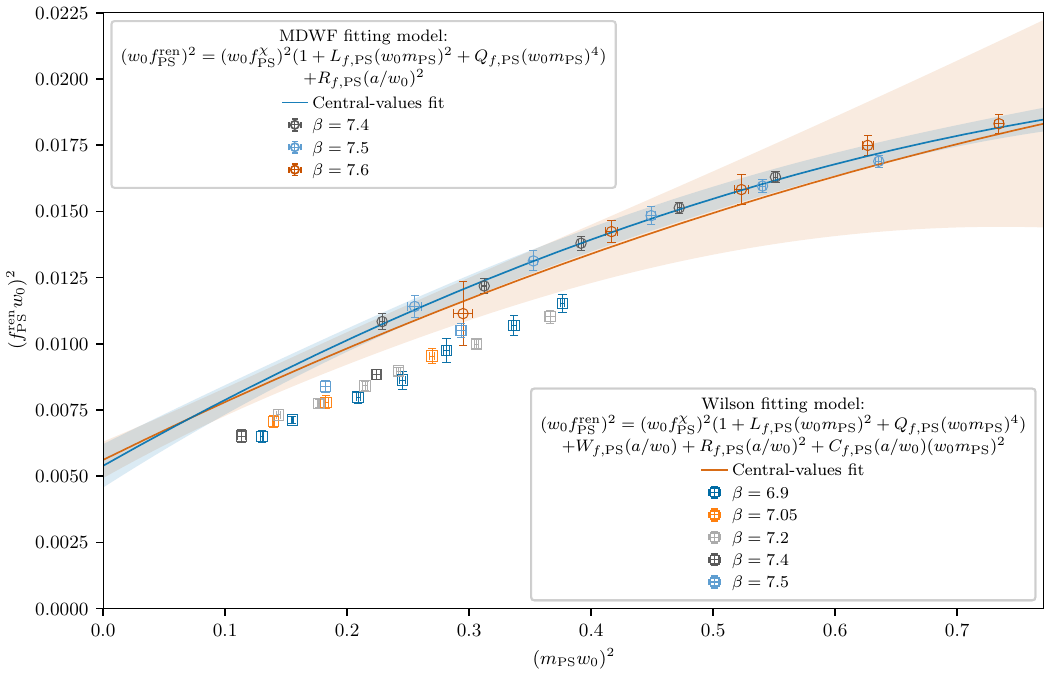}
\caption{ \label{fig:fmps_mps} 
         Numerical results for the measurement of the (renormalised) decay constant of ${\rm PS}$ mesons, expressed in units of the Wilson scale, $w_0^2f_{\rm PS}^2$, expressed in units of the Wilson scale, $w_0$, extracted with the  reference value $\mathcal{W}_0=0.28125$, as a function of the mass of the lightest ${\rm PS}$ meson, $w_0^2m_{\rm PS}^2$. The ensembles have been generated with the optimised choices of  MDWF parameters listed in Table~\ref{tab:spectrum_ensembles}. Circle markers of different color denote the three values of the lattice coupling, $\beta = 7.4, \, 7.5, \, 7.6$, as explained in the legend. We also display, for comparison, the results of the measurements of the same observables from lattice calculations using Wilson fermions, taken from Ref.~\cite{TELOS:2026alk}, represented by squared markers of different color, representing lattice couplings  $\beta = 6.9, \, 7.05, \, 7.2, \, 7.4, \, 7.5$, as quoted in the legend. The plot shows also the two continuum extrapolations, represented by the continuum lines, with the shaded bands illustrating the uncertainties of our extrapolations.}
\end{figure}

We perform a direct comparison of our new spectroscopy results, obtained using the MDWF formulation with our optimised choices of MDWF parameters, with existing, published results, taken from Ref.~\cite{TELOS:2026alk} (see also Ref.~\cite{Bennett:2019jzz}), in which the analysis employs Wilson fermions, for the same $Sp(4)$ gauge theory with $N_{\rm f} = 2$ Dirac fermions transforming in the fundamental representation. Our results are shown in Figs.~\ref{fig:mv_mps} and~\ref{fig:fmps_mps}, expressed  in units of the Wilson flow scale $w_0$, measured with the reference value $\mathcal{W}_0=0.28125$,  for the flavored ${\rm V}$-meson mass squared, $(w_0 m_{\rm V})^2$, and the renormalised {\rm PS}-meson decay constant, $(w_0 f_{\rm PS})^2$, respectively. We show both these quantities as a function of  the ${\rm PS}$-meson mass squared, $(w_0 m_{\rm PS})^2$. 

In both figures we also present the result of continuum extrapolations, obtained by taking the $a/w_0 \to 0$ limit after fitting the lattice measurements to an expansion inspired by Wilson chiral perturbation theory (W$\chi$PT)~\cite{Sheikholeslami:1985ij,Rupak:2002sm} (see also Ref.~\cite{Sharpe:1998xm}, as well as the literature on lattice improvement~\cite{Symanzik:1983dc,Luscher:1996sc}). For the Wilson fermion data taken from Ref.~\cite{TELOS:2026alk}, we use the following functional forms:
\begin{align}
(w_0 m_{\rm V})^2_{\rm{W}}
&= \left(w_0 m^{\chi,\rm{W}}_{\rm V}\right)^2
\Big(
1 + L^{ \, \rm{W}}_{m, \rm{V}} (w_0 m_{\rm PS})^2
+ Q^{ \, \rm{W}}_{m, \rm{V}} (w_0 m_{\rm PS})^4
\Big)
\nonumber \\
&\qquad
+ W^{\, \rm{W}}_{m, \rm{V}} \, (a/w_0)
+ R^{\, \rm{W}}_{m, \rm{V}} \, (a/w_0)^2
+ C^{\, \rm{W}}_{m, \rm{V}} \, (a/w_0)(w_0 m_{\rm PS})^2 \, , 
\label{eq:chiPT_W_mv} \\
(w_0 f_{\rm PS}^{\rm ren})^{2}_{\rm{W}}
&= (w_0 f^{\chi,\rm{W}}_{\rm PS})^2
\Big(
1 + L^{ \, \rm{W}}_{f, \rm{PS}} (w_0 m_{\rm PS})^2
+ Q^{ \, \rm{W}}_{f, \rm{PS}} (w_0 m_{\rm PS})^4
\Big)
\nonumber \\
&\qquad
+ W^{\, \rm{W}}_{f, \rm{PS}} (a/w_0)
+ R^{\, \rm{W}}_{f, \rm{PS}} (a/w_0)^2
+ C^{\, \rm{W}}_{f, \rm{PS}} (a/w_0)(w_0 m_{\rm PS})^2 .
\label{eq:chiPT_W_fps}
\end{align}
which include the leading terms of a double expansion in the small parameters $(w_0 m_{\rm PS})^2$ and $(a/w_0)$.  Since the MDWF data extend to a region of parameter space with larger mass for the ${\rm PS}$ mesons, with respect to the measurements with Wilson fermions, we include in the ansatz higher-order corrections proportional to $(w_0 m_{\rm PS})^4$, $(a/w_0)^2$, and the mixed term $(a/w_0)(w_0 m_{\rm PS})^2$, in order to allow us to extrapolate the fits over the whole mass range of interest.

For the MDWF measurements, as for the case of Wilson fermions, the continuum extrapolation is informed by a double expansion in both the  mass squared of the flavored ${\rm PS}$ mesons, $(w_0 m_{\rm PS})^2$, and the squared lattice spacing, $(a/w_0)^2$~\cite{Kelly:2010oxx}. The major difference is that the lattice spacing enters only at ${\cal O}(a^2/w_0^2)$. Hence, one expects faster (improved) convergence towards the continuum limit, and smaller discretisation artefacts to be visible in the individual lattice measurements. Guided by available measurements, we find a good fit of our data by using the following parameterisations:
\begin{gather}
\label{eq:eq:chiPT_DWF_mv}
\left(w_0 m_{\mathrm{V}}\right)^2_{\mathrm{DW}} =
\left(w_0 m^{\chi,\mathrm{DW}}_{\mathrm{V}}\right)^2
\Big(
1
+ L^{\,\mathrm{DW}}_{m,\,\mathrm{V}} \left(w_0 m_{\mathrm{PS}}\right)^2
+ Q^{\,\mathrm{DW}}_{m,\,\mathrm{V}} \left(w_0 m_{\mathrm{PS}}\right)^4
\Big)
+ R^{\,\mathrm{DW}}_{m,\,\mathrm{V}}(a/w_0)^2, \\
\label{eq:eq:chiPT_DWF_fps}
\left(w_0 f_{\mathrm{PS}}^{\rm ren}\right)^2_{\mathrm{DW}} =
\left(w_0 f^{\chi,\mathrm{DW}}_{\mathrm{PS}}\right)^2
\Big(
1
+ L^{\,\mathrm{DW}}_{f,\,\mathrm{PS}} \left(w_0 m_{\mathrm{PS}}\right)^2
+ Q^{\,\mathrm{DW}}_{f,\,\mathrm{PS}} \left(w_0 m_{\mathrm{PS}}\right)^4
\Big)
+ R^{\,\mathrm{DW}}_{f,\,\mathrm{PS}} (a/w_0)^2.
\end{gather}
We checked that the inclusion of additional terms, suppressed by higher powers of the expansion parameters, does not affect the extrapolation results in the region of parameter space of interest, and that the coefficients of additional, odd powers of $(a/w_0)$ are compatible with zero.

\begin{table}[t]
\centering
\caption{Result of the fits of $\left(w_0 m_{\mathrm{V}}\right)^2$ and $\left(w_0 f_{\mathrm{PS}}\right)^2$ to Eqs.~(\ref{eq:chiPT_W_mv}) and~(\ref{eq:chiPT_W_fps}), respectively, for Wilson fermions, and  Eqs.~(\ref{eq:eq:chiPT_DWF_mv}) and~(\ref{eq:eq:chiPT_DWF_fps}), for the MDWF formulation employed in this study. The quality of each fit is quantified by the corresponding reduced chi square, $\chi^2/N_{\rm d.o.f.}$. \\}
\begin{tabular}{|lcccccccc|}
\hline\hline
Discretisation & $(w_0 m^{\chi}_{\rm V})^2$ & $L_{m,\rm V}$ & $Q_{m,\rm V}$ & $W_{m,\rm V}$ & $R_{m,\rm V}$ & $C_{m,\rm V}$ & $\chi^2/\mathrm{d.o.f.}$ & $N_{\rm d.o.f.}$ \\
\hline
Wilson & 0.370(15) & 2.93(33) & -0.6(5) & -0.334(29) & 0.125(23) & -0.12(7) & 0.75 & 13 \\
MDWF & 0.40(5) & 2.78(86) & -0.5(6) & — & -0.19(6) & — & 0.32 & 11 \\
\hline\hline
Discretisation & $(w_0 f^{\chi}_{\rm PS})^2$ & $L_{f,\rm PS}$ & $Q_{f,\rm PS}$ & $W_{f,\rm PS}$ & $R_{f,\rm PS}$ & $C_{f,\rm PS}$ & $\chi^2/\mathrm{d.o.f.}$ & $N_{\rm d.o.f.}$ \\
\hline
Wilson & 0.0056(7) & 4.04(1.27) & -1.4(2.0) & -0.0039(13) & 0.0021(11) & -0.000(4) & 0.55 & 13 \\
MDWF & 0.0054(8) & 4.84(1.40) & -2.2(1.1) & — & 0.0000(13) & — & 0.45 & 11 \\
\hline\hline
\end{tabular}

\label{tab:chiPT_extrapolation}
\end{table}

The resulting best-fit values of the constants obtained from fitting the two data sets using Eqs.~(\ref{eq:chiPT_W_mv})-(\ref{eq:eq:chiPT_DWF_fps}) are summarised in Tab.~\ref{tab:chiPT_extrapolation}. We characterise the results by means of the error of the individual fitting parameter, yet we alert the reader that there are significant off-diagonal terms in the covariance matrix, which we report for the MDWF fits in  Appendix~\ref{Sec:cov}.  In addition, we perform a bootstrap analysis as a consistency check of the fit results and their associated uncertainties, with the corresponding results for the fit parameters are reported in Appendix~\ref{Sec:boot}. Systematic effects are assessed by varying the number of terms included in the fitting ansatz until stable results are obtained, at which point we estimate them to be smaller than the statistical ones. 

The continuum extrapolated values obtained with the two data sets, obtained with Wilson and MDWF formulations, are compatible with one another,  within uncertainties, as can be seen in Figs.~\ref{fig:mv_mps} and~\ref{fig:fmps_mps}. Furthermore, for finite lattice spacing the MDWF formulation exhibits a substantially milder sensitivity to discretisation effects. In particular, in the MDWF fits, the leading
lattice-spacing dependence is parametrised by the quadratic coefficients $R^{\rm DW}_{m,\rm V}$ and $R^{\rm DW}_{f,\rm PS}$ which is compatible with zero within uncertainties. By contrast, the Wilson fermion data sets display a clearly resolved $\mathcal{O}(a)$ dependence, parametrised by
$W^{\rm W}_{m,\rm V}$ and $W^{\rm W}_{f,\rm PS}$, and require additional $\mathcal{O}(a^2)$ and mixed mass--lattice-spacing corrections to obtain a satisfactory description. This is consistent with the reduced lattice artefacts observed for the MDWF formulation, for which data obtained at different values of $\beta$ are correspondingly closer to the continuum extrapolation. The analysis summarised and displayed in Figs.~\ref{fig:mv_mps} and~\ref{fig:fmps_mps} visually suggests that the adoption of the MDWF formulation might allow us to use only a small number of choices of $\beta$ to obtain reliable extrapolations to the continuum. This behaviour is in sharp contrast to what was found with Wilson fermions.

The striking effect of the improvement built into the MDWF formulation is expected, as discretisation effects appear at $\mathcal{O}(a^2)$, in contrast to the $\mathcal{O}(a)$ effects associated with unimproved Wilson fermions. The fact that, for the two data sets compared there, such effects result in an order-of-magnitude improvement in the approach to the continuum limit, depends also on the range of lattice parameter space we explored. The values of the lattice couplings, $\beta$,  required to use the MDWF formulation are comparatively large, and so are our choices for the value of the bare mass, $am_0$, and the associated mass of the ${\rm PS}$ mesons. Extending the study towards regions of parameter space with lighter fermions would therefore be particularly interesting, both to test whether the mild sensitivity to discretisation effects persists, and to better constrain the approach to the massless limit. At smaller bare masses, maintaining a sufficiently good degree of global symmetry may require increasing the extent of the fifth dimension, $N_5$, in order to keep the residual mass under control. For instance, in Sect.~\ref{sec:dwf_par_optimization} we have already reported the results of tuning the parameters for the choice of $N_5=12$ which could provide a suitable starting point for such an extension toward lower masses.

\section{Summary and outlook} 
\label{Sec:outlook}

We presented  a new numerical lattice study of the $Sp(4)$ gauge theory coupled to $N_{\rm f}=2$ Dirac fermions transforming in the fundamental representation of the group.  We applied the MDWF formulation, optimised for this theory, adopted the Wilson flow as a scale-setting procedure, and performed our measurements on a set of ensembles chosen to allow us to extrapolate our results towards the continuum limit, by means of a simplified analysis inspired by W$\chi$PT that includes ${\cal O}(a^2)$ corrections.  We focused our attention on the measurement of masses of the PNGB and lightest spin-1 mesons, and (renormalised) decay constants of the PNGBs, which are of interest for model-building purposes, both in the context of CHMs and SIMPs.  In the whole range of parameter space we analysed all the bound states of interest are stable.

Our main physics results can be condensed into two groups. First, of interest to the phenomenology community,  is the publication of the most precise available measurements to date of the properties (masses and decay constants) of the lightest spin-0 and spin-1 flavoured mesons, superseding the systematic precision of existing measurements in the literature. Compared to earlier publications~\cite{Bennett:2019jzz,TELOS:2026alk}, the main qualitative difference resides  in the improved control over  lattice artefacts due to finite spacing effects, which ultimately descends from the adoption of the MDWF formulation, supplemented by PV fields, and its built-in, ${\cal O}(a^2)$, improvement.

The second group of results is of more direct interest to lattice practitioners. The comparison we performed shows that our continuum extrapolation of the MDWF results are compatible with those in Ref.~\cite{Bennett:2019jzz,TELOS:2026alk}, based on using Wilson fermions. Yet, our new measurements are close enough to the continuum limit that the extrapolation itself might be unnecessary, finite spacing effects being strongly suppressed.  At the price of performing an initial tuning exercise on the algorithmic parameters entering the MDWF formulation---which is a theory-dependent process---we demonstrated that it is possible to perform high-precision, extensive lattice studies by using publicly available code, and moderate amounts of computational resources. 

Our collaboration plans to adapt the MDWF formulation to explore in the future more challenging regions of parameter space,  and our present results suggest that ambitious goals can be set for such a programme. We are going to test our approach in  the low-mass range of the $Sp(4)$, $N_{\rm f}=2$ theory, accessing the regime in which the spin-1 resonances are close to threshold, or even unstable. This should allow us also to extract information about  3-point and 4-point correlation functions, that enter such observables as the decay rate as well as the scattering amplitude of PNGBs---see Ref.~\cite{TELOS:2026alk}. We will also consider other theories. We can revisit the status of the $Sp(4)$ theory with $N_{\rm as}=3$ fermions transforming in the antisymmetric representation, and better characterise its continuum limit extrapolation, with respect to the current state-of-the-art, based on Wilson fermions~\cite{Bennett:2024tex}. On the algorithm side, it might prove insightful to study the low-lying eigenvalues and their distribution for M\"{o}bius fermions in $Sp(4)$ gauge theories. Last but not least, we will attempt to apply this technology to the $Sp(4)$ theory proposed in Ref.~\cite{Barnard:2013zea}, which has $N_{\rm f}=2$ and $N_{\rm as}=3$ fermions, and which is known to have a much richer dynamics and physics potential. Doing so would require overcoming the technical difficulties that limit current studies---with the use of Wilson fermions,  only a small number of ensembles have been generated so far, with comparatively large masses for both fermion species, and on comparatively  coarse lattices~\cite{Bennett:2022yfa,Bennett:2024cqv,Bennett:2024wda}---and explore larger regions of parameter space of this theory, that has a rich potential for phenomenological applications.

While these are all exciting opportunities for future research, they represent just  the tip of the iceberg of what can be achieved with the technology we   tested for the purposes of this paper. The encouraging results we obtained with the MDWF formulation in the study of the relatively simple $Sp(4)$, $N_{\rm f}=2$ theory can be replicated in other classes of theories within the BSM lattice context. As a case in point, we mention the study of $SU(2)$ theories with fermion matter field content transforming in the adjoint representation. This class of  theories has the potential to provide a non-trivial  realisation of near-conformal dynamics at long distances, and lattice studies with domain-wall fermions might be able to characterise them successfully~\cite{Athenodorou:2014eua,Athenodorou:2021wom,Athenodorou:2024rba}. Similar arguments can be applied to a plethora of gauge theories of interest for BSM physics, even beyond the compositeness framework---see the review in Ref.~\cite{Rummukainen:2022ekh} and references therein for a discussion of the current state of the art in the field.

\begin{acknowledgments}

We kindly thank Chulwoo Jung for his help and discussions on domain wall fermions.

The work of E.~B. is supported by the STFC Research Software Engineering Fellowship EP/V052489/1 and has been supported in part by UKRI via the Computational Science Centre for Research Communities (CoSeC), under Grant No.~UKRI497. The work of A.~V.-P., B.~L., E.~B., G.~S., M.~P., and N.~F. is supported by the STFC Consolidated Grant No.~ST/X000648/1. B.~L., L.~D.~D., and M.~P. received funding from the ERC under the European Union's Horizon 2020 research and innovation program under Grant Agreement No.~813942. L.~D.~D. is also supported by an STFC Consolidated Grant (ST/T000600/1, ST/X000494/1). The work of N.~F. has been supported by the STFC Doctoral Training Grant No.~ST/X508834/1. D.~K.~H. is supported by Basic Science Research Program through NRF funded by the Ministry of Education (NRF-2017R1D1A1B06033701). J.-W.~L. is supported by IBS under the project code, IBS-R018-D1. C.-J.~D.~L. acknowledges support from NSTC Taiwan through grant number 112-2112-M-A49-021-MY3. C.-J.~D.~L. is also supported by the Taiwanese MoST Grant No.~109-2112-M-009-006-MY3. C.-J.~D.~L. is also supported by Grants No.~112-2639-M-002-006-ASP and No.~113-2119-M-007-013-. B.~L. is supported in part by the STFC Consolidated Grant No.~ST/X00063X/1. The work of A.~V.-P. and G.~S. is supported by a studentship funded by the Faculty of Science and Engineering, Swansea University, and by the School of Physics and Astronomy, The University of Edinburgh. The work of D.~V. is supported by the STFC Consolidated Grant No.~ST/X000680/1.

This work used the DiRAC Extreme Scaling service (Tursa) at the University of Edinburgh, managed by the EPCC on behalf of the STFC DiRAC HPC Facility (www.dirac.ac.uk). The DiRAC service at Edinburgh was funded by BEIS, UKRI and STFC capital funding and STFC operations grants. DiRAC is part of the UKRI Digital Research Infrastructure.

Numerical computations were performed using Supercomputing Wales SUNBIRD at Swansea University and AccelerateAI at Swansea University.

Supercomputing Wales and AccelerateAI are supported by the European Regional Development Fund via Welsh Government.

\vspace{1.0cm}
{\bf Research Software Availability statement}---Raw data were generated using Grid~\cite{Boyle:2015tjk,Boyle:2016lbp,Yamaguchi:2022feu,Grid-repo} commit \texttt{48bab11} and Hadrons~\cite{antonin_portelli_2023_8023716,Hadrons-repo} commit \texttt{28d2806}. To better cope with large ensemble runs we used \texttt{hmcdj}~\cite{hmcdj}, which is wrapper built on top of Grid. The workflow used to analyse these data is available at Ref.~\cite{workflowrelease}. It is implemented using Snakemake~\cite{Molder:2021} and makes use of the Python libraries \texttt{gvar}~\cite{Lepage_gvar}, \texttt{lsqfit}~\cite{Lepage_lsqfit}, and \texttt{corrfitter}~\cite{Lepage_corrfitter}.

\vspace{1.0cm}

{\bf Research Data Availability Statement}---The raw data generated in support of this work, and processed data derived from it, are available in machine-readable format at Ref.~\cite{datarelease}. See also Ref.~\cite{Bennett:2025neg}, for a description of our approach to reproducibility and open science. 

\vspace{1.0cm}

{\bf Open Access Statement}---For the purpose of open access, the authors have applied a Creative Commons  Attribution (CC BY) license to any Author Accepted Manuscript version arising.

\end{acknowledgments}

\begin{appendix}

\section{Gauge invariance tests on Grid}
\label{Sec:gauge_invariance}

To the best of our knowledge, this is the first lattice study to deploy the MDWF formulation in $Sp(2N)$ lattice gauge theories, hence we performed a number of checks to validate the algorithms, aimed at testing that the implementation is consistent, in particular in view of the fact that the Grid framework~\cite{Boyle:2015tjk,Boyle:2016lbp,Yamaguchi:2022feu}  has only recently been supplemented with the necessary tools for treating $Sp(4)$ theories~\cite{Bennett:2023gbe}. In this brief Appendix, we report some of the results of such tests, with particular emphasis on gauge invariance.

Our first verification step pertains to the Fourier transform of the Shamir implementation of the kernel of domain-wall fermions, $D_{\rm S}$, introduced in Eq.~(\ref{eq:dwf_kernel}). The Fourier transform of $D_{\rm S}$ over the four spatial dimensions (excluding the fifth dimension, along which we label the sites as $s,\,s^{\prime} = 0, \dots, N_5 - 1$) is denoted as $\tilde{D}_{\rm S}$, and given by the following expression~\cite{Shamir:1993zy}:
\begin{equation}
\label{eq:fourier_dwf_formula}
\begin{gathered}
\tilde{D}_{\rm S}\left(s, s^{\prime}, p\right) = P_R \delta_{s+1, s^{\prime}} + P_L \delta_{s-1, s^{\prime}} - \left[1 - m_5 + \frac{1}{a} \sum_{\mu = 0}^{3} \left(1 - \cos(p_\mu a)\right) + \frac{i}{a} \sum_{\mu = 0}^{3} \gamma_\mu \sin(p_\mu a) \right] \delta_{s, s^{\prime}} \,.
\end{gathered}
\end{equation}
To test this expression, we used the Fast Fourier Transform (FFT) algorithm, which is integrated within the Grid framework through the FFTW library~\cite{Frigo:2005zln}, to compute this quantity numerically and compare it to the analytical expression in Eq.~(\ref{eq:fourier_dwf_formula}). We adopted the unitary gauge (\texttt{unit}), and the free-field Dirac kernel. Furthermore, we applied random gauge transformations (\texttt{xform}) to the gauge links and recomputed the Fourier transform using FFT. The results for the \texttt{unit} and \texttt{xform} cases in Eq.~\eqref{eq:fourier_dwf_formula} should match within machine precision. This can be confirmed by calculating the difference between the two results and evaluating the squared norm of their difference, defined as $\text{sqnorm}(A,B) \equiv |D^{(A)}_{\rm S} - D_{\rm S}^{(B)}|^2 $, where $|\dots|^2$ represents the inner product summed over the indices, $\sum_{s,s^\prime, p}$. The results, presented in Tab.~\ref{table:cross_check_fftw}, provide a stringent implementation check at the level of individual gauge links.

\begin{table}[t]
\centering
\caption{Representative result of the test of the Shamir kernel, $\tilde{D}_{\rm S}$, as in Eq.~(\ref{eq:fourier_dwf_formula}). Squared norm of the difference ($\text{sqnorm}$) between Dirac kernels computed from unitary gauge (\texttt{unit}), random gauge transformation (\texttt{xform}), and analytical momentum-space results (\texttt{textbook}). The discrepancies are small, comparable with the numerical machine precision. The numerical calculation uses the FFTW library, for a choice of parameters that is discussed in the main text. \\}
\begin{tabular}{ |c|c|c| }
\hline \hline
$~~~~$sqnorm(\texttt{unit},  \texttt{textbook}) $~~~~$&$~~~~$ sqnorm(\texttt{xform}, \texttt{textbook}) $~~~~$& $~~~~$sqnorm(\texttt{unit},  \texttt{xform}) $~~~~$   \\
\hline
$1.89940 \times 10^{-22}$ & $9.91608 \times 10^{-8}$ & $9.91608 \times 10^{-8}$ \\
\hline \hline
\end{tabular}

\label{table:cross_check_fftw}
\caption{Representative result of the tests of the free propagator. Squared norm of the differences ($\text{sqnorm}$) between free field propagators obtained from a unitary gauge configuration (\texttt{unit}), a random gauge transformation (\texttt{xform}), and the CG-based result (\texttt{CG}). The discrepancies are small, comparable with the numerical machine precision. The numerical calculation uses the FFTW library, for a choice of parameters that is discussed in the main text.  \\}
\begin{tabular}{ |c|c|c| }
\hline \hline
sqnorm(\texttt{unit}, CG) & sqnorm(\texttt{xform}, CG) & sqnorm(\texttt{unit}, \texttt{xform})\\
\hline
$~~~~$$1.00866 \times 10^{-19}$ $~~~~$&$~~~~$ $9.80618 \times 10^{-14}$ $~~~~$&$~~~~$ $9.80617 \times 10^{-14}$$~~~~$  \\
\hline \hline
\end{tabular}

\label{table:cross_check_fftw2}
\end{table}

Next, we proceeded to validate the implementation of the free propagator. To this aim, we analytically inverted Eq.~\eqref{eq:fourier_dwf_formula}, followed by the application of a backward FFT step, to compute the position-space free propagator. The same result was expected from the conjugate gradient (CG) algorithm~\cite{saad2003iterative}. These checks were conducted both in a configuration obtained in the unitary gauge (\texttt{unit})  as well as in one obtained after a random gauge transformation was applied to the unitary gauge configuration (\texttt{xform}). We computed the squared norm of the difference between propagators as $\text{sqnorm}(A,B) \equiv |S^{(A)}_{\rm S} - S_{\rm S}^{(B)}|^2 $, where the inner product is summed over the indices, $\sum_{s,s^\prime, x}$. The results are displayed in Tab.~\ref{table:cross_check_fftw2}.

The numerical tests reported here have been generated on a test configuration with lattice volume $\tilde{V}=32\times16^3$, fifth-dimensional extent $N_5=32$, bare fermion mass $am_0=0.01$, and domain-wall height $a m_5=1.2$, and are carried out using the aforementioned FFTW library. These tests allow us to verify both the implementation of the Dirac kernel and the associated propagator, including the treatment of non-trivial gauge links, as well as the correct functionality of the CG solver for domain-wall fermions. Overall, they provide a consistency check that the numerical implementation reproduces the expected momentum-space results.

\section{Finite size effects}
\label{Sec:finite_size_effects}

\begin{table}[t]
\centering
\caption{Characterisation of the ensembles generated to study finite-volume effects in the $Sp(4)$ lattice gauge theory coupled to $N_{\rm f} = 2$ Dirac fermions transforming in the fundamental representation, realised with the MDWF formalism. We used the optimised choices of MDWF parameters discussed in Sec.~\ref{sec:dwf_par_optimization}, $am_5 = 1.8, \, a_5 / a= 1.0,  \, am_{\rm PV} = 1.0$ for every ensemble, and $\alpha=1.75, \, 1.625, \, 1.5$, for ensembles with $\beta=7.4, \, 7.5, \, 7.6$, respectively. For each ensemble, we tabulate the lattice coupling, $\beta$, the bare mass, $a m_0$, the number of sites in the lattice directions, $N_5$, $N_t$, and $N_s$,  the MDWF parameters, $\alpha$, $a_5/a$, and $a m_5$,  the residual mass, $am_{\rm res}$, and the mass of the lightest ${\rm PS}$ mesons, $am_{\rm PS}$, expressed in units of the lattice spacing. The uncertainties quoted are purely statistical. \\}
\begin{tabular}{|l|c|c|c|c|c|c|c|c|c|c|}
\hline\hline
Ensemble & $\beta$ & $am_0$ & $N_t$ & $N_s$ & $N_5$ & $\alpha$ & $a_5/a$ & $am_5$ & $am_{\rm res}$ & $am_{\rm PS}$ \\
\hline
MB74M6Sc3 & 7.4 & 0.06 & 32 & 16 & 8 & 1.75 & 1 & 1.8 & 0.00278(4) & 0.4780(23) \\
MB74M2Sc3 & 7.4 & 0.02 & 32 & 16 & 8 & 1.75 & 1 & 1.8 & 0.00271(5) & 0.284(8) \\
MB76M6Sc2 & 7.6 & 0.06 & 32 & 16 & 8 & 1.5 & 1 & 1.8 & 0.001476(14) & 0.4273(28) \\
MB76M2Sc2 & 7.6 & 0.02 & 32 & 16 & 8 & 1.5 & 1 & 1.8 & 0.001462(17) & 0.279(11) \\
MB746M6Ns12 & 7.4 & 0.06 & 24 & 12 & 8 & 1.75 & 1 & 1.8 & 0.00282(8) & 0.511(5) \\
MB746M6Ns20 & 7.4 & 0.06 & 40 & 20 & 8 & 1.75 & 1 & 1.8 & 0.00271(4) & 0.4728(22) \\
MB74M2Ns12 & 7.4 & 0.02 & 24 & 12 & 8 & 1.75 & 1 & 1.8 & 0.00242(10) & 0.351(29) \\
MB74M2Ns20 & 7.4 & 0.02 & 40 & 20 & 8 & 1.75 & 1 & 1.8 & 0.00274(5) & 0.272(4) \\
MB74M2Ns32 & 7.4 & 0.02 & 64 & 32 & 8 & 1.75 & 1 & 1.8 & 0.002711(22) & 0.272(6) \\
MB76M6Ns12 & 7.6 & 0.06 & 24 & 12 & 8 & 1.5 & 1 & 1.8 & 0.00157(4) & 0.507(8) \\
MB76M6Ns20 & 7.6 & 0.06 & 40 & 20 & 8 & 1.5 & 1 & 1.8 & 0.001482(9) & 0.4173(16) \\
MB76M6Ns28 & 7.6 & 0.06 & 56 & 28 & 8 & 1.5 & 1 & 1.8 & 0.001489(9) & 0.4149(9) \\
MB76M2Ns12 & 7.6 & 0.02 & 24 & 12 & 8 & 1.5 & 1 & 1.8 & 0.001505(28) & 0.482(16) \\
MB76M2Ns20 & 7.6 & 0.02 & 40 & 20 & 8 & 1.5 & 1 & 1.8 & 0.001431(8) & 0.2452(21) \\
MB76M2Ns28 & 7.6 & 0.02 & 56 & 28 & 8 & 1.5 & 1 & 1.8 & 0.001454(16) & 0.2318(22) \\
MB76M2Ns32 & 7.6 & 0.02 & 64 & 32 & 8 & 1.5 & 1 & 1.8 & 0.001474(8) & 0.2321(10) \\
MB74M2Sp & 7.4 & 0.02 & 48 & 24 & 8 & 1.75 & 1 & 1.8 & 0.002682(9) & 0.2717(13) \\
MB74M6Sp & 7.4 & 0.06 & 48 & 24 & 8 & 1.75 & 1 & 1.8 & 0.002742(8) & 0.4733(6) \\
MB76M2Sp & 7.6 & 0.02 & 48 & 24 & 8 & 1.5 & 1 & 1.8 & 0.001470(8) & 0.233(3) \\
MB76M6Sp & 7.6 & 0.06 & 48 & 24 & 8 & 1.5 & 1 & 1.8 & 0.001495(3) & 0.4150(9) \\
\hline\hline
\end{tabular}

\label{tab:finite_volume}
\end{table}

    \begin{figure}
\centering
\begin{tabular}{c}
\includegraphics[width=0.48\textwidth]{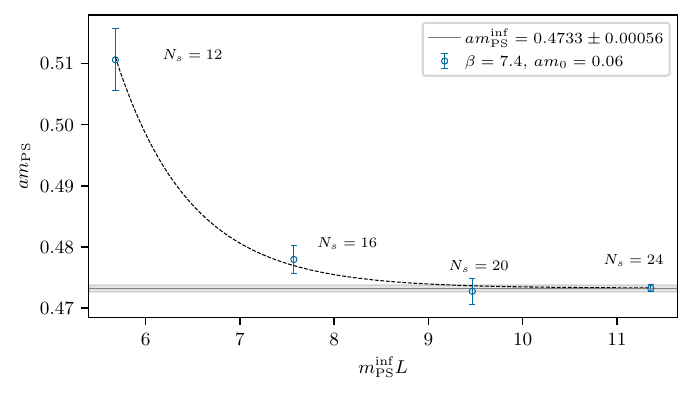} \hspace{0.0cm}
\includegraphics[width=0.48\textwidth]{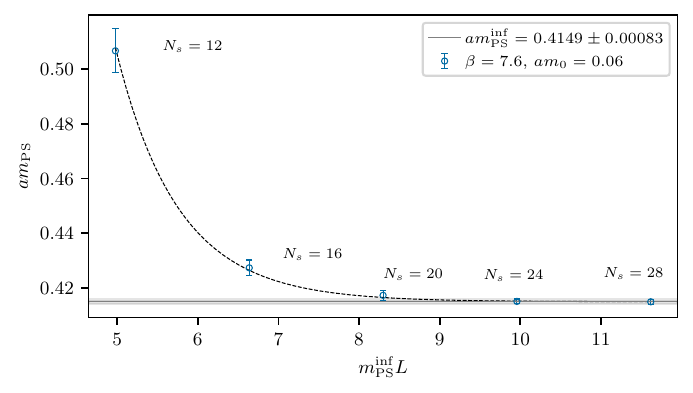} \hspace{0.0cm} \\
\multicolumn{1}{c}{
\includegraphics[width=0.48\textwidth]{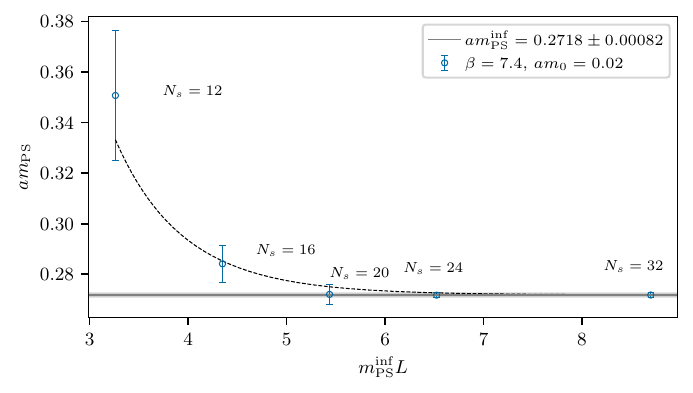} \hspace{0.0cm}
\includegraphics[width=0.48\textwidth]{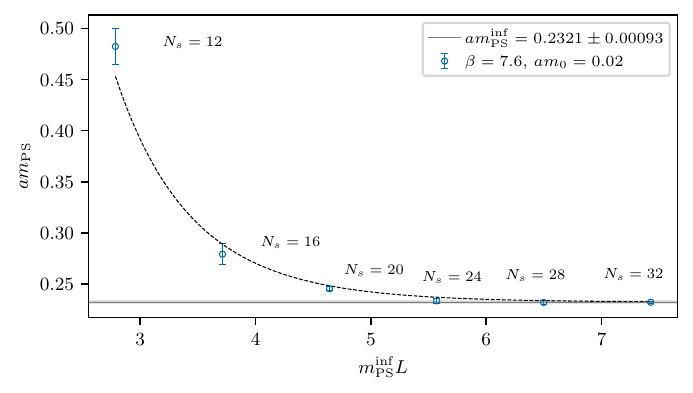}
} \\
\end{tabular}
\caption{ \label{fig:finite_size}     Representative examples of results of the finite-size study in the $Sp(4)$ gauge theory coupled to $N_{\rm f}=2$ Dirac fermions, treated with the MDWF formulation. The ensembles considered have fifth-dimension extent $N_5 = 8$, lattice coupling $\beta = 7.4$ (left panels), or $\beta= 7.6$ (right),   bare mass  $am_0 = 0.06$ (top panels), or $a m_0=  0.02$. All ensembles are isotropic in the space directions, with $N_s$ sites, and the time direction is chosen to have $N_t=2N_s$ sites (see also Table~\ref{tab:finite_volume}). We show the dependence of the mass of the PNGBs, $a m_{\rm PS}$, on the spatial extent of the lattice size, $N_s=L/a$, by looking at the dependence of the spectroscopy measurements on the combination $m_{\rm PS}^{\rm inf} L$.  The plots show also a best-fit curve, based on Eq.~(\ref{eq:finite_size_effects_eq}), and the asymptotic value of the mass of the ${\rm PS}$ meson, denoted $am_{\rm PS}^{\rm inf}$, for reference.}
\end{figure}

Numerical results obtained on finite lattices are always affected by systematic uncertainties arising from the finite volume of the lattice. In confining gauge theories, finite-volume effects on masses and decay constants are expected to be exponentially suppressed as a function of the spatial lattice extent, $L=a N_s$, provided this extent is significantly larger than the inverse of the longest physical length in the theory, which usually is the Compton wavelength associated with the ${\rm PS}$ meson. This can be written as the requirement that  $m_{\rm PS} L \gg 1$. The magnitude of these finite-volume effects can be systematically estimated using chiral perturbation theory---see for example Refs.~\cite{RBC-UKQCD:2008mhs,Golterman:2009kw}. For the mass of the ${\rm PS}$ mesons, this is expected to take the form:
\begin{equation}
m_{\rm PS}(L) = m_{\rm PS}^{\rm inf} \bigg( 1 +
\frac{\left(m_{\rm PS}^{\rm inf}\right)^2}
{\left(4\pi f_{\rm PS}^{\rm inf}\right)^2}
\frac{24}{\left(m_{\rm PS}^{\rm inf} L\right)^{3/2}}
\sqrt{\frac{\pi}{2}}\,
e^{-m_{\rm PS}^{\rm inf} L}
+ O\left(e^{-\sqrt{2}\,m_{\rm PS}^{\rm inf} L}\right)
\bigg)\,.
\end{equation}
where $m_{\rm PS}^{\rm inf}$ and $f_{\rm PS}^{\rm inf}$ denote the
infinite-volume pseudoscalar mass and decay constant, respectively.
Since our measurements are obtained from ensembles that are not
particularly close to the massless limit, we find it convenient, for
the purpose of assessing finite-volume effects, to introduce a
model-dependent constant, $\mathcal{A}$, that parametrises the effects
of discretisation and finite-mass artefacts, and adopt the following
relation:
\begin{equation}
\label{eq:finite_size_effects_eq}
m_{\rm PS}(L) = m_{\rm PS}^{\rm inf} \left[ 1 + \mathcal{A}
\left(\dfrac{1}{(m_{\rm PS}^{\rm inf}L)^{3/2}}
e^{-m_{\rm PS}^{\rm inf} L} \right)\right] \,,
\end{equation}
where we retain only the leading term in the large-$m_{\rm PS}^{\rm inf}L$
expansion.

The ensembles generated to study the size of  finite-volume effects are characterised in Tab.~\ref{tab:finite_volume}, and representative examples are plotted in Fig.~\ref{fig:finite_size}, where the mass of the ${\rm PS}$ mesons is analysed as a function of the spatial extent, $L = N_s a$.  We present results for different lattice extents and compare choices of  the bare mass, $am_0 = 0.02, \, 0.06$, and lattice coupling, $\beta = 7.4, \, 7.6$.  As shown in the plots, finite-volume effects are negligible (smaller than the statistical uncertainties) for $m_{\rm PS}^{\rm inf} L \gsim \FveLargestMass$ for $am_0=0.06$ and $m_{\rm PS}^{\rm inf} L \gsim \FveSmallestMass$ for $am_0=0.02$.  We fit the numerical data using Eq.~(\ref{eq:finite_size_effects_eq}). In all the examples we show, the resulting infinite-volume extrapolations, computed at different values of $\beta$ and $am_0$, are compatible, within statistical uncertainties, with the mass of the ${\rm PS}$ mesons obtained at $N_s=24$. We therefore adopt $N_s=24$ and $N_t=48$ for the spectral study performed in the main body of the paper, as by doing so finite-volume effects can be considered negligible compared to the statistical uncertainties.


\section{Fitting details}
\label{Sec:fitting_details}
In this appendix, we describe explicitly the fitting method used to extract the residual mass, $am_{\rm res}$, and, similarly, the renormalisation constant $Z_A$. We then provide additional technical details about the bootstrap procedure used in the continuum extrapolation of our lattice measurements and report the covariance matrices of the fitting parameters entering our numerical analysis.

\subsection{Residual mass}

All correlation functions entering the analysis are constructed from gauge configurations that are treated as statistically uncorrelated. For the determination of the residual mass, we monitor both the integrated autocorrelation time of the plaquette, $\tau_{\rm int}^{\rm plaq.}$, and that of the residual-mass estimator, $\tau_{\rm int}^{am_{\rm res}}$, evaluated at the first time slice of the plateau region used in the fit. The configurations are then selected from the Monte Carlo history with a separation of approximately twice the larger of these two autocorrelation times.

We therefore assume that we have recorded $N_{\rm cfg}$ thermalised and effectively uncorrelated gauge configurations and extract the residual mass by performing a correlated fit to a constant of the ratio of the folded correlators $C_{\rm PS,PS}(t)$, defined in Eq.~\ref{eq:corr_1}, and $C_{\rm PS,PS,q}(t)$,
\begin{equation}
    C_{\rm PS,\,PS,q}(t) =
    \sum_{\vec{x}} \left\langle
    O_{\rm PS}(\vec{x},t)\,
    O_{\rm PS,q}^\dagger(\vec{0},0)
    \right\rangle\,.
\end{equation}
For each configuration, labelled by $k=1,\dots,N_{\rm cfg}$, we fold the two correlators about the temporal midpoint:
\begin{align}
    \tilde{C}^k_{\rm PS,\,PS}(t) &=
    \frac{1}{2}\Big(C^k_{\rm PS,\,PS}(t)
    + C^k_{\rm PS,\,PS}(N_t-t)\Big)\,, \\
    \tilde{C}^k_{\rm PS,\,PS,q}(t) &=
    \frac{1}{2}\Big(C^k_{\rm PS,\,PS,q}(t)
    + C^k_{\rm PS,\,PS,q}(N_t-t)\Big)\,,
\end{align}
where $t=0,1,\dots,N_t/2$. The ratio entering the fit is constructed from the configuration-averaged folded correlators,
\begin{equation}
    \bar{R}(t)=
    \frac{\frac{1}{N_{\rm cfg}}\sum_{k=1}^{N_{\rm cfg}}
    \tilde{C}^k_{\rm PS,\,PS,q}(t)}
    {\frac{1}{N_{\rm cfg}}\sum_{k=1}^{N_{\rm cfg}}
    \tilde{C}^k_{\rm PS,\,PS}(t)}\,.
\end{equation}
We then identify a plateau window, and denote its time slices as $\{\tau_1,\dots,\tau_{n_p}\}$. We use it to introduce  the data vector
\begin{equation}
    \mathbf{y}
    \equiv
    \Big(
    \bar{R}(\tau_{1}),
    \bar{R}(\tau_{2}),
    \dots,
    \bar{R}(\tau_{n_p})
    \Big)\,,
\end{equation}
given by the mean sample of the ratio,  $   \bar{R}(t)$, evaluated at $t=\tau_i$.

Statistical uncertainties are estimated via bootstrap resampling of the gauge configurations. For each bootstrap replica, label by  $b=1,\,\dots,\,N_b$, we draw, with replacement, $N_{\rm cfg}$ configurations from the original ensemble. For each replica we compute the quantity
\begin{equation}
    \bar{R}^{(b)}(\tau)=\frac{\frac{1}{N_{\rm cfg}} \sum_{k=1}^{N_{\rm cfg}} \tilde{C}_{ \mathrm{PS,\, PS,q}}^{b(k)}(\tau)}{ \frac{1}{N_{\rm cfg}} \sum_{k=1}^{N_{\rm cfg}} C_{\mathrm{PS,\,PS}}^{b(k)}(\tau)},
\end{equation}
where $b(k)$ denotes the index of the configuration drawn at position $k$ in replica $b$. We also define the bootstrap data vectors,   $ \mathbf{y}^{(b)}$,  and the mean bootstrapped data vector, $  \bar{\mathbf{y}}^{(b)}$, as
\begin{equation}
    \mathbf{y}^{(b)} = \Big(
    \bar{R}^{(b)}(\tau_{1}),
    \bar{R}^{(b)}(\tau_{2}),
    \dots, \bar{R}^{(b)}(\tau_{n_p})
    \Big)\,, \qquad 
    \bar{\mathbf{y}}^{(b)} = \frac{1}{N_b}\sum_{b=1}^{N_b}\mathbf{y}^{(b)}\,.
\end{equation}
The covariance matrix of the estimator, $\bar{R}(t)$, is then computed as the bootstrap ensemble
\begin{equation}
    \Sigma_{ij}
    \equiv
    \mathrm{Cov}(y_i,y_j) =
    \frac{1}{N_b-1}
    \sum_{b=1}^{N_b}
    \Big(y^{(b)}_i-\bar{y}_i^{(b)}\Big)
    \Big(y^{(b)}_j-\bar{y}_j^{(b)}\Big)\,.
\end{equation}
We use it to perform a correlated constant fit with a constant function, $am_{\rm res}(t)=am_{\rm res}$, by minimising the chi square function: 
\begin{equation}
    \chi^2(am_{\rm res}) \equiv (\bar{\mathbf{y}}-m\mathbf{1})^{T}
    \Sigma^{-1}
    (\bar{\mathbf{y}}-m\mathbf{1})\,,
\end{equation}
where $\mathbf{1}=(1,\,1,\,\cdots,\,1)$, which yields
\begin{equation}
    am_{\rm res} = \frac{\mathbf{1}^{T}\Sigma^{-1}\bar{\mathbf{y}}}
         {\mathbf{1}^{T}\Sigma^{-1}\mathbf{1}}, \qquad
    \sigma_m= \sqrt{\frac{1}
    {\mathbf{1}^{T}\Sigma^{-1}\mathbf{1}} }\,.
\end{equation}

The same procedure is applied to the extraction of $Z_A$, which is obtained from the appropriate ratios of the correlation functions defined in Eq.~\eqref{eq:ren_const_dwfs}. Indeed, at sufficiently large Euclidean times, this ratio is expected to approach a plateau, and $Z_A$ can therefore be determined by adopting the aforementioned procedure.

\subsection{Correlations in the continuum extrapolations}
\label{Sec:cov}

Here, we report the covariance matrices associated with the continuum extrapolations of the ${\rm V}$-meson mass, $\mathrm{Cov}_{m_{\rm V}}$, and of the renormalised decay constant of the ${\rm PS}$ mesons, $\mathrm{Cov}_{f_{\rm PS}}$. These are obtained by computing the covariance matrix in the maximum likelihood analysis described in the body of the paper. For the ${\rm V}$-meson mass extrapolation, the covariance matrix, in the basis $\bigl((w_0 m_{\rm V}^{\chi})^2, L_{m,\rm V}, Q_{m,\rm V}, W_{m,\rm V}\bigr)$,  is
\begin{equation}
\label{eq:cov_mV}
\mathrm{Cov}_{m_{\rm V}} = \begin{pmatrix}
2.446\times 10^{-3} & -4.062\times 10^{-2} & 2.537\times 10^{-2} & -1.625\times 10^{-4} \\
-4.062\times 10^{-2} & 7.379\times 10^{-1} & -4.946\times 10^{-1} & -1.191\times 10^{-2} \\
2.537\times 10^{-2} & -4.946\times 10^{-1} & 3.51\times 10^{-1} & 1.423\times 10^{-2} \\
-1.625\times 10^{-4} & -1.191\times 10^{-2} & 1.423\times 10^{-2} & 3.808\times 10^{-3}
\end{pmatrix}

\end{equation}
For the renormalised decay constant of the ${\rm PS}$ mesons, written in the basis $\bigl((w_0 f_{\rm PS}^{\chi})^2, L_{f,\rm PS}, Q_{f,\rm PS}, W_{f,\rm PS}\bigr)$, we obtain
\begin{equation}
\label{eq:cov_fPS}
\mathrm{Cov}_{f_{\rm PS}} = \begin{pmatrix}
6.722\times 10^{-7} & -1.104\times 10^{-3} & 7.806\times 10^{-4} & -1.469\times 10^{-7} \\
-1.104\times 10^{-3} & 1.965 & -1.46 & -2.247\times 10^{-4} \\
7.806\times 10^{-4} & -1.46 & 1.126 & 3.355\times 10^{-4} \\
-1.469\times 10^{-7} & -2.247\times 10^{-4} & 3.355\times 10^{-4} & 1.676\times 10^{-6}
\end{pmatrix}

\end{equation}

By inspection of the off-diagonal entries one can identify combinations of the parameters that are strongly correlated. As can be seen from Eqs.~\eqref{eq:cov_mV} and~\eqref{eq:cov_fPS}, the most pronounced correlation is observed between the coefficients $L$ and $Q$, which control the leading and higher-order dependence on the ${\rm PS}$ mass in the corresponding fit ansatz. Their covariance is negative in both channels, indicating that variations of one coefficient can be partially compensated by variations of the other. This behaviour is therefore indicative of a partial degeneracy between these two contributions over the range of  masses included in the fits.

To quantify this effect independently of the overall normalisation of the parameter uncertainties, we also compute the corresponding correlation coefficients, which are defined in general as
\begin{equation}
\rho_{ij} \equiv
\frac{\mathrm{Cov}_{ij}}
{\sqrt{\mathrm{Cov}_{ii}\mathrm{Cov}_{jj}}}\,.
\end{equation}
We find
\begin{equation}
\label{eq:rho_LQ}
\rho_{L,Q}^{m_{\rm V}} = -0.972\qquad \rho_{L,Q}^{f_{\rm PS}} = -0.982
,
\end{equation}
The values reported in Eq.~\eqref{eq:rho_LQ} confirm that $L$ and $Q$ are strongly anti-correlated in both the extrapolation of the mass of the  ${\rm V}$ meson and of the decay constant of the PNGBs.  The covariance matrices should therefore be taken into account when interpreting the uncertainties and the numerical values of the individual fit coefficients, and the individual errors we quote for the coefficients are provided only for illustration purposes. As intermediate results of our continuum extrapolation, they should not be used without knowledge of the complete covariance matrix, which is included in our extrapolations.

\subsection{Continuum extrapolation with bootstrap}
\label{Sec:boot}

In this section, we describe the bootstrap procedure used as a consistency check of the fit results in Sec.~\ref{Sec:chiPT_masses_fpi}, and report the corresponding results for the fitting parameters and their uncertainties. We  start by recalling the expressions of the fitting ansatz employed for the continuum extrapolations of the vector mass, $w_0 m_{\rm V}$, and the pseudoscalar decay constant, $w_0 f_{\rm PS}$, expressed in units of the Wilson flow scale, $w_0$:
\begin{gather}\label{eq:fps_fit}
\left(w_0 f_{\mathrm{PS}}\right)^2_{\mathrm{DW}} =
\left(w_0 f^{\chi,\mathrm{DW}}_{\mathrm{PS}}\right)^2
\Bigg(
1
+ L^{\mathrm{DW}}_{f,\,\mathrm{PS}} \left(w_0 m_{\mathrm{PS}}\right)^2
+ Q^{\mathrm{DW}}_{f,\,\mathrm{PS}} \left(w_0 m_{\mathrm{PS}}\right)^4
\Bigg)
+ R^{\mathrm{DW}}_{f,\,\mathrm{PS}} (a/w_0)^2, \\
\label{eq:mps_fit}
\left(w_0 m_{\mathrm{V}}\right)^2_{\mathrm{DW}} =
\left(w_0 m^{\chi,\mathrm{DW}}_{\mathrm{V}}\right)^2
\Bigg(
1
+ L^{\mathrm{DW}}_{m,\,\mathrm{V}} \left(w_0 m_{\mathrm{PS}}\right)^2
+ Q^{\mathrm{DW}}_{m,\,\mathrm{V}} \left(w_0 m_{\mathrm{PS}}\right)^4
\Bigg)
+ R^{\mathrm{DW}}_{m,\,\mathrm{V}}(a/w_0)^2.
\end{gather}

\begin{table}[t]
\centering
\caption{Fit parameters and corresponding uncertainties obtained using the direct fit in Sec.~\ref{Sec:chiPT_masses_fpi}, together with the results of the bootstrap analysis, defined in Eq.~\ref{eq:bootstrap_param}, for the MDWF extrapolations of $\left(w_0 m_{\mathrm{V}}\right)^2$ and $\left(w_0 f_{\mathrm{PS}}\right)^2$. For the bootstrap analysis, we use 200 bootstrap samples. \\}
\begin{tabular}{|l@{\hspace{0.6em}}|@{\hspace{0.6em}}cccc@{\hspace{0.6em}}|@{\hspace{0.6em}}cccc|}
\hline\hline
Fit & $(w_0 m^{\chi}_{\rm V})^2$ & $L_{m,\rm V}$ & $Q_{m,\rm V}$ & $R_{m,\rm V}$ & $(w_0 f^{\chi}_{\rm PS})^2$ & $L_{f,\rm PS}$ & $Q_{f,\rm PS}$ & $R_{f,\rm PS}$ \\
\hline
Central values fit & 0.40(5) & 2.78(86) & -0.5(6) & -0.19(6) & 0.0054(8) & 4.84(1.40) & -2.2(1.1) & 0.0000(13) \\
Bootstrap & 0.40(5) & 2.89(92) & -0.6(6) & -0.19(6) & 0.0054(8) & 5.02(1.38) & -2.3(1.0) & 0.0000(12) \\
\hline\hline
\end{tabular}

\label{tab:mv_fps_extrapolation_bootstrap_compare_mdwf}
\end{table}

For each ensemble, labelled by the index $i$ in the equations below and corresponding to a given set of parameters $(\beta, am_0)$, we generate $N_b$ bootstrap replicas. For each bootstrap sample, $b=1,\,\cdots,\,N_b$, we compute the observables:
$
\left(am_{\rm PS},\, am_{\rm V},\,af_{\rm PS},\,w_0/a\right)_b
$.
From these, we construct the dimensionless quantities entering the fit:
\begin{equation}
\alpha_i^{(b)} = \left(a/w_0\right)_i^{2\,(b)}\,,
\qquad
x_i^{(b)} = \left(w_0 m_{\mathrm{PS}}\right)_i^{2\,(b)}\,,
\qquad
y_{f,i}^{(b)} = \left(w_0 f_{\mathrm{PS}}\right)_i^{2\,(b)}\,, 
\qquad
y_{m,i}^{(b)} = \left(w_0 m_{\mathrm{V}}\right)_i^{2\,(b)}\,.
\end{equation}
Since there are no correlations between different ensembles, the covariance matrix is diagonal:
\begin{equation}
\mathrm{Cov}(i,j) = \sigma_{\mathcal{E},i}^2 \, \delta_{ij},
\end{equation}
where $\sigma_{\mathcal{E},i}$ denotes the statistical uncertainty associated with the observable $\mathcal{E}$ for the $i$-th ensemble. For each bootstrap replica, $b$, we perform a fit by minimising the chi-square function
\begin{equation}
\chi_b^2(\theta^{(b)}) \equiv
\sum_{i=1}^n
\frac{
\left[
y_{\mathcal{E},i}^{(b)} - F_{\mathrm{fit}}\big(x_i^{(b)}, \alpha_i^{(b)}; \theta^{(b)}\big)
\right]^2
}{
\sigma_{\mathcal{E},i}^2
}\,,
\end{equation}
and  $\theta^{(b)}$  denotes the set of fit parameters defined in Eqs.~\eqref{eq:fps_fit} and~\eqref{eq:mps_fit}, for each bootstrap replica, $b$. The final estimate of the fit parameters  and associated uncertainties is obtained as
\begin{equation}\label{eq:bootstrap_param}
\overline{\theta} = \frac{1}{N_b} \sum_{b=1}^{N_b} \theta^{(b)},
\qquad
\sigma_\theta^2 =
\frac{1}{N_b - 1}
\sum_{b=1}^{N_b}
\left(
\theta^{(b)} - \overline{\theta}
\right)^2.
\end{equation}

In Tab.~\ref{tab:mv_fps_extrapolation_bootstrap_compare_mdwf}, we report the results obtained with the bootstrap procedure described above, using $N_b=200$ bootstrap samples, together with those obtained from the direct fit. All the fitting parameters are compatible within the quoted uncertainties. The largest differences are observed for the coefficients $L$ and $Q$, which are also found to be strongly correlated, as discussed in Appendix~\ref{Sec:cov}.

\end{appendix}

\bibliographystyle{JHEP} 
\bibliography{ref}

\end{document}